\documentclass[twocolumn]{aastex6}

\usepackage[title]{appendix}
\usepackage{setspace}
\usepackage{ulem}
\usepackage{color,ulem}
\usepackage{multirow}
\usepackage{amsfonts}
\usepackage{booktabs}
\usepackage{natbib}
\usepackage{lipsum}
\usepackage{subfig}
\include{epsf}
\include{epstopdf}
\DeclareGraphicsExtensions{.png,.jpg,.pdf,.eps}
\usepackage{graphicx}
\usepackage{amsmath}
\usepackage{upgreek}

\begin{document}

\newcommand{\apz}{\overline{\alpha_{\phi z}}}
\newcommand{\apr}{\overline{\alpha_{r\phi}}}
\newcommand{\tr{}}{\textcolor{red}{}}

\title{Depletion of Moderately Volatile Elements by Open-System loss \\ in the Early Solar Nebula}

\author{Debanjan Sengupta\altaffilmark{1,2,4}}
\author{Paul R. Estrada\altaffilmark{1}}
\author{Jeffrey N Cuzzi\altaffilmark{1}}
%\author{Uma Gorti\altaffilmark{1}}
\author{Munir Humayun\altaffilmark{3}}

\altaffiltext{1}{NASA Ames Research Center; Mail Stop 245-3, Moffett Field, CA 94035, USA}
\altaffiltext{2}{Universities Space Research Association; 7178 Columbia Gateway Drive, Columbia, MD 21046, USA}
\altaffiltext{3}{National High Magnetic Field Laboratory and Department of Earth, Ocean and Atmospheric Science, Florida State University, 1800 E. Paul Dirac Drive, Tallahassee, Florida, 32310, USA}
\altaffiltext{4}{NASA Postdoctoral Program (NPP) Fellow}

\begin{abstract}

Rocky bodies of the inner solar system display a systematic depletion of the ``Moderately Volatile Elements" (MVEs) that correlates with the expected condensation temperature of their likely host materials under protoplanetary nebula conditions. In this paper, we present and test a new hypothesis in which open system loss processes irreversibly remove vaporized MVEs from high nebula altitudes, leaving behind the more refractory solids residing much closer to the midplane. The MVEs irreversibly lost from the nebula through these open system loss processes are then simply unavailable for condensation onto planetesimals forming even much later, after the nebula cooled, overcoming a critical difficulty encountered by previous models of this type. We model open system loss processes operating at high nebula altitudes, such as resulting from disk winds flowing out of the system entirely, or layered accretion directly onto the young sun. We find that mass loss rates higher than found in typical T-Tauri disk winds, lasting short periods of time, are most satisfactory, pointing to multiple intense early outburst stages. Using our global nebula model, incorporating realistic particle growth and inward drift for solids, we constrain how much the MVE depletion signature in the inner region is diluted by the drift of undepleted material from the outer nebula. We also find that a significant irreversible loss of the common rock-forming elements (Fe, Mg, Si) can occur, leading to a new explanation of another longstanding puzzle of the apparent ``enhancement" in the relative abundance of highly refractory elements in chondrites.

\end{abstract}

\section{Introduction}\label{sec:intro}

%\begin{obeylines}

\subsection{The signature of the depletion of Moderately Volatile Elements (MVEs): }\label{sec:signature} 
\begin{figure*}[t]
%\centering
\vspace{-.7cm}
\includegraphics[width=\textwidth]{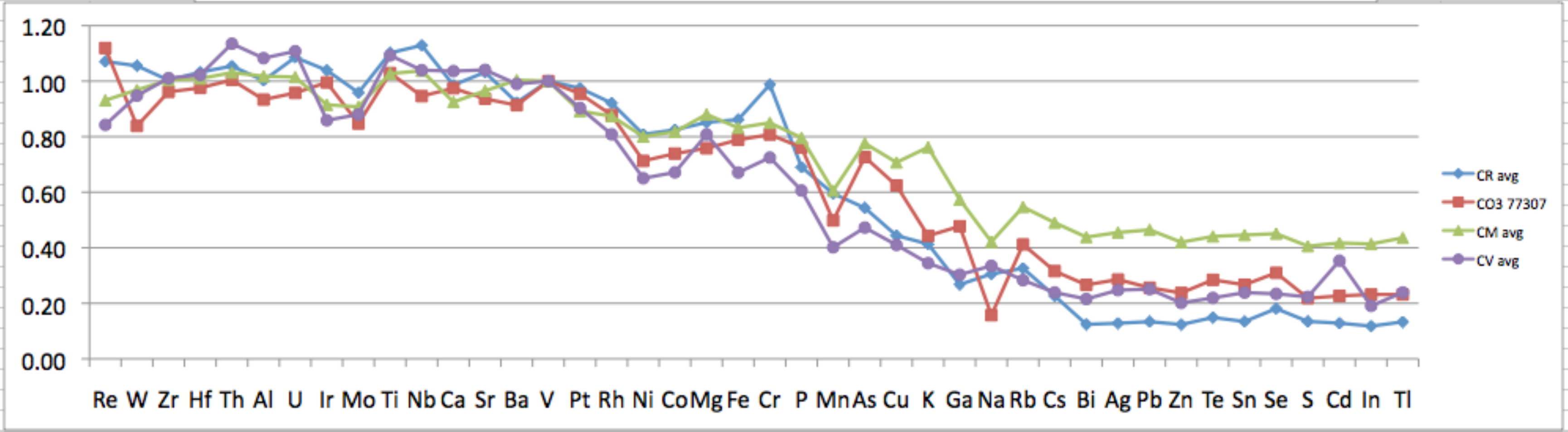}
\vspace{0 cm}
\caption{Elemental abundances of bulk CM, CR, CO, and CV carbonaceous chondrites, normalized to CI chondrites and to Vanadium. The elements are ordered from left to right in order of decreasing condensation temperature, or increasing volatility. Adapted from \citet{Braukmulleretal2018}.} 
\label{fig:braukmuller}
\end{figure*}

Rocky bodies of the inner solar system - from  100km diameter, never-melted  parent planetesimals of chondritic meteorites, to Mars and the Earth - display a systematic depletion of the  ``Moderately Volatile Elements" (MVEs), each denoted $i$,  that clearly correlates with the expected condensation temperature $T_{ci}$ of their most likely {\it single} host material under protoplanetary nebula conditions \citep{Palmeetal88,Palme2001}.   The roughly two dozen elements in question have $T_{ci}$ ranging from roughly 600K (Zn, S) to 1350K (Ni, Fe, Mg, Si). The amount of depletion and other details of the $T_{ci}$ dependence vary from object to object but the systematic signature itself is ubiquitous (see Figure \ref{fig:braukmuller}). 

To avoid the complications of  volatility-dependent losses associated with massive melting, as in the Earth-Moon system or even early achondrite parent bodies, and because the non-Carbonaceous chondrites (NCs) may have been affected by subsequent processing, we will focus here on the Carbonaceous Chondrites (CCs), which show the simplest MVE depletion patterns \citep[][see section \ref{sec:other_proc} for more discussions on NCs]{Alexander2019CC,Alexander2019NC}.  The most primitive (CI) carbonaceous chondrites lack this MVE depletion and generally mirror the composition of the Sun (the primordial nebula) fairly well across this range of volatility. CI chondrites are often linked to comets and truly primordial solids, but relative to true cosmic abundance they are significantly depleted in carbon \citep{Jessbergeretal1988, MummaCharnley2011, Woodwardetal2020}, and perhaps even water  \citep{Tenneretal2015,Marrocchietal2018, Hertwigetal2018, Hertwigetal2019,McCubbinBarnes2019,  Alexander2019CC,  Woodwardetal2020, Fukudaetal2021}. Nevertheless, they provide a good ``primordial" sample in the 600-1350K volatility range relative to which the MVE depletions in other bodies can be normalized. 

Figure \ref{fig:braukmuller} shows an example of the effect. The plotted relative  abundances are taken from the elemental mass fractions in samples of CM, CO, CR, and CV carbonaceous chondrites, as normalized to the corresponding mass fractions in CI chondrites, and again normalized at a particular element (here, the refractory element Vanadium).\footnote{Most plots of this type are normalized to Si, Mg, or Fe, and show the same general overall trend; we discuss the different normalizations further in section \ref{sec:normalization}.} The elements are ordered from left to right by decreasing $T_{ci}$. A large number of the more refractory (higher $T_{ci}$) elements are in agreement with CI abundances when normalized this way, but starting around Pt, all these different carbonaceous chondrite groups are  depleted relative to CI, as $T_{ci}$ decreases. Relative to CI, the so-called  ``slope elements" between Pt and Cs ($T_{ci}$ from 1408K-799K) are increasingly depleted, and the so-called ``plateau elements" from Cs to Tl (799-532K) are depleted by a constant fraction. The $T_{ci}$ of the plateau elements ranges over 215K, about the same as the range between K and P, so the lack of volatility-dependent depletion in the plateau elements is significant. This  signature is echoed in the Earth's mantle \citep{Palme2001,Braukmulleretal2019}. 
Some of the $T_{ci}$ differences between adjacently placed elements in figure \ref{fig:braukmuller} are only 1-2K  \citep{Lodders2003}, so the various dips and bumps in the curves may be due to uncertainties about mineral hosts, and the chemistry and pressure of the local environment, which all affect calculations of  $T_{ci}$.
Furthermore, processes other than volatility may cause some of the more dramatic bumps and dips, and overall the relative values of $T_{ci}$ are more well known than their absolute values that depend on nebular pressure and temperature. Thus here we focus on the general trends (For more discussion see section \ref{sec:equilib}).

The ``non-carbonaceous" (mostly Ordinary and Enstatite) chondrites, or NCs,  have a more complicated MVE depletion pattern, suggesting complex processing \citep{Alexander2019CC,Alexander2019NC}, that we will  only mention in the context of speculation on the broader process (See Discussion section \ref{sec:disc}).  In this paper we will treat only the general MVE depletion pattern observed in the CCs.  

\subsection{Alternate MVE depletion scenarios:} Here we note two alternate MVE depletion scenarios that do {\it not} occur during the early, hot inner nebula stage.
One such hypothesis has been advanced in which  the MVE signature is inherited from the parent molecular cloud, inspired by astronomical observations of volatility-dependent signatures  \citep{Yin2005}. However, the observations of grain properties relied on by this suggestion are from the hot, diffuse ISM, and the immediate parent cloud of the solar system is so cold and dense that the MVEs would all condense.  Moreover, if MVE depletion were inherited from the ISM, then U-Pb ages of volatile-depleted bodies (e.g. the Earth) would systematically be older by about the residence time of matter in the ISM ($\sim$100 Myr) than the U-Pb ages of refractory inclusions, presumed to be the first solids formed in the solar system \citep[eg.,][]{Amelinetal2002, WoodHalliday2005} and this is  not seen. For other discussion see \citet{Ciesla2008} and \citet{Jacquet2014}. 

A different process that  operates at high temperatures - but much later - is chondrule formation. Whatever that process actually was, %It is well known that the different  compositions of chondrules, and the so-called ``matrix" of micron and submicron sized  enveloping grains, are chemically {\it close} to adding up to CI composition regarding major element redistributions (Mg, Fe, Si), that is, are ``complementary". This is usually ascribed to the more volatile  Fe and the siderophile elements leaving the molten chondrules, as vapor or immiscible  liquid drops, for a cooler neighborhood  \citep[see][and references  therein]{Palmeetal2015}. 
it's natural to wonder if it might have produced the observed MVE depletion (figure \ref{fig:braukmuller}). MVE depletion is observed within metal particles in CR chondrites that apparently formed within molten chondrules   \citep{Connollyetal2001, Humayun2012}.  However, as discussed below, there are some indications that the chondrule formation process, while producing some volatility fractionation, was not responsible for most of the observed MVE depletion in bulk chondrites. In any case, actually {\it reproducing} the observed chondrule and chondrite MVE signatures from initially CI material has never been attempted experimentally or theoretically \citep{Alexander2005}. The goal of this paper is to model the MVE depletion process quantitatively as it might have occurred in an earlier, inner hot nebula, stage. 

\subsection{Models of MVE depletion in a hot inner early nebula:} Attempts to understand the MVE depletion signature go back more than 50 years \citep{Anders1964, Larimer1967, LarimerAnders1967, WassonChou1974, Anders1977, WaiWasson1977, Wasson1977}, and the original debates have been reviewed and updated in more recent papers. An early consensus emerged that the cause was most plausibly related to differential condensation and retention by planetesimals in a hot, cooling, early nebula \citep{HumayunClayton1995}, supported by early  physical and thermal models of full-scale, evolving protoplanetary disks by  \citet{Cassen1996, Cassen2001}. More recently, \citet{Ciesla2008} presented a Cassen-like model with some improvements, such as adding some fraction of solid ``migrators" that can drift radially inwards due to gas drag with respect to the evolving gas/small particle mix. These migrators ultimately evaporate when they cross an Evaporation Front (or ``EF") of some specific material   \citep{CieslaCuzzi2006}. In doing this, they augment both the vapor phase in that  material and the solid particle phase in everything more refractory. In that sense the composition is open to radial redistribution. However, the results were somewhat discouraging when other constraints were folded in,  as discussed more below.  For other good introductions see \citet{Palmeetal88}, \citet{Palme2001}, and \citet{Blandetal2005}.

All  previous early inner nebula models are what we  call ``closed system" models. That is, a parcel of the nebula containing gas and small particles may heat up or cool down, or evolve radially, but the elements in it merely change from solid phase to vapor phase and back unless the solids are decoupled from the evolving nebula entirely by being converted at some assumed rate  into  planetesimals that do not migrate radially from where they form. So, they retain a cumulative record of the materials that have been solid at that location, as the gaseous nebula evolves into the sun, loses mass and opacity, and cools.  It is the composition of these decoupled planetesimals that the Cassen and Ciesla models track. As the nebula cools, the inner regions viscously evolve into the sun, carrying all vaporized material and associated small particles, and leaving behind only the immobile planetesimal component. By the time the nebula cools enough for more volatile species to condense onto existing planetesimals, there is less of each of them, producing the MVE depletion signature in the left-behind planetesimals. Basically, the models of \citet{Cassen1996, Cassen2001} and \citet{Ciesla2008} agree that plausible MVE depletion signatures can be produced in a hot, cooling inner nebula for a range of parameters.

Unfortunately, as described by \citet{Ciesla2008}, the parameters leading to satisfactory solutions are in disagreement with other meteoritical constraints. First, to preserve the depletion signature (figure \ref{fig:braukmuller}), the decoupled planetesimals all have to form while the nebula is cooling through the range 1350-600K - a very early stage that lasts at most a fraction of a Myr \citep{Cassen1996,Cassen2001, Ciesla2008}. This timeframe is in disagreement with formation age dates for the chondrules from which all chondritic planetesimals that  dominate the asteroid belt are made (2-4 Myr after the first solids). Moreover, such early accretion would imply that more than enough live $^{26}$Al would be accreted to completely melt any planetesimal larger than 10km diameter \citep{WoolumCassen1999, SandersScott2012a}, inconsistent with the large, widespread, mostly or entirely unmelted chondrite parent bodies that dominate the asteroid belt \citep{Scott2006, Vernazzaetal2014, Johansenetal2015}.  While those  early, molten bodies that have been studied show (often rather extreme) MVE depletion \citep{PringleMoynier2017}, it is not hard to imagine that their molten nature,  differentiation, and outgassing could itself have produced MVE depletion even from initially CI-like accreted material \citep{NorrisWood2017} - so this evidence is inconclusive regarding MVE depletion of the initial material from which they  %and perhaps the chondrites after them  
accreted, without more study. 
% deleting this because we are NOT discussing the NCs: An independent problem with these previous early-depletion models is that the necessarily incremental accretion of increasingly volatile material on top of originally more refractory planetesimals as the nebula cooled, would lead to strong ``onion shell" chemical gradients with depth that are not observed in at least ordinary chondrite parent bodies \citep[see][Section 3.3]{Johansenetal2015}. 
 While there may be {\it some} evolutionary pathway from early-formed, molten planetesimals to actual chondrite parent bodies forming at 2-4Myr after CAI \citep{SandersScott2012a}, the situation  left \citet{Ciesla2008}  discouraged that any hot early inner nebula model could explain the MVE signal. 

Since that time, several other studies have appeared, adding sophistication in various ways. Several of these focus on the MVE signatures of the terrestrial planets themselves, mapping condensation models into planetesimals formed at some specific times in nebula evolution, and allowing these planetesimals to mix dynamically as they accrete terrestrial planets \citep{Bondetal2010, Elseretal2012,Moriartyetal2014}. These models differ in some ways with each other but all conclude that planetary compositions can only be matched if the ``decoupling time" of planetesimals is in the $10^5$ year timeframe, encountering the same problem found by \citet{Ciesla2008} with the models by \citet{Cassen1996,Cassen2001}. Moreover, the compositions of the planetesimals {\it themselves} would have step-functions of elemental abundance corresponding to EFs near their formation locations  (elements would be either fully present or completely absent), as can be seen in the models shown by \citet{Moriartyetal2014}, and no planetesimals with such properties are known. The gradual slope in MVE depletions the above models {\it do} find in terrestrial planets \citep[see also][]{McDonoughSun1995, YoshizakiMcDonough2020} is, we suspect, contributed only by the subsequent dynamical mixing stage due to gravitational scattering and collisions. So, these models do not explain the MVE depletions in actual primitive bodies. Another set of models by \citet{Lietal2020, Lietal2021} do address planetesimal (chondrite parent body) properties, adding more sophisticated condensation models and ongoing infall to those of \citet{Cassen1996,Cassen2001} and \citet{Ciesla2008}, but still end up facing the same  rapid-formation problem with  ``decoupling" times essentially equal to those of \citet{Cassen1996,Cassen2001} - in the $10^5$ year timeframe. Like the Cassen and Ciesla models, {\it all} these models are ``closed system" models, and this aspect leads directly to the short-timescale problem. 

Another new line of argument has emerged \citep{SandersScott2012a, SandersScott2012b,  Johnsonetal2014, CarterStewart2020, SheikhHumayun2021} that chondrules formed in connection with massively disruptive collisions between pre-existing planetesimals, which one might argue could be MVE-depleted survivors from an early stage as modeled by \citet{Cassen1996,Cassen2001} and \citet{Ciesla2008}. Such a scenario might  circumvent the objections of Cisela (2008). While forming {\it normal} chondrules other than in CB and CH chondrites by collisions is still debated and has yet to be studied in any detail, we note the possibility here for completeness.

In this paper, we will describe an early hot nebula model which is ``open system", allowing material to be irreversibly lost along the way, which avoids the obstacles noted by \citet{Ciesla2008} (sections \ref{sec:diskmodel} - \ref{sec:disc}). The model is, however,  best regarded as only part of a much larger context scenario, as we sketch in section \ref{sec:other_proc}.  
%%%%%%%%%%%%%%%%%%%%%%%%%%%%%%%%%%%%
%%%%%%%%%% delete or move elsewhere 
%%%%%%%%%%%%%%%%%%%%%%%%%%%%%%%%%%%%
%{\bf Constraints on when MVE depletion occurred:} The earliest-forming age-dated objects are carbonaceous and noncarbonaceous achondrites and irons (Kruijer et al 2017). If we could confidently establish a chondritic MVE depletion signature relative to CI in these bodies, that would argue that the depletion happened very early, well before the chondrule formation era. Conversely, if their original pre-melting MVE composition were CI, it would allow the MVE depletion  to occur much later, perhaps during the chondrule formation process itself.  Unfortunately, the very fact of their thorough melting and differentiation, with the possible losses to space that might entail, makes it difficult currently to establish what the as-accreted MVE content of theseearly planetesimals might have been in any detail. \citet{Palme2001}  made an attempt to do this for achondrite rocks, based on Ag/Pd ratios and Ag age-dating, but the temporal resolution of this raadiometric pair is too crude). Iroo meteorites - from the cores of early differentiated planetesimals -are better dated, but their MVE compositions are complex and possibly subject to other processes. Some work along these lines would be a desirable goal for the future, though.   

\subsection{The message of the Matrix:}\label{sec:matrix} The ``flat" MVE distribution of the plateau elements is CI-like (section \ref{sec:signature}), and indeed, various studies confirm that the plateau elements are carried primarily in the matrix of micron and submicron material in which the chondrules are embedded \citep[][and references therein]{Zandaetal2006, Alexander2019CC}.  \citet{Hussetal2003} studied the matrices of 8 very primitive carbonaceous chondrites, looking specifically at several tracers of nebula or parent body thermal or aqueous processing, including SiC grains, graphite, and three different isotopic patterns in Xe that is trapped in nanodiamonds. They noted that these materials would not survive the chondrule-forming process/environment. Indeed, they concluded that the matrices of at least the CM chondrites are essentially CI-like in all of their metrics, even after some amount of aqueous alteration. This result is in good agreement with the results of \citet{Zandaetal2011_Paris} and \citet{Hewinsetal2014} who were able to study portions of the Paris CM chondrite that have {\it not} been aqueously altered, and found it to be essentially CI in showing no volatility-dependent MVE depletion trend \citep[see also][for more discussion]{Zandaetal2018}. 
%These results have led some to conclude that matrix in at least CM chondrites, and possibly all chondrites, is  essentially, mostly CI material \citep[see also][]{Alexander2005}. The intimate mixture of heavily thermally processed chondrules,  and variable amounts of both processed and  unprocessed (CI) matrix, has long suggested a prominent role for some kind of post-chondrule formation mixing process \citep[see more recently][]{Jacquetetal2016, Jacquetetal2019, Pignataleetal2018, Pignataleetl2019}, and we will return to this idea in section ??.. 

%%% MOVE TO after BLAND et al Zanda also showed that regions of Paris matrix that {\it were} aqueously altered showed an MVE depletion pattern echoing that of the bulk, similar to the results of Bland et al (2005) for several CM2 (aqueously altered) chondrites. Zanda concluded that the original accreted matrix of all CM2s, including those studied by Bland et al, probably showed no MVE depletions, and the signal detected by Bland et al (and reproduced by herself in altered regions of Paris) was an artifact of parent body alteration - the undepleted matrix exchanged elements with highly MVE depleted chondrules. Whatever temperature  Paris achieved, it was enough when combined with aqueous alteration to mobilize these elements. Zanda's finding that CM matrix was originally CI seems to validate the work of Huss et al, which was less direct, but looked at tracers that would be less affected by low-temperature aqueous alteration then the usual MVEs themselves (MUNIR: IS THIS TRUE?)

However, \citet{Hussetal2003} and \citet{ Huss2004} went on using the same techniques to find significant differences between CI and the matrices of CO, CR, and CV chondrites, that {\it are} associated with moderate-to-high temperature thermal alteration (670-970K). That is, the depletions in the  matrix tracers, which escaped chondrule formation, overlap the lower end of the thermal range covered by the MVEs and are correlated with the depletions experienced by the chondrules in the parent chondrites. In several cases these studies detected multiple components of tracers, suggesting a mixture of a ``high" temperature processed matrix, with inferred processing temperatures comparable to those inferred from the bulk MVEs themselves, and a ``low" temperature processed (not-quite-CI) matrix, which even approached the properties of CI material in the CMs \citep[consistent with][]{Zandaetal2011_Paris}. %What we take away from these results is that, while primordial CM2 matrix does not differ significantly from CI, the matrix of these other chondrule groups is not consistent with CI, and shows the signature of thermal processing similar to that associated with MVE depletion in the chondrules of those chondrites. 

\citet{Blandetal2005}  studied matrix and bulk, comparing with published studies of chondrules, for four different carbonaceous chondrite groups (CV, CO, CM, and CR) and found that matrix {\it itself} is MVE depleted in all of them, to differing degrees. In some instances, they   detect a true complementarity in {\it a few} (siderophile or metal-loving) elements, which is a clue that chondrule formation did play {\it some} role transferring some elements from chondrules to matrix, %either by associated extrusion of metal alloy from molten silicate chondrules, or co-condensation of evaporated siderophiles   - 
but this transfer apparently did  not apply to the lithophile or ``rock-loving" elements - their abundances in chondrules and matrix are the same regardless of volatility. This all suggests that the entirety of the parent material of these chondrites was MVE-depleted before chondrule formation began. However, a caveat is required because many of the matrices sampled by \cite{Blandetal2005} had been aqueously altered (see below). 

Another piece of evidence along these lines has been seen in metal particles found external to chondrules in CR chondrites, some having apparently formed inside molten chondrules and then been extruded, and others which  apparently formed outside the chondrule by condensation of evaporated volatiles. These objects contain not only Fe but other siderophile elements covering the MVE volatility range.  While the apparently internally formed and extruded metal particles contain a full CI-like abundance of the refractory siderophiles, and the metal particles formed from the evaporated vapor are lacking in these refractories, the volatile siderophiles Au and P are noticeably underabundant in {\it both} kinds of metal particles \citep{Connollyetal2000,Connollyetal2001}. If the Au and P had been lost from molten chondrules into the surrounding gas, causing their depletion in the metal drops that formed within chondrules, these elements should be  {\it overabundant} in metal grains that formed from that external vapor - but indeed they are depleted there, too. The implication is that the chondrule precursors themselves were already depleted in Au and P, and presumably the other MVEs, prior to chondrule formation, as part of an earlier MVE depletion. Similar conclusions, although with slightly different emphasis, were reached by \citet{Humayun2012} and \citet{Jacquetetal2013}. One {\it might} argue that the CR chondrites formed later than the bulk of chondrule-forming events, leaving time for earlier chondrule-formation processes to deplete the MVEs; perhaps studies along the lines of the above studies could be done in other chondrite groups. 

Some of the results of \citet{Blandetal2005} should be taken with caution. \citet{Zandaetal2011_Paris, Zandaetal2018} \citep[see also][]{Hewinsetal2014} found that CM matrix (in the very primitive Paris chondrite)  was aqueously altered on the parent body to {\it show} an MVE-depletion signature that was an artifact of the alteration. Since some of the other chondrite groups studied by \citet{Blandetal2005} were affected by alteration, the question is, could their pre-alteration matrices also have been CI-like? 

The  MVE-depletion effects \citet{Blandetal2005}  saw in CO and CV chondrites were larger than in CM  though, and (some) matrix MVE depletion was also seen in highly primitive (unaltered) ungrouped chondrites like Acfer094.  \citet{Hussetal2003} and \citet{ Huss2004} would agree with \citet{Zandaetal2011_complementarity} that CM matrix is CI-like. However, \citet{Hussetal2003} and  \citet{Huss2004} also saw matrix processing effects in CO and CV chondrite matrices that had {\it avoided} chondrule formation, that were insensitive to aqueous alteration, and indicated moderate-$T$ thermal processing at temperatures comparable to those needed for MVE depletion, consistent with \citet{Blandetal2005} for these groups. \citet{Hussetal2003} concluded that this more processed matrix material was co-genetic with the chondrule precursor material in these chondrites, and already MVE-depleted. Along these lines, \citet{Alexander2005} also concludes that matrix, in general, contains 10-30\% ``processed" (ie MVE-depleted) fine-grained material, quite likely to be co-genetic with chondrule precursors, and predating the chondrule formation process. These findings support the conclusions of \citet{Blandetal2005}, regarding the non-CM chondrite groups, at least. If a batch of material started near CI composition, and the chondrule formation process depleted MVEs from chondrules into surrounding cooler regions, matrix in general should have been {\it enhanced} in MVEs in a complementary fashion to the depletions seen in chondrules, and this is never observed regardless of how much  matrix is accreted with the chondrules. 
%%%% made invisible for now! Some means would need to remove all the complementary, post-CF, MVE-enhanced Matrix material before unrelated, relatively unprocessed CI material mixes into the region, to be accreted with chondrules into a planetesimal. 

So with a few caveats, the results of  \citet{Hussetal2003}, \citet{ Huss2004} and \citet{Alexander2005} support the results of \citet{Zandaetal2011_Paris} and \citet{Hewinsetal2014} for CM chondrites but also the results of  \citet{Blandetal2005} for MVE depletions in CV, CO, and CR chondrite matrix that are not only pre-accretionary but also  pre-chondrule formation. In view of possible subsequent stages (section \ref{sec:other_proc}), the same hot early inner nebula process could have applied to all non-CI carbonaceous chondrite parent material, with the final  bulk differences coming to some degree from  how much in-mixing occurred of unprocessed CI material prior to parent body accretion, with CM getting the most (see section \ref{sec:slowmixing}). 
\citet{Zandaetal2006, Zandaetal2009, Zandaetal2011_complementarity, Zandaetal2012,Zandaetal2018} have also emphasized the mixing of chondrules,  which  formed ``hot", with essentially unprocessed CI material, but it is not clear that these studies can rule out fractional amounts of admixed matrix material that was co-genetic with chondrule precursors  \citep{Hussetal2003, Huss2004, Blandetal2005,Alexander2005}.  However, it is important to mention that the new work by  \cite{vanKooten2019} suggested that even CV matrix (in the form of fine grained rims) is close to CI. On the other hand, \cite{Simonetal2018} note that only a fraction of CV chondrules do have Fine Grain Refractories.

%\textcolor{red}{Jeff - remember to add somewhere in here new work by Van Kooten et al that suggests even CV matrix (in the form of fine grained rims) is close to CI. On the other hand, Simon et al 2018note that only a fraction of CV chondrules DO have FGRs.} 

\citet{Hussetal2003} conclude by inferring that ``.. the parent material was heated to different degrees in the nebula before chondrule formation and accretion, producing the basic chemical characteristics of the different chondrite groups. Chondrules then formed, mechanical mixing of components and, in some cases, metal-silicate fractionation, took place, meteorite parent bodies accreted, and parent body processes further altered the material." 
Overall, it seems that the results of \citet{Hussetal2003,  Alexander2005, Zandaetal2011_Paris} and \citet{Hewinsetal2014} regarding CM chondrites, {\it and} also the conclusions of \citet{Connollyetal2001,Hussetal2003, Huss2004,Alexander2005} {\it and} \citet{Blandetal2005} support some nontrivial amount of matrix material being MVE-depleted and cogenetic with chondrule precursor material, at least in CR, CO, CV, and other ungrouped primitive chondrites, but eventually mixed with a group-dependent amount (70-90\%) of CI material that  carries the ``plateau" MVEs. Also, isotopic evidence indicates that Allende matrix and chondrules bear opposing $^{183}$W nucleosynthetic isotope anomalies that indicate a separate provenance of the material in matrix from that in chondrules \citep{Buddeetal2016}. The question of admixing of matrix and chondrules is surely not completely resolved though, and further observations and models will be needed. 

This general scenario agrees fairly well with  one recently advanced by \citet{Nanneetal2019}, only the first part of which we address in this paper (an early, hot inner nebula). Subsequent stages, discussed below, include (2) expansion past the snowline, (3) isolation by a jovian core,  (4) radial mixing within the now-isolated regions, and (5) ongoing infall. For the CCs, this radial mixing brought in the less-processed or un-processed CI-like matrix. In the case of CMs, any co-depleted matrix material may have been swamped by a larger amount of in-mixed CI material. %In this paper we will only describe and explore stage (1), but 
We will sketch some possible roles of subsequent stages below.

\subsection{Early nebula stages and processes of potential interest:}\label{sec:other_proc} %Below we sketch some of the early nebula processes that offer open system behavior, and then discuss other meteoritically motivated scenarios to provide a more general context within which our limited study would fit. 

Astronomical observations and models provide limited insight into this early stage of nebula history. The earliest recognizable stage of a ``protostar" comes perhaps $10^4$ years after infall begins in the parent cloud. Over the next few $10^5$ years (the so-called ``embedded" or class 0/1 stage) as infall proceeds, the mismatch between mass infall (most of which falls onto the growing disk) and the rate at which the disk can drain this material into the star, can lead to multiple outbursts of varying strength; the most extreme of the observed outbursts are the so-called FU Orionis outbursts, but there is a spread of outburst energies, repetitions, and durations, covering objects known generally as YSOs \citep{Audardetal2014}. These outbursts represent brief stages of intense mass accretion onto the star, probably also associated with high temperatures, rapid expansion of the inner disk,  and strong, perhaps violently fluctuating, radial mixing \citep{VorobyovBasu2005, VorobyovBasu2006, Boss2008, Boss2012}. Many previous models have treated this stage with smooth 1D ``$\alpha$-models", but reality is almost surely much more complicated. It is worth noting here, that a considerable amount of meteoritical evidence carries the suggestion of these intense, high-temperature, fluctuating processes. In particular, the earliest solids, called ``Calcium-Aluminum-Rich Inclusions" or CAIs, or at least the ones remaining for us to study, formed over a very short period of time, are frequently seen to display intensely heated rims, odd combinations of isotopic properties, and  variations in the isotopic compositions of their surroundings. For general discussions see \citet{ Krotetal2009} and \citet{ Kitaetal2013}, and for discussion of early short-timescale fluctuations in particular see  \citet{Simonetal2011, Simonetal2019, Macpherson2017,Marrochietal2019}, and \citet{Krotetal2020}. 

The exact way by which material {\it does} flow from the disk into the star remains uncertain. One of the favored processes, that has some bearing on the stage and process we model here, is called ``layered accretion" \citep{Gammie1996,Zhuetal2010}. In this process, the densest midplane regions only transfer material to the star slowly, but the upper, low density layers of the disk can undergo MHD-driven turbulent viscosity that move mass much faster, directly into the star. There are temperature and/or density thresholds determining when and where layered accretion can operate.  Also during the few $10^5$ year ``embedded" timeframe of star formation, infall from the parent cloud continues at a range of radii in the disk \citep{CassenMoosman1981, HuesoGuillot2005,Visseretal2009, YangCiesla2012}, and direct {\it outflows} from the young system may also be occurring, in the form of stellar and/or disk winds and/or jets (section \ref{subsec:diskwind}).  These processes are all likely to be more complex than any models existing in the current literature.  
	
Several processes in this early epoch could provide  ``open system" or irreversible loss or gain behavior. Infall, of course, continually adds fresh gas and solids, the composition of which may well change with time  due to imperfect mixing in the parent cloud. Layered accretion carries preferentially the vapor and tiny grains from rarified upper altitudes of the disk  into the star. Disk winds and jets carry disk material from high altitudes  out of the system entirely. Materials can be fractionated in this way (differentially depleted or augmented) by removal to, or addition from, ``outside", but this ``open system" behavior has not been previously applied to MVE depletion. 

Another important stage in the story of the chondrites may be a ``truncation" of the disk, preventing  material from either side from mixing, and allowing an originally homogeneous  mixture to diverge compositionally. The motivation for this is the very strong isotopic and elemental clustering that separates the carbonaceous chondrites or CCs (and their melted elder siblings, the carbonaceous achondrites) from the equally primitive (and in many cases, contemporaneously forming) ``Non-carbonaceous chondrites" (NCs) \citep{Warren2011}. The most popular hypothesis for this disk truncation is the emergence of Jupiter's core, probably at the snowline. This truncation began as early as  0.2 Myr and lasted at least 2 Myr \citep{Kruijeretal2017, Deschetal2018}. A jovian core is only one hypothesis for preventing mixing of these two ``reservoirs" of material, but the almost complete \citep[but not 100\% complete,][]{Williamsetal2020} separation of the CC and NC populations testifies to the high effectiveness of whatever the isolating  process was. 

 \citet{Nanneetal2019}  proposed a sequence of several of the early nebula processes mentioned above, to explain {\it qualitatively} certain isotopic properties of the CCs (which might have formed beyond the snowline but are closely tied isotopically to CAIs, that probably formed less than an AU or so from the sun) and NCs (which are not connected closely with CAIs, but yet probably formed into planetesimals closer to the sun). In this scenario, a well-mixed inner disk expands rapidly out past the snowline before Jupiter's core forms, and is then trapped outside it by truncation. Ongoing infall from the parent cloud has an outer limit (in current simple models) that {\it might} not have been far from the snowline at this time, and continued to evolve the composition of what became the NC parent material inside the truncation boundary, while the material outside the truncation boundary, that became the CCs, interacted by radial mixing with unprocessed material further outwards. %The essential elements of this story are (a) a dramatic expansion of a well-mixed inner disk, (b) truncation, (c) radial mixing outside snowline and ongoing infall inside snowline. 
None of the elements of this scenario are well understood, but there are plausible models for most of them.  \citet{Pignataleetal2019} and \citet{Jacquetetal2019} have also modeled a scenario of time-varying infall in an early nebula as explaining the very early formation and narrowly-defined $^{26}$Al abundance of CAIs, and several other stable isotope differences between different meteoritical groups. In their models, hot inner nebula processes are  also followed by outward advection and turbulent mixing. The spirit of these latter models is  consistent with the scenario we envision. 

We mention these early nebula complexities here, and will return to them in section \ref{sec:disc}, because the model we present here cannot exist in isolation from the broader picture, even though the broader picture remains to be modeled quantitatively. 
\vspace{0.2in} 

%6. Can we prove HOW Early MVE depletion occurred? Not yet. 

%The earliest planetesimals (the irons and achondrites) at 0.2-3Myr have already been segregated NC/CC by (current consensus) a Jup core 
	
%But their MVE depletions are even MORE dramatic, maybe DUE to early melting so not clear yet whether a global MVE depletion predated them or not
	
%Can mention Palme argument but it's not discriminating enough

%Good topic for future work

\vspace{0.2in}
\subsection{The goal of the paper} Our goal is to explore several early stage, open system processes  (FU Ori events, early disk winds, and layered accretion) in an idealized fashion. We use a sophisticated model of the growth, drift, and opacity of nebula solids  and their evaporation fronts, in which particles can be solid or porous aggregate in nature, to explore the possibilities. 

We will assess the degree to which {\it early, pre-expansion, pre-truncation} inner nebula processes \citep[in the spirit of][see section \ref{sec:other_proc}]{Nanneetal2019} may produce a single reservoir of MVE-depleted material that can cool and evolve for 2Myr without subsequent condensation of accompanying MVE vapor washing out the depletion signature or creating an onion-shell planetesimal compositional structure. 

The questions we will address include: Can a nebula condensation model explain the first-order MVE depletions without resorting to chondrule formation? How much MVE loss can occur in chondrule precursor material in the early, hot disk stage well before the  chondrule formation epoch and even before or during the formation of achondrite planetesimals? What kind of early nebula environment can produce the observable MVE depletions? And, to a limited degree, what other subsequent early disk processes acted on the depleted material, moving or mixing it with less processed material?  

As in \citet{Nanneetal2019} it is our premise that the current CC/CAI complex most closely resembles the composition of the very early disk, as independently concluded by \citet{Pignataleetal2019} and \citet{Jacquetetal2019}.  However, surely there was {\it some} subsequent radial mixing outside the Jupiter barrier  contributing unprocessed material such as diamonds, organics, presolar grains, plateau MVEs, and water. It is beyond the scope of this paper to explore this later process quantitatively, but we discuss it in assessing the success of our depletion models in section \ref{sec:globaldist}. 

MVE patterns in NCs are distinct from those in CCs \citep{Alexander2019CC, Alexander2019NC}. Here, chondrule formation and/or a more oxidizing environment might be involved \citep{Blandetal2005}, perhaps on top of  ongoing infall after disk truncation by a growing Jupiter, all happening after the common process we start to explore here. We defer discussion of the CC-NC divergence to future work and focus on the plausibility of creating this initial pre-expansion, pre-truncation, MVE-depleted material. 

%	(Note to self: re-run Cuzzietal2010 diffusion model)
%It is emerging that a significant amount of meteoritic puzzlement may originate in complex early nebula processes that have yet to be studied in detail. 

%\end{obeylines}

%\section{DISK WINDS \& MVE DEPLETION: THE HYPOTHESIS}\label{sec:hypothesis}

\section{Disk Model}\label{sec:diskmodel}

The investigation of the MVE depletion problem described in section \ref{sec:intro} requires a complex and detailed model of  global nebula evolution  capable of simulating the early hot inner region self-consistently. In order to build our numerical framework, we adopted the 1+1D global nebula evolution model from \citet[][E16 hereafter]{Estradaetal2016} and updated the code with several new features relevant to our current work. Using a vertically integrated 1D model forces us to adhere to several assumptions regarding the original hypothesis. We ask the reader to see  section \ref{sec:1Dlimitations} where we discuss the possible limitations of our current model. We first mention the main components of the E16 model in brief before moving on to the new components implemented. The reader is referred to E16  for details.

\begin{itemize}
    \item The model by E16 calculates the $1$D time evolution of the gas disk surface density $\Sigma(R,t)$ and radial velocity $v_g(R,t)$ with a turbulent viscosity prescription $\nu=\alpha c H_g$, where $c$ is the local sound speed and $H_g=c/\Omega$ is the gas scale-height. Here,
\begin{equation}
    \Omega=\sqrt{\frac{GM_{\star}}{R^3}}
\end{equation}
with $M_{\star}$ is the stellar mass and $G$ is the gravitational constant. The equations for the radial evolution and velocity of the nebular gas, derived from the continuity and angular momentum conservation equations, are given by \citet{Pringle1981}:
\begin{equation}
\frac{\partial\Sigma}{\partial t}=\frac{3}{R}\frac{\partial}{\partial R}\left[R^{1/2}\frac{\partial}{\partial R}\left(R^{1/2}\nu\Sigma\right)\right]
\label{eqn:sigmaevolution}
\end{equation}
\begin{equation}\label{eqn:gasvelocity}
    v_g=-\frac{3}{R^{1/2}\Sigma}\frac{\partial}{\partial R}(R^{1/2}\nu\Sigma).
\end{equation}
\item The growth model for particles is implemented by a moment method \citep{EstradaCuzzi2008} to enhance the numerical efficiency. The collision model includes experimental collision outcomes, such as sticking, fragmentation, bouncing and mass-transfer. Relative velocities of collisions incorporate contributions from Brownian motion, gas pressure gradient and turbulence \citep{OrmelCuzzi2007,Ormeletal2007, Senguptaetal2019}. E16 also allows  so-called ``lucky particles``, also called migrators, to grow beyond the fragmentation threshold mass and drift inwards by headwind drag.  After drifting inwards some distance, these migrators will partially evaporate at {\it Evaporation Fronts} (see below).

\item The original E16 code only accounted for particle growth as solid (zero porosity). The updated code now also allows for porous particles by including new routines for aggregate growth that implement a fractal growth and compaction recipe developed by other workers \citep{Suyamaetal2012,Okuzumietal2012,kataokaetal2013}. The implementation is described in detail  by  \citet{Estradaetal2021}.

\item The E16 model incorporates Evaporation Fronts (EFs) for an arbitrary number of species. At these EFs, the local midplane temperature exceeds the sublimation temperature of one of the  constituent condensed volatiles, and that part of the solids contained in this specific material is converted to vapor. Each element that we track is assumed to be present in only one mineral host, each of which has a given 50\% condensation/evaporation temperature $T_{ci}$ (see Table \ref{tbl:mves}). Inward drifting particles crossing each EF can significantly alter the abundances of the solid and volatile species in the region. The numerical model then redistributes the solids and vapors by subsequent advection, diffusion, and re-condensation to/in regions both inside and outside of the EF (see E16 and section \ref{subsec:mves} for more details). The vapor and solid abundances are tracked by the advection-diffusion equation:

\begin{equation}
\frac{\partial \Sigma_{v,p}^i}{\partial t}=\frac{1}{R}\frac{\partial}{\partial R}\left[R\Sigma D_{g,p}^i\frac{\partial f_{v,p}^i}{\partial R}-Rv_{g,p}\Sigma_{v,p}^i\right]+S^i,
\label{eqn:advdiff}
\end{equation}
where $\Sigma_{v,p}^i$ is the surface density and $f_{v,p}^i$ is the vertically integrated mass fraction of particles ($p$) and vapor ($v$) of the $i^{th}$ species \citep{Deschetal2017}. Also, $v_{g,p}$ is the net radial velocity of the vapor or particle species, and $D_{g,p}^i$ is the corresponding diffusivity. All the different vapor species share the local gas velocity (equation \ref{eqn:gasvelocity}) and diffusivity (the latter assumed to be given by the viscosity $\nu$), but each particle size has its own values of these properties (E16). The term $S^i$ represents sources and sinks for particles and vapor of species $i$ at an EF and/or in winds or layered accretion. Large particles tend to drift inward by headwind drag, and small particles tend to diffuse radially and advect with the gas, while molecules and sufficiently tiny grains can be carried off irreversibly by disk winds or layered accretion at high altitudes.

\item The model from E16 calculates the local temperature of the nebula ($T_{neb}$) assuming energy inputs from both internal viscous heating and a temporally varying stellar illumination which heats the upper disk layer. The particle distribution obtained from the growth and advection-diffusion module is resolved in the vertical direction, balancing vertical settling and upward diffusion, giving the particle model a $1+1$D configuration. The midplane temperature is calculated using both   Rosseland and Planck  mean opacities $\kappa_R$ and $\kappa_P$, respectively, which are a function of the composition and size distribution of the solid components. The opacities are calculated for aggregates with arbitrary porosity, composition and size from the monochromatic opacities $\kappa_{\lambda}$  using the `{\em Utilitarian opacity model}` from \citet{Cuzzietal2014}. For regions with low optical depth, $\kappa_P$ is preferable \citep{NakamotoNakagawa1994}. However, \citet{Pollacketal1994} and others have shown that for small solid particles the distinction between the Rosseland and Planck opacities is not large. 

\end{itemize}
We apply the above numerical framework in a disk model whose initial condition is derived from the analytical expression from \citet{LyndenBellPringle1974}, and later generalized by \citet{Hartmannetal1998}. The gas surface density $\Sigma$ and the radial velocity $v_g$ are initiated as
\begin{equation}
    \Sigma(R,0)=\frac{M_D}{\pi R_0^2}\left(\frac{2-\beta}{2}\right)\left(\frac{R}{R_0}\right)^{-\beta}e^{-(R/R_0)^{-\beta}};
\end{equation}
\begin{equation}
    v_g(R,0)=-\frac{3\nu(R_0)}{2R}\left(\frac{R}{R_0}\right)^{\beta}\left[1-\left(4-2\beta \right)\left(\frac{R}{R_0}\right)^{2-\beta}\right].
\end{equation}
where we choose the disk mass $M_D=0.2M_{\odot}$, a characteristic radius $R_0=20$AU and $\beta=1$. $\nu(R_0)$ is the viscosity evaluated at $R_0$. The initial luminosity of the central object is assumed to be $\sim 10$ times higher than the typical main sequence value during the first million years of evolution \citep{DantonaMazzitelli1994, Siessetal02}. This formulation leads to a hotter, denser and more compact nebula than the Minimum Mass Solar Nebula, and closer to the model proposed by \citet{Desch2007}.

%\textcolor{red}{Debanjan, note the term $\nu(R_0)$ in the equation. I am not sure if $\nu$=constant, or is allowed to scale in some way with $c$ and $H$. If the latter (like, through $\beta$), we should be more specific.  Also, if here is no $H$, then we can drop the subscript $g$ from $H_g$ for simplicity and consistency. I have left it alone. } 

\subsection{The disk wind model}\label{subsec:diskwind}

One specific process leading to an open system is a disk wind, where an outflow of gas is launched from the upper layers of the disk and mass is lost irreversibly. The flux of the gaseous component escaping the disk in a wind is governed by the strength and geometry of the magnetic fields and the wind is centrifugally supported. Several such wind models exist in the literature  in the context of global nebula evolution \citep{teitler2011,Bai2016, Suzukietal2016}. Although the diagnostics of the wind in all these models are similar, we adopt the model by \citet[][S16 hereafter]{Suzukietal2016} which suits well  the structure of our numerical model. 

Before we move on to the details of the wind model, it is important to remind the reader that infall is also a major event going on during the class 0/I phases of the nebula. As the observations are still limited at these stages, it is hard to assess the effect of material loss through wind in the presence of infall. So, understanding the overall effect will require a sophisticated model for infall added to our current numerical framework which is beyond the scope of the present work, and is postponed for a future study.

S16 solves the continuity and the angular momentum conservation equations for the disk, leading to the evolution equation
\begin{equation}\label{eqn:suzukievolution}
    \frac{\partial \Sigma}{\partial t}+\frac{1}{R}\frac{\partial}{\partial R}\left(R\Sigma v_g\right)+\dot{\Sigma}_W,
\end{equation}
where $v_g$ is the radial gas velocity, and is given by
%\begin{equation}\label{eqn:suzukigasvelocity} R\Sigma v_g=-\frac{2}{R\Omega}\left[\frac{\partial}{\partial R}\left(R^2\Sigma \bar{\alpha}_{r\phi}c^2\right)+R^2\bar{\alpha}_{\phi z}\left(\rho c^2\right)_{mid}\right].\end{equation}
\begin{equation}\label{eqn:suzukigasvelocity}
    R\Sigma v_g=-\frac{2}{R\Omega}\left[\frac{\partial}{\partial R}\left(R^2\Sigma \overline{\alpha_{r\phi}}c^2\right)+R^2\overline{\alpha_{\phi z}}\left(\rho c^2\right)_{mid}\right].
\end{equation}
In equation \ref{eqn:suzukigasvelocity} the basic $\alpha-$parameter of earlier equations is replaced by the quantity $\overline{\alpha_{r\phi}}$  in the first term, representing the usual viscous torque (equation \ref{eq:visctorque});  the second term, parameterized by the analogous quantity $\overline{\alpha_{\phi z}}$, represents the torque exerted on the disk by the wind (equation \ref{eq:windtorque}). Note that the viscous terms in equation \ref{eqn:gasvelocity} and \ref{eqn:suzukigasvelocity}  are basically the same except for a factor of $2/3$, because S16, similar to \citet{Bai2016}, parameterize $\overline{\alpha_{r\phi}}$ directly from $T_{r\phi}=\rho v_r v_{\phi}-B_rB_{\phi}/4\pi$. The gas velocity in equation \ref{eqn:gasvelocity}, on the other hand, is calculated using  $T_{r\phi}=\nu \rho (d\Omega/dR)$. %where $\nu$ is the kinematic viscosity.  
The extra factor arises from the term $d\Omega/dR$. 

The term $\dot{\Sigma}_W$ in equation \ref{eqn:suzukievolution} depicts the wind mass loss which is written following S16 (see below) as 
\begin{equation}\label{eqn:diskwind}
    \dot{\Sigma}_W = (\rho v_z)_W = C_W(\rho c)_{mid}.
\end{equation}
Here $v_z$ is the vertical component of the gas velocity at the wind base $z_{wl}$,  while $c$ and $\rho$ are the thermal speed and gas density at the midplane. $C_W$ is a dimensionless parameter determining the wind mass flux (see below). It is worth noting that, because of the radial dependence of $\rho$ and $c$, the mass loss rate $\dot{\Sigma}_W$ is much larger at small radii.

S16 considers three different contributions to mass loss: the usual viscous accretion $(\dot{M}_{r\phi})$ due to the viscous stress parameterized by $\overline{\alpha_{r\phi}} (=\alpha)$, the mass loss due to the disk wind $(\dot{M}_z)$, and the mass accretion due to the  torque exerted by the disk wind at the wind base $z_{wl}$ (parameterized by $\overline{\alpha_{\phi z}})$. Using equation \ref{eqn:diskwind}, the total disk wind mass loss from $R$ to the outer edge of the disk $R_{\rm out}$ can be written as:
\begin{equation}\label{eqn:windmassflux}
    \dot{M}_z(R) = 2\pi\int_{R}^{R_{\rm out}}R'\,C_W(\rho c)_{mid}\,dR'.
\end{equation}
The other two terms, corresponding to the two terms on the RHS of equation (\ref{eqn:suzukigasvelocity}), are written by S16 as
\begin{equation}\label{eq:visctorque}
    \dot{M}_{r\phi}(R) = -\frac{4\pi}{R\Omega}\frac{\partial}{\partial R}(R^2\Sigma \overline{\alpha_{r\phi}}c^2)
\end{equation}
and
\begin{equation}\label{eq:windtorque}
    \dot{M}_{\phi z}(R) = -\frac{4\pi}{\Omega}R\overline{\alpha_{\phi z}}(\rho c^2)_{mid}.
\end{equation}

The value of  $(\overline{\alpha_{\phi z}})$ that parameterizes the wind torque is  estimated as a power-law in the surface density $\Sigma$. The functional form used by S16 is 
\begin{equation}
    \overline{\alpha_{\phi z}}(R)=10^{-5}\left[\frac{\Sigma(R)}{\Sigma_{init}(R)}\right]^{-0.66},
\end{equation}
where $\Sigma_{init}$ is the surface density at $t=0$. \citet{Bai2016} presented a similar variation in terms of vertical magnetic fields as $\overline{\alpha_{\phi z}} \propto (B_z^2/8\pi\rho c_s^2)^{-0.66}$.

The model by S16 constrains the  energetics of the wind by a non-dimensional parameter $C_{W,e}$ which is written as 
\begin{equation}\label{eqn:cwe}
    C_{W,e}=(1-\epsilon_{rad})\left[3\sqrt{\pi/2}h^2\overline{\alpha_{r\phi}} + \sqrt{2}h\overline{\alpha_{\phi z}}\right].
\end{equation}
Here, $\epsilon_{rad}$ is the fraction of energy lost through radiation and $h=H_g/R$ is the disk aspect ratio. %$\apr$ in equation \ref{eqn:cwe} is essentially the same as Shakura-Sunyaev turbulence efficiency $\alpha$ as described in section \ref{sec:diskmodel}.
S16 also constrain the wind mass flux by another parameter $C_{W,0}$ from local simulations, which depends on the gas density and thermal speed at the wind base. In their model, this dimensionless constant basically translates the DW model parameters to the respective midplane values. The value of $C_{W,0}$ generally hovers around the factor by which the gas density falls between the midplane and the wind base. Finally then, the wind flux $C_W$ is written as
\begin{equation}
    C_W={\text{min}}(C_{W,0}, C_{W,e}).
\end{equation}

\begin{figure}
%\centering
%\vspace{-.7cm}
%\includegraphics[width=\textwidth]{windloss-profile.pdf}
\includegraphics[width=0.45\textwidth]{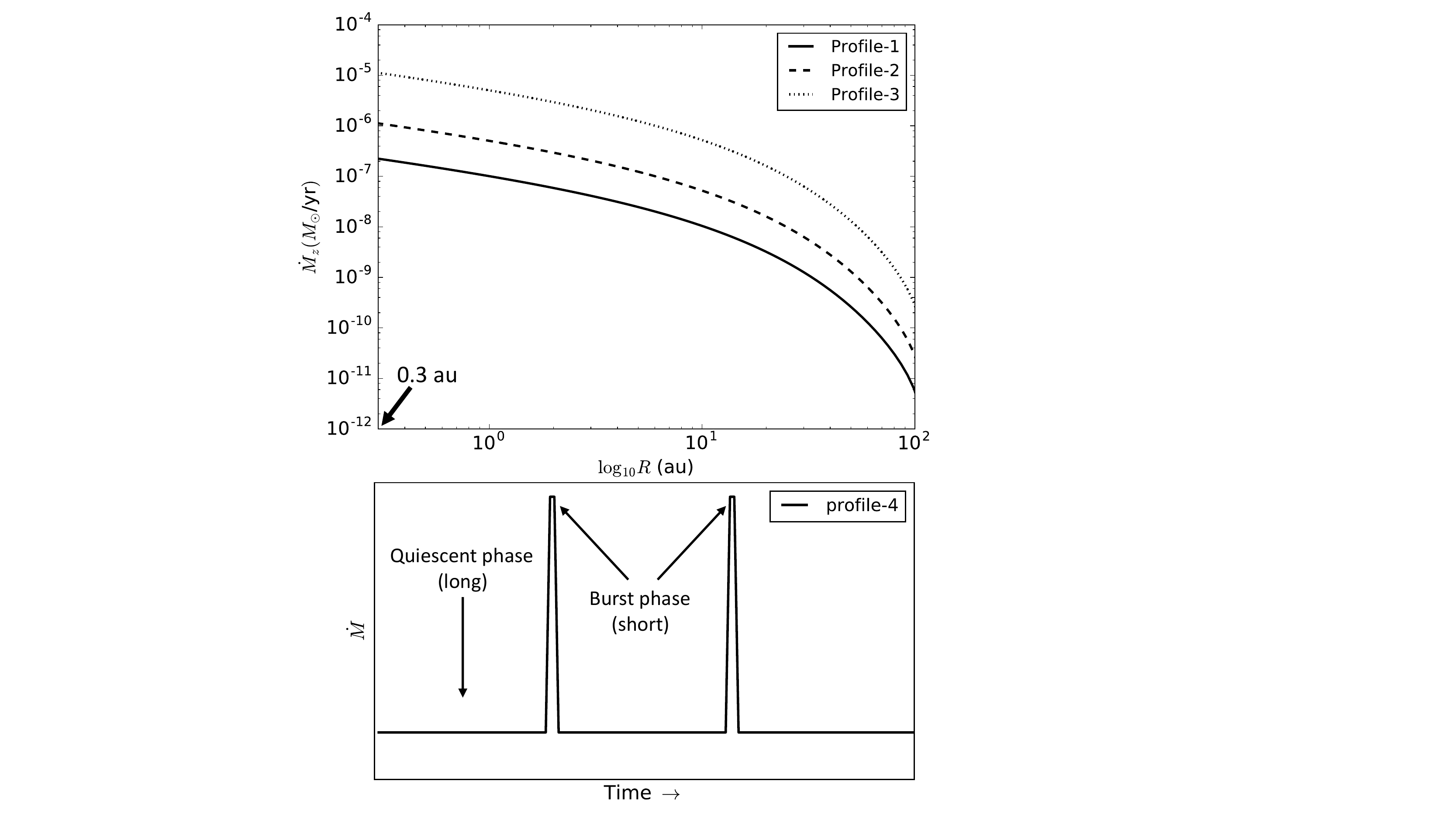}
\vspace{0 cm}
\caption{ {\bf Top panel:} The radially integrated mass loss rate from $R$ outwards, from equation \ref{eqn:windmassflux}. The solid curve  represents the lowest wind mass loss rate (profile-1) used in our simulations using parameters as shown in table \ref{tbl:simulations}. The wind parameters for every $\apr$ are carefully chosen to achieve profile-1 in each case.  The dashed and dotted curves (profile-2 \& 3) are the two higher mass loss rates used in our investigation.   {\bf Lower panel:} A customized time profile that we call profile-4, with mass losses having the {\it radial} dependence of the panel above. The quiescent phase with $\apr=10^{-3}$ and the low mass loss rate of profile-1 spans about $40$k years, and is followed by a short intense mass loss phase lasting $1$k years with the mass loss rate of profile-3.}
\label{fig:windprofile}
\end{figure}

In our simulations, we have used four different wind profiles, as depicted in figure \ref{fig:windprofile}. The top panel shows radial profiles of mass loss at three different intensities. The solid curve (profile-1) in the top panel is the one with the lowest mass loss rate with $\sim 10^{-7}M_{\odot}$~yr$^{-1}$ in the inner nebula. This profile matches with the weak disk wind model from S16 for an MRI-inactive case. For profile-1, we have used $\epsilon_{rad}=0.9$, $\apz=10^{-3}$ and $C_{W,0}=3\times 10^{-6}$ for both $\apr=10^{-3}$ and $10^{-4}$ cases.  Note that, this mass loss profile can be achieved using several different combinations of parameters as long as they are physically realistic. For example, in order to achieve a loss rate similar to profile-1, S16 used $\apr=8\times 10^{-5}$ with a slightly higher value for $C_{W,0}=10^{-5}$ and a torque $(\apz)$ dependent on $\Sigma_g$.

The mass loss rate from profile-2 (dashed line) is similar to the wind model in an MRI-active disk as depicted in S16, early in their disk evolution. However, our model, in accordance with our hypothesis, represents an even earlier stage when infall and outbursts are  probably still ongoing. This earlier timescale and stronger activity may slightly increase the mass loss rate over what we are assuming in profile-2. In addition to that, our disk model is more massive in the very inner part than the one used by S16, a factor that would enhance the wind mass flux further. Hence, we believe that profile-2 is  perfectly reasonable, or perhaps a little subdued, as a wind model for class 0/I phases of the solar nebula.

It is important to remark that we have not used any parameters to obtain profile-2 for our simulations. With the finding that profile-1 and profile-2 bear the same radial dependence,  profile-1 has been scaled by a constant factor in order to achieve the profile-2 mass loss. In the MRI-active case corresponding to profile-2, S16 used $\apr=8\times 10^{-3}$ as a conservative approach. However, we have used both $\apr=10^{-3}$ and $10^{-4}$ in equation \ref{eqn:advdiff} in order to encompass the effect of a dead zone along with an MRI-active upper zone where from the wind is launched.  It is important to note that given the small timescale of our disk evolution, this difference in $\apr$ will not affect our overall findings. In addition to that, the torque term in simulations with profile-2 is turned off $(\apz=0)$. With a nebula evolution time $\sim 10^5$~yrs, S16 found no difference in the surface density profiles in the evolution with and without wind torque beyond $0.1$~AU (See figure 1 from S16).  

Our profile-3 (dotted line) would require a set of parameters which are not   immediately justified by current models, but we explore for completeness as an extreme case. As in profile-2, profile-3 is also obtained by a scaling applied to the profile-1 mass loss profile, and also is treated assuming $\apz=0$. It is worth mentioning that we do not identify profile-3 as a canonical MHD wind or layered accretion flow, but rather with a short intense outflow during outburst  events. (see section \ref{depletionpathways} and \ref{sec:sec7.1} for more discussion.)

%However, we will still use the overall S16 formalism, \textcolor{red}{``while regarding profile-3 mass loss rates as representing a different mass loss process, such as layered accretion onto the Sun (see section \ref{sec:a3w2-a3w3} for more discussion)..." but, we now believe layered accretion per se is not able to deliver such large mass loss rates.} Hence, as in profile-2, profile-3 is also obtained by a scaling applied to the profile-1 mass loss profile.}

%\textcolor{red}{In section ??, we discuss how we deal with these higher mass loss cases.} 

The bottom panel of figure \ref{fig:windprofile} shows a customized wind mass-loss time profile (profile-4) intended to mimic a YSO-type repetitive  outburst behavior typical of many young star-disk systems (see section \ref{sec:intro} and \citet{calvetetal93,Zhuetal2010} or \citet{kadametal2020} for details). This mass loss profile alternates in time between  profile-1 and profile-3, and will approximate a slightly more complex pathway to MVE depletion in the early solar nebula (see section \ref{depletionpathways} for more discussion). Note that, in our simulations with profile-4 mass loss we use $\apr=10^{-3}$ throughout. Several previous results suggested an increase in $\apr$ during outbursts due to gravitational instability through accumulation of mass in the inner nebula \citep{Zhuetal2010,kadametal2020}. However, in those models, the outbursts are actually modeled and the increase in $\apr$ was a natural outcome of the simulations. We, on the other hand, impose the outburst effects generically in our simulation and do not need to elevate the value of $\apr$ in order to get those. Moreover, the overall short duration of the outburst phase will not significantly alter the outcome in terms of gas evolution and solid growth.

%\textcolor{red}{The above DW model is used to calculate the sink terms, or loss rates, for the various MVEs  $S_i$ as follows: blah blah, an equation. An assumption of the model is that vaporized MVEs are vertically mixed from the midplane up to the loss altitude $nH_g$ on a timescale short compared to the evolution time of the operative loss process. It is easy to show that the vertical mixing time for vapors and tiny grains is on the order of $n^2/\alpha$ orbit periods. This assumption is checked retroactively for each set of model parameters ... SO, do we have trouble anywhere?? What I remembered from before was the timescale only to one scale height but the factor of $n^2$ might matter! Since writing this I sent you a whole page or two of equations, this would be the place for a discussion of this topic, justifying (even vaguely) our assumption of instantaneous mixing.}

\subsection{Updated Opacity Prescription}\label{subsec:newopacity}

%Disk wind, a key component of our `open system' modeling for the current work

Testing our hypothesis on MVE depletion due to open system behavior required a realistic model of the hot inner disk, where temperature can reach as high as $5000$ K at early times, and all the solids evaporate. We have updated the opacity module in E16 following \citet{Fre14}, to include  tabulated Rosseland and Planck mean opacities for gases over higher temperatures, wider pressure ranges, finer temperature resolution and higher metallicities than in E16. The tabulated values are applicable for  pressures ranging from $1\mu$bar to $300$ bar and temperatures between $75$ to $4000$ K. The model from \citet{Fre14} also provides closed form prescriptions for mean opacities with weighting functions valid up to $7000$K. The  model includes opacity contributions from several molecular species, of which CH$_4$, CO$_2$, CO, and H$_2$O, H$_2$S, and H$_2$ are particularly relevant to our work.

\subsection{The MVE Species}\label{subsec:mves}

In order to investigate the problem of systematic depletion of MVEs in chondritic parent bodies, we have added the tracking of solid and volatile phases of a representative sample of the species shown in Fig. \ref{fig:braukmuller} to the code. Figure \ref{fig:braukmuller} includes many refractories (Al-V) and ``ultra refractories'' (Re-Th) in addition to MVEs. %, and we select several of these and normalize our results to V as in \citet{Braukmulleretal2019} to facilitate comparison.
\citet{Lodders2009} lists 22 species classified as moderately volatile, but often Ni, Co, Mg, Si and Fe have been included in this group \citep[see, for instance][]{Cassen1996} given that their condensation temperatures are comparable to that of the least volatile MVE (Ni). In total, we choose 17 species (as listed in  Table \ref{tbl:mves}) within this group ($T_{ci}$ between 664K (S) - 1353K (Ni)) which provides a sufficient range in condensation temperatures to capture the trend seen in Fig. \ref{fig:braukmuller}. Our sample of the {\it refractories} covers a range of $T_{ci}$ from 1408K (Pt)-1764K (Zr). The selected species are  chosen such that there is
%There are a total of $23$ MVE species listed in \citet{Cassen1996}, of which we have picked $11$ species ranging from Si to S in descending order of their condensation temperature, as is listed in table \ref{tbl:mves}. We have chosen elements with 
a reasonable difference in $T_{ci}$ between species to avoid the problem of having the EF of more than one species in the same numerical code radial grid point, at least initially. Several of the ``ultra-refractories" have a condensation temperature very close to 1659 K \citep{Lodders2009}. These include Th, Sc, Y, Gd, Tb, Dy, Ho, Er, Tm, and Lu. We combine them and and label the set as ``UR'' in Table \ref{tbl:mves}. Finally, our model also includes mass fractions of ``moderately refractory" organics (CHON or Tholin-like materials), water, and the supervolatiles CO, methane and CO$_2$. We do not  track the supervolatiles for the purpose of this paper, but their depletion also seems significant in inner solar system bodies,  and will provide a good future test of the predictability of the model \citep[see section \ref{subsec:watercarbon}, and][] {Estradaetal2022}. Chondrites also contain smaller amounts of even more refractory amorphous or even graphitic C-species (IOM), that we also do not track for the purpose of this paper.

% Using X might get confused with that representation typically used for the fraction of H
The initial mass fraction $\eta_{i,0}$ for any species $i$ is calculated approximately as
\begin{equation}\label{eqn:massfraction}
\eta_{i,0}=\frac{N_i}{N_{H}+N_{He}}\frac{M_i}{1.26},
\end{equation}
where $N_i$ is the number of atoms or molecules in species $i$ and $M_i$ is the molecular weight of species $i$. The factor $1.26$ arises from the correction due to the combined presence of both H and He. The initial number of atoms of the selected species, using the elemental abundances of the proto-Sun scaled to $10^6$ Si atoms, are adapted from \citet[][see also \citealt{Lodders2021}]{Lodders2003}, and are detailed in Table \ref{tbl:mves}. 

Our mass fractions for all $i$ are constructed by starting with the initial refractory organics fraction in \citet{Pollacketal1994} which in our treatment accounts for $\sim 26$\% of carbon atoms \citep[see][]{MummaCharnley2011, Woodwardetal2020}. The remaining carbon is partitioned equally amongst CO, CH$_4$ and CO$_2$ \citep[see][]{Estradaetal2021}. The silicate rock fraction is a combination of magnesium-rich orthopyroxene and olivine ({\it i.e.} incorporates all Mg and Si), while all of the aluminum is contained in corundum (Al$_2$O$_3$). The remaining oxygen is placed in H$_2$O. Iron-rich silicates are not included, but rather the mass fraction of refractory iron is what is left after all the S is placed in troilite (FeS).
The above fractions overwhelmingly account for the total initial solids mass fraction, or metallicity $Z$, of our disk models. Although all of the refractories and MVEs we adopt are, according to \citet{Lodders2003}, initially present in a mineral host which determines their $T_{ci}$, for simplicity the mass fractions of the refractories (with the exception of Al, which is treated as part of the Al$_2$O$_3$ molecule)  are calculated just from their (very small) elemental abundances (though, for Ca see below). %\textcolor{red}{But Ca also is abundant from table 1, and carries some O with it... how to handle this? maybe don't bring it up?} 

The contribution of the MVEs to the calculation of the Rosseland (and Planck) mean opacity, is unevenly handled but fortunately, not large. The MVEs are often referred to as ``trace elements" because of their small abundances. Optical constants for these species are in many cases either not available, or have only been determined over a very limited range of wavelength, whereas we do have fairly well determined optical constants for our major solid species like iron, silicates, refractory organics, and the major ices, \citep[see][]{Cuzzietal2014}, as well as corundum \citep{Koikeetal1995}. Given the very small contribution of the MVEs to the overall mass fraction, we then generally ignore the opacities of most of the species in Table \ref{tbl:mves} in our calculation of the temperature except when an element has a reasonable range of wavelength coverage, or the host has optical constants similar to  those of our major species. For example, the initial host of Zn and Mn has properties similar to silicates (Forsterite and Enstatite, e.g. [Zn,Mn]SiO$_3$), while Na and K are initially in Feldspars (e.g., [Na,K]AlSi$_3$O$_8$). We treat all of these as ``silicates''. On the other hand, for Ca which is relatively abundant, we use Gehlenite (Ca$_2$Al$_2$SiO$_7$) as the host for which some optical properties are available \citep{Burshtein:93,Speck_etal_2011}. 
In most cases the number of, e.g., O atoms contained in these hosts is minimal so that assigning mass fractions from their elemental abundances is reasonable. If all the Ca were contained in Gehlenite, this would amount to $\sim 1.5\%$ of all O atoms (and 3/4 the available Al), but for this work we ignore this additional complexity.
%\textcolor{red}{and, what.., the properties are known?}. 

The total initial metallicity including all species is $Z\simeq 0.014$. The evolution of refractories, MVEs, and volatiles  are tracked in both the solid and vapor form following equation \ref{eqn:advdiff}.

\begin{deluxetable}{c c c c} [!hbt]
\tabletypesize{\scriptsize}
\setlength{\tabcolsep}{10pt}
\renewcommand{\arraystretch}{1.2}
\tablecolumns{0} 
%\tablenum{1}
\tablewidth{\columnwidth} 
%\tablecaption{MVE species included in our model, along with their condensation temperatures $(T_{ci})$ and initial abundances. \label{tbl:mves}} 
\tablecaption{Refractory and MVE species \label{tbl:mves}}
\tablehead{
%\colhead{MVE}  & \colhead{$T_{ci}$} & \colhead{Number of} & \colhead{Initial mass} \\
  & \colhead{$T_{ci}$} & \colhead{Number of} & \colhead{Initial mass} \\
 \vspace{-0.6cm} \\
%\colhead{species} & \colhead{$(K)$} & \colhead{atoms $(N)$\tablenotemark{a}} &
 & \colhead{$(K)$} & \colhead{atoms $(N)$\tablenotemark{a}} &
\colhead{fractions $\eta_0$}}
\startdata 
\vspace{-0.4cm}\\
Zr & 1764 & 11.33 & $4.16\times 10^{-8}$ \\
Hf & 1703 & 0.170 & $1.06\times 10^{-9}$ \\
Al\tablenotemark{b} & 1665 & $8.41\times 10^{4}$ & $1.28\times 10^{-4}$ \\
UR\tablenotemark{c} & 1659 & 40.0 & $9.26\times 10^{-8}$ \\
Mo & 1590 & 2.601 & $7.43\times 10^{-9}$ \\
Ti & 1584 & 2422 & $3.45\times 10^{-6}$ \\
Ca & 1517 & $6.287\times 10^4$ & $7.50\times 10^{-5}$ \\
Sr & 1464 & 23.64 & $6.17\times 10^{-8}$ \\
V & 1429 & 288.4 & $4.37\times 10^{-7}$ \\
Pt & 1408 & 1.357 & $7.88\times 10^{-9}$ \\
\hline
%Si & $1450$ & $1\times 10^6$ & $3.01\times 10^{-3}$  \\
Ni,Co & $1353$ & $5.012\times 10^4$ & $8.77\times 10^{-5}$  \\
Fe\tablenotemark{d} & 1350 & $8.38\times 10^5$ & $6.54\times 10^{-4}$ \\
Mg\tablenotemark{e} & 1336 & $1.02\times 10^6$ & $3.01\times 10^{-3}$ \\
Cr & $1296$ & $1.286\times 10^4$ & $1.99\times 10^{-5}$  \\
P & $1229$ & $8.373\times 10^3$ & $7.72\times 10^{-6}$ \\
Mn & $1158$ & $9.168\times 10^3$ & $1.50\times 10^{-5}$  \\
As & 1065 & 6.089 & $1.36\times 10^{-8}$ \\
Cu & $1037$ & $5.270\times 10^2$ & $9.97\times 10^{-7}$   \\
K & $1006$ & $3.692\times 10^3$ & $4.30\times 10^{-6}$  \\
Ga & 968 & 35.97 & $7.47\times 10^{-8}$ \\
Na & $958$ & $5.751\times 10^4$ & $3.81\times 10^{-5}$  \\
Ge & $883$ & $1.206\times 10^2$ & $2.61\times 10^{-7}$ \\
Rb & 800 & 6.572 & $1.67\times 10^{-8}$ \\
Bi & 746 & 0.139 & $8.64\times 10^{-10}$ \\
Zn & $726$ & $1.226\times 10^3$ & $2.38\times 10^{-6}$  \\
S\tablenotemark{d} & $664$ & $4.449\times 10^5$ & $1.16\times 10^{-3}$  \\
%\vspace{-0.2cm}\\
%\hline
\vspace{-0.4cm}\\
 %\vspace{-0.2cm}\
 \enddata 
%\vspace{-1.8cm} 
%\tablecomments{Science simulation set: }
%\tablenotetext{a}{The number of atoms, normalized by $1\times 10^6$ Si atoms, are adopted from Lodders 2003, Table 2, third column.The initial mass fractions are calculated using equation \ref{eqn:massfraction}.}
\tablenotetext{a}{Scaled to $10^6$ Si atoms, adopted from Lodders 2003}
\tablenotetext{b}{Al is contained in Al$_2$O$_3$}
\tablenotetext{c}{``Ultra refractories'', Lodders (2003), Table 8}
\tablenotetext{d}{All S is in FeS, remaining Fe is refractory}
\tablenotetext{e}{All Mg (and Si) in silicates}
\end{deluxetable}

\begin{deluxetable}{c c c c c c c} [!hbt]
\tabletypesize{\footnotesize} 
%\tabletypesize{0.8\scriptsize}
\tablecolumns{0} 
%\tablenum{1}
\tablewidth{0pt} 
\tablecaption{List of simulations and relevant parameters. \label{tbl:simulations}} 
\tablehead{ 
\colhead{Simulation} \vspace{-0.2cm} & \colhead{} & \colhead{} & \colhead{} & \colhead{$R_{in}$} & \colhead{$\dot{M}$} & \colhead{Particle} \\
 \vspace{-0.4cm} \\
\colhead{} & \colhead{$\apr$} & \colhead{$\apz$} & \colhead{$C_{W,0}$} & \colhead{} & \colhead{} & \colhead{} \\ 
\vspace{-0.6cm}\\
\colhead{Name} & \colhead{} & \colhead{} & \colhead{} & \colhead{(au)} & \colhead{profile \tablenotemark{a}} & \colhead{type} }
\startdata 
\vspace{-0.2cm}\\
A3W1 & $10^{-3}$ & $10^{-3}$ & $3\cdot 10^{-6}$ & $0.3$ &  1 & porous  \\
A3W2 & $10^{-3}$ & \nodata & \nodata & $0.3$ &  2 & porous   \\
A3W3 & $10^{-3}$ & \nodata & \nodata & $0.3$ &  3 & porous  \\
A3W2S & $10^{-3}$ & \nodata &  \nodata & $0.3$ &  2 & solid  \\
A3W3S & $10^{-3}$ & \nodata & \nodata & $0.3$ &  3 & solid  \\
\vspace{-0.2cm}\\
\hline
%\vspace{-0.2cm}\\
%A2W1 & $10^{-2}$ & $10^{-2}$ & $3\cdot 10^{-6}$ & $0.3$ &  1 & porous \\
%A2W2 & $10^{-2}$ & \nodata & \nodata & $0.3$ &  2 & porous  \\
%A2W3 & $10^{-2}$ & \nodata & \nodata & $0.3$ &  3 & porous  \\
 %\vspace{-0.2cm}\\
% \hline
 \vspace{-0.2cm}\\
A4W1 & $10^{-4}$ & $3\cdot 10^{-4}$ & $3\cdot 10^{-6}$ & $0.3$ & 1 & porous \\
A4W2 & $10^{-4}$ & \nodata & \nodata & $0.3$ & 2 & porous  \\
A4W3 & $10^{-4}$ & \nodata & \nodata & $0.3$ & 3 & porous \\
 \vspace{-0.2cm}\\
\hline
 \vspace{-0.2cm}\\
 A3W4 \tablenotemark{a} & \nodata & \nodata & \nodata & $0.3$ & 4 & porous  \\
%  A3W401 \tablenotemark{b} & \nodata & \nodata & \nodata & $0.1$ & 4 & porous  \\
 \vspace{-0.2cm}\\
 \hline
 \vspace{-0.2cm}\\
A3W3R01 & $10^{-3}$ & \nodata & \nodata & $0.1$ & 3 & porous   \\
A3W3R02 & $10^{-3}$ & \nodata & \nodata & $0.2$ & 3 & porous   \\
 \vspace{-0.2cm}\\
\hline
 \vspace{-0.2cm}\\
 %\vspace{-0.2cm}\
\enddata 
%\vspace{-1.8cm} 
%\tablecomments{Science simulation set: }
\tablenotetext{a}{See Figure \ref{fig:windprofile} for respective wind profiles.}
\tablenotetext{}{See sections \ref{subsec:diskwind} and \ref{sec:results} for discussion of  $\apr$ and other wind parameters.}
\vspace{0.2in}
\end{deluxetable}

\subsection{Turbulent intensities}

The intensity of turbulence in a planet forming disk is a matter of active research and there are significant uncertainties regarding the value of $\apr$, a measure of turbulence intensity. On the theoretical end, the community has reached a consensus that magnetorotational instability (MRI) \citep{balbus1991}, producing $\apr \sim 10^{-2}$, will not be active throughout {\it most} of the disk, except for the very top layers with low gas density and fewer particles \citep{Gammie1996, Zhuetal2010}, and it has even been claimed that it is not active there \citep{Bai2013}. Meanwhile, several non-MHD, purely hydrodynamic instabilities have been proposed in the last two decades \citep{lovelace1999, Nelsonetal2013, Lyra2014, Turneretal2014, Stoll_Kley_2014,  Marcusetal2015}, estimating values of $\apr$ in the range of $\sim 10^{-4}-10^{-3}$. 

On the observational end, \citet{flaherty2015, flaherty2017}, using high resolution CO line spectra from ALMA , found a spectral broadening less than 3\% of the local thermal speed, corresponding to an $\apr$ between $10^{-4}$ and $10^{-3}$. Meanwhile, \citet{pinte2016} analyzed vertical settling of $100~\upmu$m particles and estimated $\apr\sim 10^{-4}$. Finally, \citet{Dullemondetal2018}, from modeling of particle belts observed by ALMA, give a lower limit of $\apr$ as  $10^{-4}$, but state that much {\it larger} values of $\apr$ are also consistent with their data. With these recent developments in mind, we chose a fiducial value of $\apr = 10^{-3}$ (section \ref{sec:fiducialmodel}), and also performed simulations with the same mass loss profiles with $\apr=10^{-4}$, the results of which are presented in section \ref{subsec:varyalpha}.

\section{Simulations and Setup}\label{sec:simulations}

The simulations performed for this work are tabulated in table \ref{tbl:simulations}, with their names and relevant parameters. The number in the simulation name following the first letter `A' denotes the negative exponent of $\apr$. For example, A3 and A4 stand for simulations with $\apr=10^{-3}$ and $10^{-4}$, respectively. Similarly, the number following the letter `W' is indicative of the wind profile; W1 stands for profile-1 and so on. Most of the simulations are for porous aggregate particles, following \citet{Estradaetal2021}; for comparison we ran several cases for solid (zero-porosity) particles, following \citet{Estradaetal2016}. Simulation names with `S' at the end denote solid particle cases. The last two simulations in table \ref{tbl:simulations}, ending with `R01' and `R02' have inner radii $0.1$ and $0.2$~AU respectively.  The radial domain in each simulation is divided into $96$ radial grid points, equidistant in $\log$-space. We ensure that no radial bin contains the evaporation fronts of more than one MVEs at $t=0$. However, this restriction was not imposed during the evolution due to the fixed-grid nature of our model. At the inner and outer edges of the disk, the boundary condition is imposed following a combination of Neumann and Dirichlet conditions (Robin boundary condition) through mass flux of vapor (and/or solids). Ghost points outside the boundaries are used in order to make the boundary condition implicit. For details of the said implementation, the reader is referred to \citet{Estradaetal2016}.

\section{RESULTS}\label{sec:results}

In this section, we present the main results from the set of simulations tabulated in table \ref{tbl:simulations}.  In these plots, the fractional abundance of each element (in both solid and vapor phases) is normalized by the fractional abundance of V in CI material and denoted $\eta_{MVE}/\eta_V$. The depletion signatures, in each case, are presented at $0.3$, $0.5$, $0.75$ and $1$~AU as snapshots taken at different times. The models generally assume $R_{in}=0.3$AU. For simulations with $R_{in}=0.1$~AU and $0.2$~AU, additional radii are added to the figures as well. We start this section with the results from simulation A3W1 with $\apr=10^{-3}$ and wind loss profile-1, followed by results with different mass loss profiles and $\apr$.

\subsection{Profile-1 mass loss with $\apr=10^{-3}$.}\label{sec:fiducialmodel}

\begin{figure*}[!htbp]
\centering
\includegraphics[width=\textwidth]{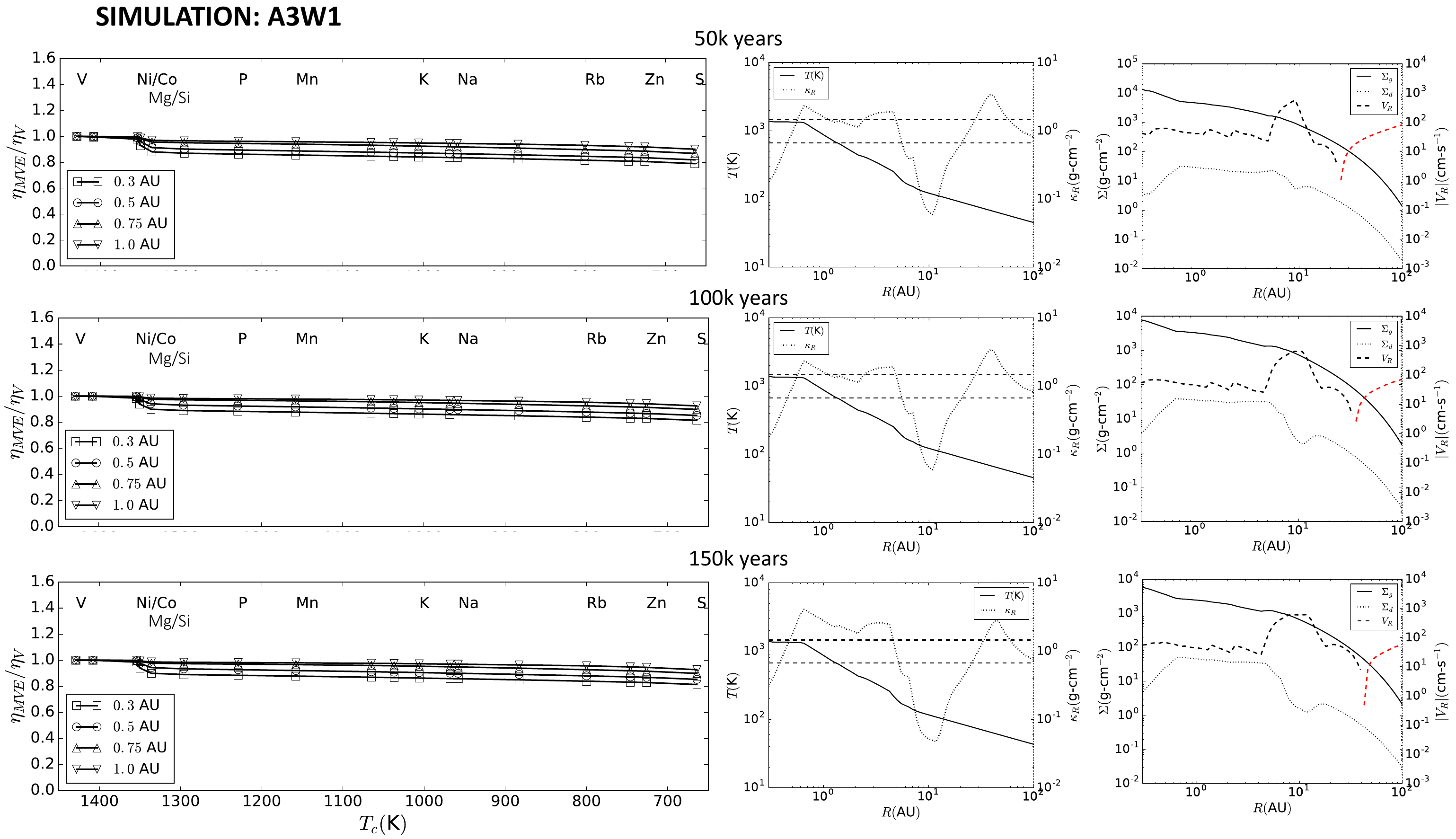}
\vspace{0 cm}
\caption{Results from simulation A3W1: $\apr=10^{-3}$ with profile-1 mass loss. The three rows correspond to the snapshots at $50$, $100$ and $150$ thousand years. The first column shows the depletion profile at $0.3$ (square), $0.5$ (circle), $0.75$ (triangle) and $1$~AU (inverted triangle) for the MVEs listed in table \ref{tbl:simulations}, normalized by the depletion of Vanadium (V) in CI material. Note that Ni/Co and Mg/Si are not ratios, but are mentioned in that way in order to fit the closely spaced elements in the plot (notation used as (Ni, Co) and (Mg, Si) in text). The figures in the second column are the corresponding radial temperature profile (solid black line, left axis) and Rosseland mean opacities $(\kappa_R)$~(black dotted curve, right axis). The two dashed black horizontal lines correspond to $T=1450$ (upper line) and $664$~K (lower line; $T_{ci}$ for S), plotted for reference. The third column shows the surface densities for gas (solid black) and solids (dotted black) on the left axis along with the magnitude of the radial drift velocities $\left|V_R\right|$ for solids (dashed curve) on the right axis. The red part of the radial drift speed plot signifies an outward drift. } 
\label{fig:alpha3-profile1}
\end{figure*}

Simulation A3W1, with $\apr=10^{-3}$ and mass loss  profile-1, is our fiducial model;  the results are shown in figure \ref{fig:alpha3-profile1}. Three snapshots at $50$k (top row), $100$k (middle row) and $150$k (bottom row) years are presented. The depletion signatures of the MVEs are shown in the left column.  The middle column sub-figures in each row show the radial temperature profile $T(R)$  on the left axis (solid black) and Rosseland mean opacities $(\kappa_R)$ on the right axis (dotted black). The two horizontal dashed lines represent $T=1450$~K and $664$~K. The figures in the third column show the surface densities for gas $\Sigma_g$ (solid black) and solids $\Sigma_d$ (dotted black) on the left axis and the magnitude of the net inward drift velocity $\left|V_R\right|$\footnote{Note that the radial velocity for solids is denoted by $v_p$ in equation \ref{eqn:advdiff}. However, for all our arguments, we are only concerned with the absolute values of the inward drifting speed of the solids and the notation $\left|V_R\right|$ will be used hereafter.} for the major mass carrying solids (dashed black) on the right axis. The red dashed part in the velocity plot represents the radial positions where solids are (on average) advected outward with the viscously evolving gas. Note in the middle column figures, the particle opacity $\kappa$ decreases dramatically inside 0.7AU, where all elements more volatile than V (including the rock-forming elements that provide most of the solids mass) are vaporized.

From the figure we can see that at $50$k years the depletion for S only reaches $0.8$ at $0.3$~AU, compared to the observed values of 0.15-0.40 (figure \ref{fig:braukmuller}). The depletions at larger radii are even smaller. The outcome does not change at $100$k and $150$k years and the profile remains constant for the full simulation time without coming close to the desired level of depletion. 

A sharp step or drop is noted for elements slightly more volatile than  (Ni,Co)\footnote{In all our depletion plots, (Ni,Co) is written as Ni/Co in order to fit closely spaced elements. The same is done for (Mg,Si). Note that, they do NOT mean elemental abundance ratios.}, after which the pattern remains relatively flat to higher volatilities.  This step feature is ubiquitous in all our results, and gets more prominent as more depletion is achieved. The observations (figure \ref{fig:braukmuller}) exhibit a similar change near (Cr,P), whose adopted $T_{ci}$ differs from that of our (Ni,Co) by $\sim 100$~K. This step is attributed to the dramatic opacity change associated with evaporating most of the rock-forming elements %the opacities and relative abundances of the different solid compounds in our model 
(see section \ref{sec:discussionopacity} for a detailed discussion).

Overall, with the low mass loss rate of profile-1, MVEs in the part of the disk inside $1$~AU do not get depleted fast enough. The temperature remains above the condensation temperature of all MVE species at least out to $0.7$~AU (middle panel) and all MVEs are lost at an equal rate. Moreover, an important factor is the inward drift of solid particles, the flux of which nearly matches the mass loss rate from the inner disk.  In our model, the MVEs are lost from the inner disk only in the form of vapor, which is produced after  inward drifting MVEs cross their respective evaporation fronts (EFs) in solid form. Profile-1 translates to a gas mass-loss $\dot{M}$ of $\sim 10^{-9}$~g-cm$^{-2}$-s$^{-1}$ at $1$~AU. The loss rate of MVEs can be roughly estimated as $\dot{M}\sum_S^{Mg} \eta_i$, which amounts to $\sim 4 \times 10^{-12}$~g-cm$^{-2}$-s$^{-1}$ (Figure \ref{fig:alpha3-profile1}, right column). Meanwhile, the inward mass flux of solids can be calculated as $\Sigma_g\sum_S^{Mg} \eta_i\left|V_R\right|/2\pi R \sim 3\times  10^{-12}$g-cm$^{-2}$-s$^{-1}$ with $\left|V_R\right|\sim 10^2$~cm-s$^{-1}$ - basically the same. So, it can be safely stated that the MVEs lost from the inner disk are constantly replenished by the drifting solids, stagnating the depletion signature long before it reaches the desired level.

\subsection{Higher Mass Loss}\label{sec:a3w2-a3w3}

 With the failure of profile-1 mass loss rate to reach a depletion magnitude comparable with figure \ref{fig:braukmuller}, we explore higher mass loss rates using profile-2 and profile-3 of figure \ref{fig:windprofile}. 

\subsubsection{Profile-2}

As mentioned in section \ref{subsec:diskwind}, according to the S16 model, profile-2 is consistent with a weak disk wind in an MRI-active disk, possibly starting earlier in our models at the final stages of infall. This still constitutes ``open system" behavior, in that the more volatile MVEs are irreversibly fractionated away from the more refractory ones, which remain settled within the first vertical scale height of the nebula to be accumulated into planetesimals, although perhaps in a different time and location than when and where the depletion occurs. The physics of mass loss for layered accretion is more like standard alpha-disk viscosity in a thin upper layer than a wind that exerts a local torque on the entire disk. For convenience in this first study of the potential importance of open-system behavior,  we simply extend the mass loss rate formalism developed for the disk wind (equation \ref{eqn:suzukigasvelocity}) but set the separate torque term   (the $\apz$ term on the right of equation \ref{eqn:suzukigasvelocity}) to zero. Under such conditions, as we will see below, the loss rate here exceeds the inward mass flux due to radial drift of solids, resulting in  depleted signatures for the MVEs.

%As mentioned in section \ref{subsec:diskwind}, according to the S16 model, the profile-2 mass loss may require  values of $\apz$ and $C_{W,0}$ which are not physically consistent. However, recent theoretical considerations do not fully preclude profile-2 as a valid disk wind scenario. Following the work of \citet{Suzukietal2016}, \citet{Kunitomoetal2020} estimated the mass loss rates through MHD disk winds in protoplanetary disks. In their strong disk wind case \citep[][figure 2]{Kunitomoetal2020} the mass loss rate due to an MHD wind reaches $5\times 10^{-7}~M_{\odot}$yr$^{-1}$ which is slightly less than the mass loss rate from profile-2. So, while we identify profile-2 only as a generic mass loss, MHD disk winds could actually be consistent with these loss rates, especially in class 0/I stages of the PPD.

\begin{figure*}
%\centering
%\vspace{-.7cm}
%\includegraphics[width=\textwidth]{Braukmuller_fig.pdf}
\includegraphics[width=\textwidth]{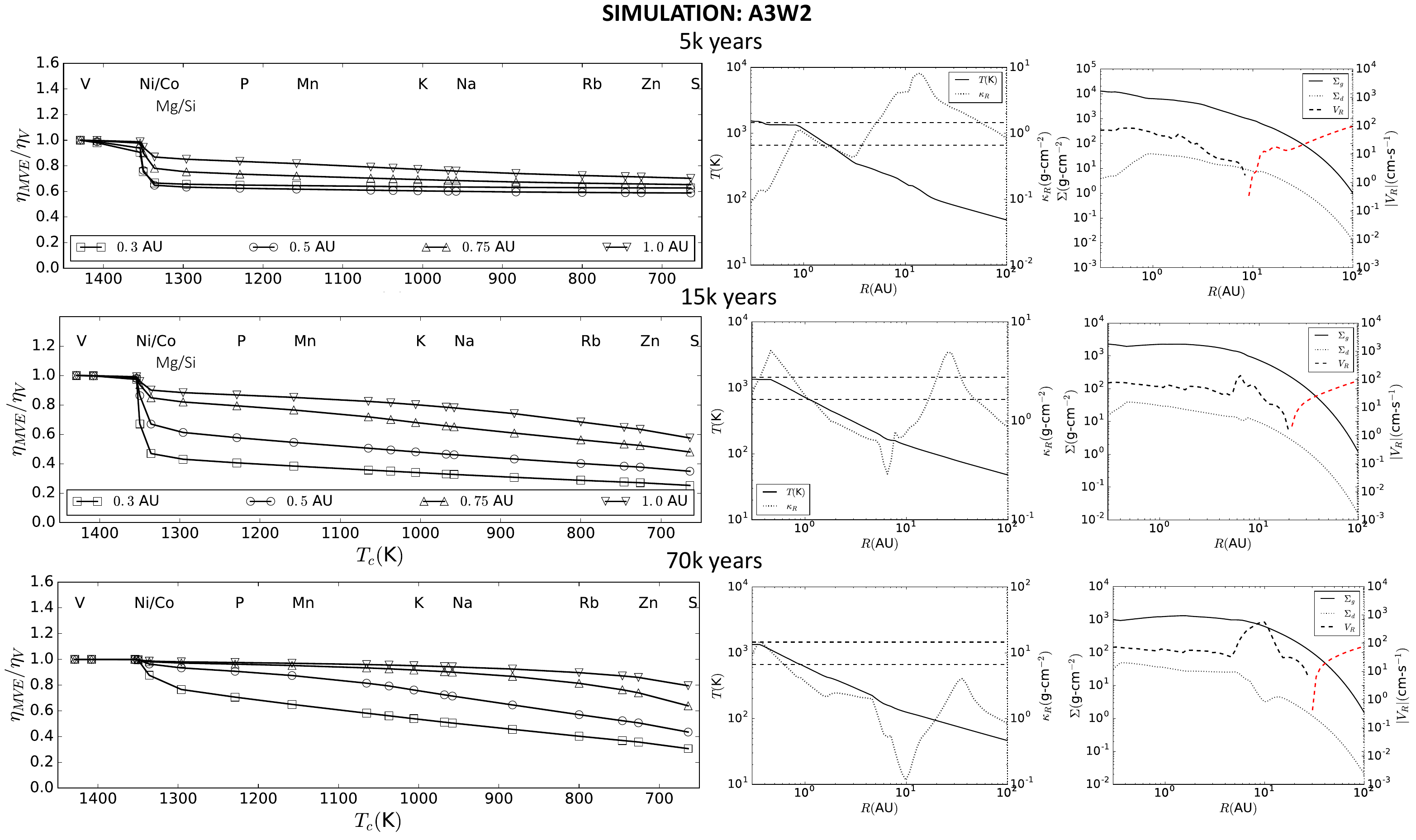}
\vspace{0 cm}
\caption{Results for simulation A3W2 with $\apr=10^{-3}$ and mass loss profile-2, presented in a similar fashion as figure \ref{fig:alpha3-profile1}. The snapshots are presented for $5$k (top row), $30$k (middle row), and $70$k (bottom row) years.} 
\label{fig:alpha3-profile2}
\end{figure*}

In figure \ref{fig:alpha3-profile2}, we present three snapshots at $5$k (top row), $30$k (middle row) and $70$k (bottom row) years from simulation A3W2 with $\apr=10^{-3}$ and profile-2. The structure of the figure is the same as figure \ref{fig:alpha3-profile1}, with the depletion signature in the first column, $T(R)$ and $\kappa_R$ in the second column, and $\Sigma_{g}$, $\Sigma_d$, and $\left|V_R\right|$ in the third column. At $5$k years, the MVE abundances drop slightly for elements more volatile than the step near (Mg,Si), but the profile remains flat in $T_{ci}$ with $\eta_{MVE}/\eta_V \sim 0.6-0.8$  at all radii shown.  The temperature at this stage remains above the $T_{ci}$'s of all MVE species (see the temperature plot for $5$k years) resulting in a loss of all species at similar rates. A large loss in solids due to evaporation inside $1$~AU can be inferred from the drop in $\kappa_R$. The signatures at different radii then start to diverge as time goes on and the disk cools down. The snapshot at $30$k years
shows the depletion for S reaching $0.2$ at $0.3$~AU,  and $0.6$ at $1$~AU. This depletion signature is consistent with figure \ref{fig:braukmuller}, but following further disk cooling (e.g., see the $70$k year snapshot), the signature starts to wash out or weaken due to the effect of inward drift. The inward drift velocity remains fairly constant over the times presented (third column)  and, following the depletion at $30$k years, because of the cooling, it is able to overcome the ongoing loss in vapor from the inner disk. 

\subsubsection{Profile-3}

  In the profile-3  range of mass loss rates, the physics of disk winds as described by both \citet{Suzukietal2016} and  \citet{Kunitomoetal2020} is inapplicable. Instead, we present this higher mass loss rate as a generic one which could be associated with the enhanced rates seen during an outburst. We outline our thoughts on the possible nature of this mass loss mechanism in section \ref{depletionpathways}. Similar to profile-2, we set $\apz=0$ here as well.
 
\begin{figure*}[t]
%\centering
%\vspace{-.7cm}
%\includegraphics[width=\textwidth]{Braukmuller_fig.pdf}
\includegraphics[width=\textwidth]{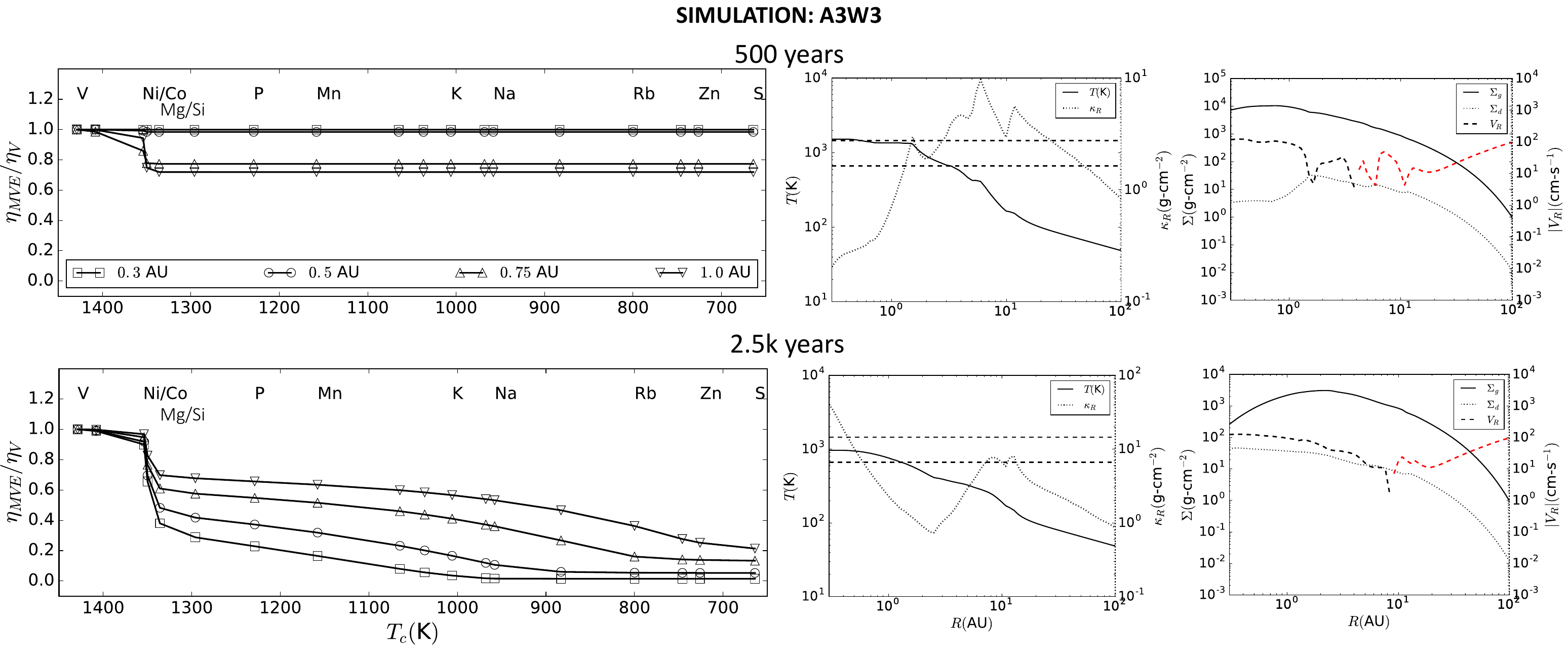}
\vspace{0 cm}
\caption{Results for simulation A3W3 with $\apr=10^{-3}$ and mass loss profile-3, presented in a similar fashion as figure \ref{fig:alpha3-profile1}. The snapshots are presented for 500~years (top row) and 2.5k~years (bottom row).} 
\label{fig:alpha3-profile3}
\end{figure*}

In figure \ref{fig:alpha3-profile3} we present the results for profile-3 mass loss with $\apr=10^{-3}$ (simulation A3W3). Two snapshots are presented at $500$ years (top row) and $2.5$k years (bottom row). It is important to note that the entire process of reaching the desired depletion   takes place in a very short period of time (here, 3k years). The radial temperature profile at $500$ years stays above $1450$~K inside $1$~AU, leading to a flat depletion profile with depletion levels of $\sim 0.7$ at both $0.3$ and $0.5$~AU on the volatile side of the ``edge" near (Mg,Si). Significant decrease in $\kappa_R$ occurs in the hot region inside $1$~AU due to the evaporation of most solids (see $\Sigma_d$ profile in third column). This results in rapid disk cooling, and along with the high rate of mass loss produces an almost complete depletion of S at $0.3$~AU by $2.5$k years. This cooling of the inner disk elevates $\Sigma_d$, and hence, $\kappa_R$ increases by more than two orders of magnitude. For this profile-3 case, runs are not executed beyond this short time as we expect the depletion signature to be time limited. However, the reader is referred to the profile-4 case in section \ref{sec:profile4} where profile-3 is used in combination with low mass loss rates in simulations executed for longer duration. 

Thus, simulation A3W3 achieves a depletion signature which is consistent with figure \ref{fig:braukmuller} in a short period of time. In fact, the magnitude of depletion in this case exceeds that of the observed values for the ``plateau" elements ($\sim 0.002$ vs $0.18$ in figure \ref{fig:braukmuller}). Moreover, the plateau element depletion pattern at $2.5$k years is quite flat between Na and S at $0.3$~AU and between Ge and S at $0.5$~AU. The corresponding temperature profile in the second column of figure \ref{fig:alpha3-profile3} shows that the temperature remains flat at $1000$~K out to $0.6$~AU, resulting in equal loss rates for all MVE species with $T_{ci}$ less than that temperature, making the profile flat. Similar flat patterns are observed at $0.75$ and $1$~AU as well, but for a shorter range of $T_{ci}$ (between Rb and S at $0.75$~AU). Consistent with this, the temperature at $0.75$~AU at $2.5$k years is $\sim 800$~K which is the condensation temperature of Rb (table \ref{tbl:mves}). 

While this flat appearance of the depletion profile in the ``plateau element" range is encouraging because the observations also show a similar and prominent feature (see  figure \ref{fig:braukmuller}), the situation is more complicated, because it appears that the observed ``plateau element" material {\it may} actually be ``flat" in this range of $T_{ci}$  merely because it is CI material remixing with depleted material (see section \ref{sec:matrix}). That is to say, a successful depletion signature should actually asymptote at lower values of  $\eta_{MVE}/\eta_V$ from Na to S than seen in figure \ref{fig:braukmuller}, in order to allow for this subsequent mixing. This situation is discussed more in section \ref{sec:slowmixing}.

\subsection{$\apr=10^{-4}$ with different mass-loss profiles}\label{subsec:varyalpha}

\begin{figure*}
%\centering
%\vspace{-.7cm}
%\includegraphics[width=\textwidth]{Braukmuller_fig.pdf}
\includegraphics[width=\textwidth]{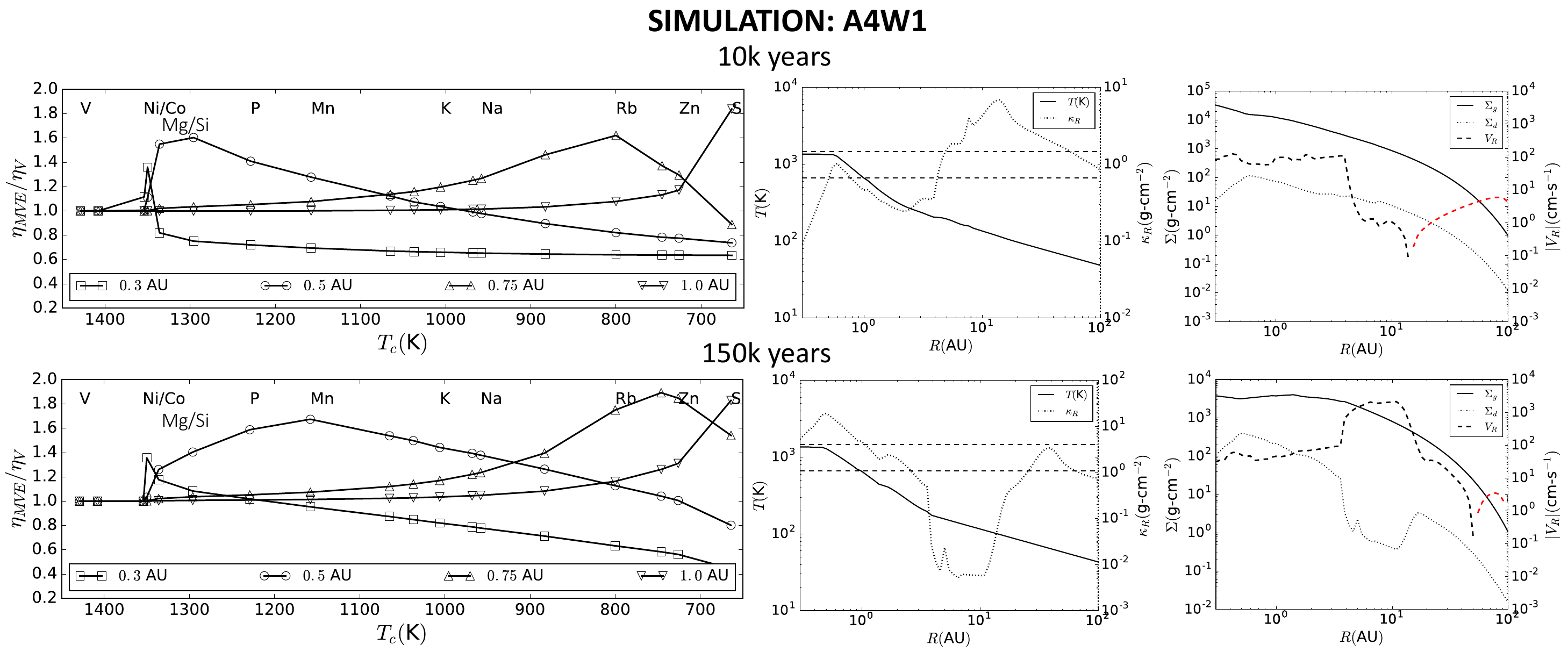}
\vspace{0 cm}
\caption{Results for simulation A4W1 with $\apr=10^{-4}$ and mass loss profile-1. Two snapshots at $10$k (top) and $150$k (bottom) are presented. The left panels show the depletion pattern. In the subpanels at center and right, the corresponding temperature, $\kappa_R$, $\Sigma_g$, $\Sigma_d$, and $V_R$ are shown as in previous figures. Interestingly, contrary to the usual  depletion signature, the MVEs show an enhancement in this simulation for radii except the very innermost ones ($0.3$ and $0.5$~AU). This enhancement is most prominent for species with EFs lying near the various radii that show peaks in these species, and can be attributed to the substantial inward drift of solid materials, seen in the ensuing deep cutouts in $\kappa$ and $\Sigma_d$ between 4-15AU, combined with the weak radial diffusion of particles and vapor, and viscous evolution of the gas disk, for this low $\apr$. %The radial velocity plot for $150$~k years clearly shows an elevated radial speed of material and a corresponding dip in the solid abundances between $4$ and $20$~AU.
} 
\label{fig:alpha4-profile1}
\end{figure*}

{\it Profile-1:} Figure \ref{fig:alpha4-profile1} shows the results from simulation A4W1 ($\apr =10^{-4}$ and mass loss profile-1) with snapshots taken at $10$k and $150$k years. The results are presented in similar fashion as figure \ref{fig:alpha3-profile1}. Interestingly, the signatures found here are markedly different than the ones obtained with $\apr=10^{-3}$. At $10$k years, instead of a depletion, we see an enhancement in MVE abundances, with the condensation temperature $T_{ci}$ showing maximum enhancement varying with radius. A flat depletion signature is seen only at $0.3$~AU to the right of a small peak at (Ni,Co) and the usual dip at (Mg,Si). The depletion at this radius reaches $0.6$ at most. A small depletion starting with Na is also observed at $0.5$~AU. 

%\textcolor{red}{\sout{??? However, at $150$k years, these depletion profiles are almost gone with all the MVE abundances getting enhanced throughout the inner disk. ???}}

With $\apr=10^{-4}$ and profile-1 mass loss, the disk evolution is slow and the inner disk retains its material longer. Moreover, radial drift, the factor that counters the MVE loss, seems to be more effective compared to simulation A3W1 (figure \ref{fig:alpha3-profile1}). In general, in the absence of bouncing events in the solid growth model, the St for the largest particle size in the growth model can be estimated as ${\rm St}_{max}=v_f^2/2\alpha c_s^2$ \citep{Senguptaetal2019}, where $v_f$ is the fragmenting threshold velocity. If fragmentation dominated the limits of growth, with all other parameters fixed, one would expect $\apr=10^{-4}$ to produce particles with St an order of magnitude higher than that for $\apr=10^{-3}$, making the major mass carrying solids more susceptible to inward radial drift. However, E16 found that when bouncing is modeled as a possible collision outcome, the maximum size of the particles is limited by the mass-dependent bouncing threshold. E16 being the basic structure of our current model, the A4W1 case is not expected to grow solids significantly bigger than that of A3W1. This can be confirmed by the fact that the radial drift velocity is similar in magnitude in both figures \ref{fig:alpha3-profile1} and \ref{fig:alpha4-profile1} until the water snow line (see the adjacent temperature plot for $\sim 170$~K) beyond which $\left|V_R\right|$ increases by almost an order of magnitude. However, the inward drift of the solids is mitigated by the diffusion which is significantly less in simulation A4W1 with $\apr=10^{-4}$ compared to A3W1. Hence, although the drift velocities do not show any striking differences, the stronger diffusion in A3W1 weakens the overall inward drift and dilution of the depletion signature. 

It is important to remark that this result does not exclude $\sim 10^{-4}$ as a possible value for $\apr$ in the context of meteoritic evidence, as a similar mass loss profile with $\apr=10^{-3}$ in figure \ref{fig:alpha3-profile1} did not show the desired depletion either. All we can infer from this result is that the profile-1 mass loss rate is probably not an ideal scenario in order to obtain a depletion signature resembling the observations.

\begin{figure*}
%\centering
\includegraphics[width=\textwidth]{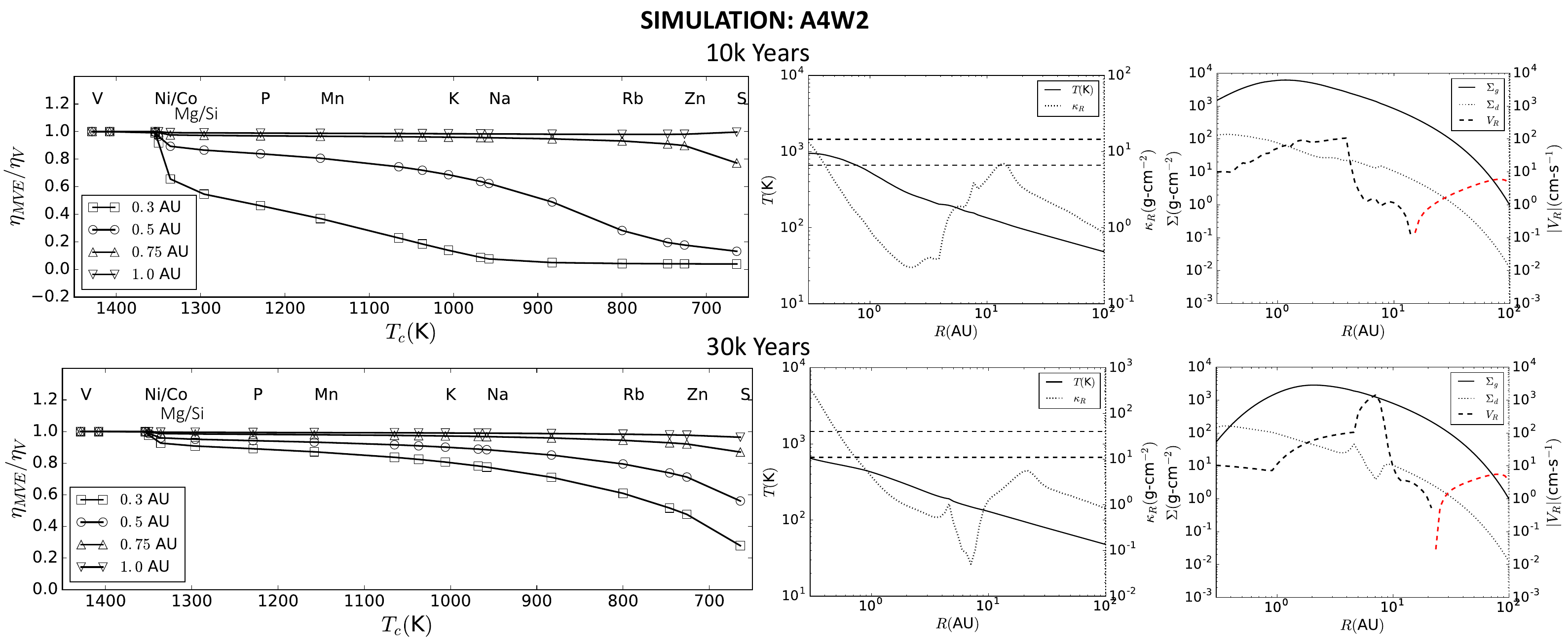}
\vspace{0 cm}
\caption{ The MVE depletion signatures for simulations A4W2 with $\apr=10^{-4}$ and a mass-loss with profile-2. The snapshots are taken at $10$ (left) and $30$k (right) years. Initially, there is a significant depletion at $0.3$~AU and $0.5$~AU. The depletion of S at $0.3$ and $0.5$~AU goes well below $0.2$ in a relatively short period of time. However, the signature gets washed away at $30$k years due to the combined effect of disk cooling and inward radial drift of materials. At $30$k years, the temperature of the inner disk radius goes below the condensation temperature of S, allowing all MVE species to condense. As a result the MVEs are unable to escape in vapor form, flattening the depletion signature eventually. The corresponding increase in $\kappa_R$ can also be seen at $25$k years due to the enhancement of solids in the inner disk.}
\label{fig:alpha4-prof2}
\end{figure*}

{\it Profile-2:} Figure \ref{fig:alpha4-prof2} shows the results from simulation A4W2 with the profile-2 mass loss. The outcome of this simulation is similar to that of simulation A3W2 (figure \ref{fig:alpha3-profile2}) initially, but as discussed above, the enhanced radial drift  washes the depletion signature out rapidly. Two snapshots at $10$k (top row) and $30$k (bottom row) years are presented.  At $10$k years, the depletion of the elements from Na to S at $0.3$~AU reaches almost zero. However, the depletion trend is shallow at $0.5$~AU and no significant depletion is seen at $0.75$ and $1$~AU. At $0.3$~AU the plateau from Na to S arises from the similar temperature effect as described in section \ref{sec:a3w2-a3w3} regarding simulation A3W3. The more rapid cooling of this case, relative to the case A4W1 above, is caused by the combination of stronger {\it gas} mass loss and smaller viscous $\apr$. By $30$k years, when the radial temperature profile goes below $T_{ci}$ of S, the depletion profile starts to flatten. The condensation of all solids in the inner disk is reflected by the increase of $\kappa_R$ (middle plot, second row of figure \ref{fig:alpha4-prof2}).

\begin{figure}
%\centering
%\vspace{-.7cm}
%\includegraphics[width=\textwidth]{Braukmuller_fig.pdf}
\includegraphics[width=0.48\textwidth]{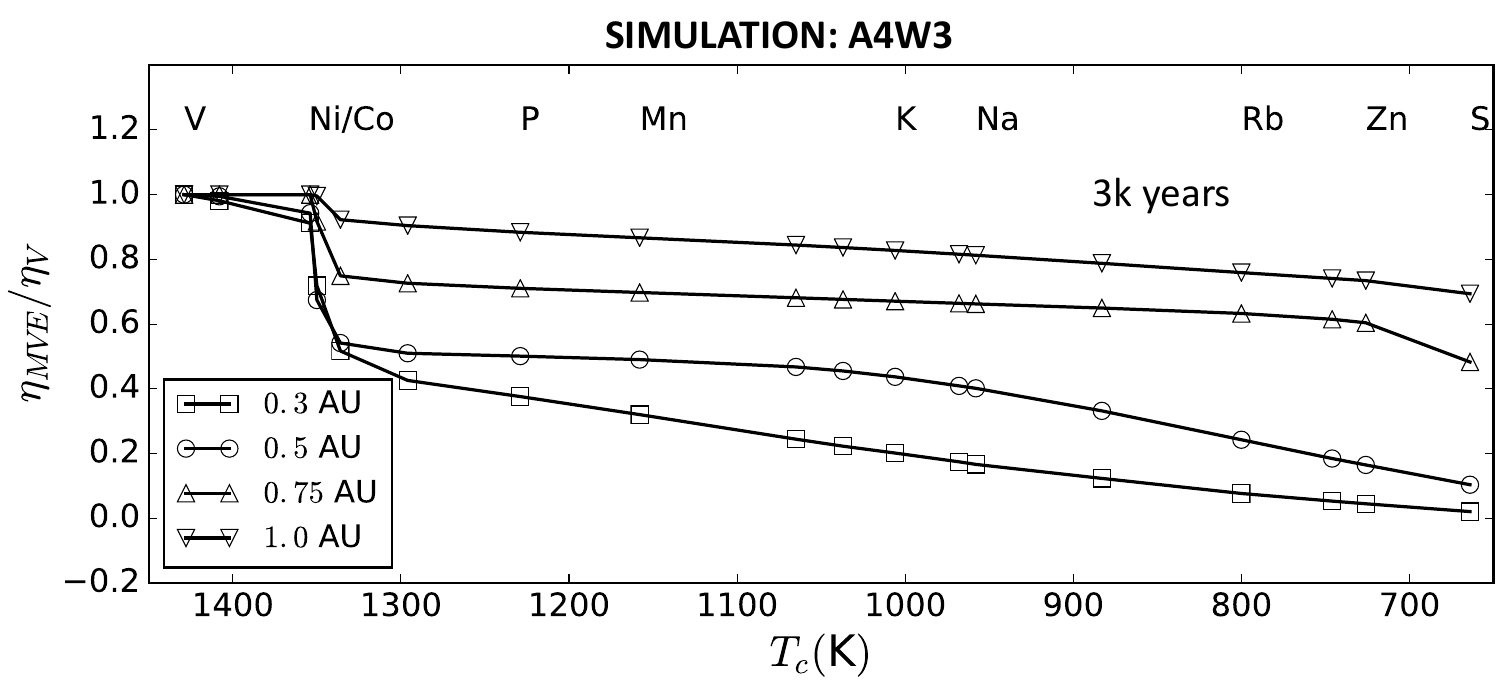}
\vspace{0 cm}
\caption{Figure from simulation A4W3 with $\apr=10^{-4}$ and and profile-3 mass loss showing the depletion profile at $3$k years.} 
\label{fig:alpha4-prof3}
\end{figure}

{\it Profile-3:} Figure \ref{fig:alpha4-prof3} illustrates the results from simulation A4W3 where rapid MVE depletion is observed in just $2.5$k years. The depletion of S approaches zero at $0.3$~AU and a fair amount of depletion is also observed at $0.5$~AU. The overall pattern of the results from simulation A4W3 is qualitatively similar to that of simulation A3W3. The fact that 0.3 and 0.5 AU here look like the signatures for 0.75 and 1 AU in the A3W3 case simply indicates this case was cooler for reasons discussed above in A4W2, with less coverage by EFs. The high mass loss rates of profile-3 effectively reduces the MVE abundances systematically and quickly, irrespective of the value of $\apr$, but the effect of inward drift and disk cooling dilutes the depletion signature  shortly thereafter.  This characteristic is the main feature of our simulations with a customized profile-4 mass loss rate, as detailed in section \ref{sec:profile4} below.

%The same wind signature but different values of $\alpha$.

%\subsection{A Higher  Mass Loss; layered accretion?}\label{subsec:highmdot}

%\textcolor{red}{Here, we shall present the high mass loss rate cases.
%We may present the variable mass loss rate simulation (the one that mimics the repeated outbursts) here as well.}

%In this range of mass loss rates, the physics of disk winds as described by \citet{Suzukietal2016} is inappropriate. Instead, we envision these higher mass loss rates to be associated with layered accretion directly onto the star \citep{Gammie1996, Zhuetal2010}. This still constitutes ``open system" behavior, in that the more volatile MVEs are fractionated away from the more refractory ones, which remain in the nebula to be accumulated into planetesimals, although perhaps in a different time and location than where the depletion occurs. The physics of mass loss for layered accretion is more like standard alpha-disk viscosity in a thin upper layer than a wind that exerts a local torque on the entire disk. For convenience in this first study of the potential importance of open-system behavior,  we simply extend the mass loss rate formalism developed for the disk wind (equation \ref{eqn:suzukigasvelocity}) but set the separate torque term   (the $\apz$ term on the right of equation \ref{eqn:suzukigasvelocity}) to zero. 

\subsection{Profile-4 Mass Loss}\label{sec:profile4}

\begin{figure}
%\centering
%\vspace{-.7cm}
%\includegraphics[width=\textwidth]{Braukmuller_fig.pdf}
\includegraphics[width=0.48\textwidth]{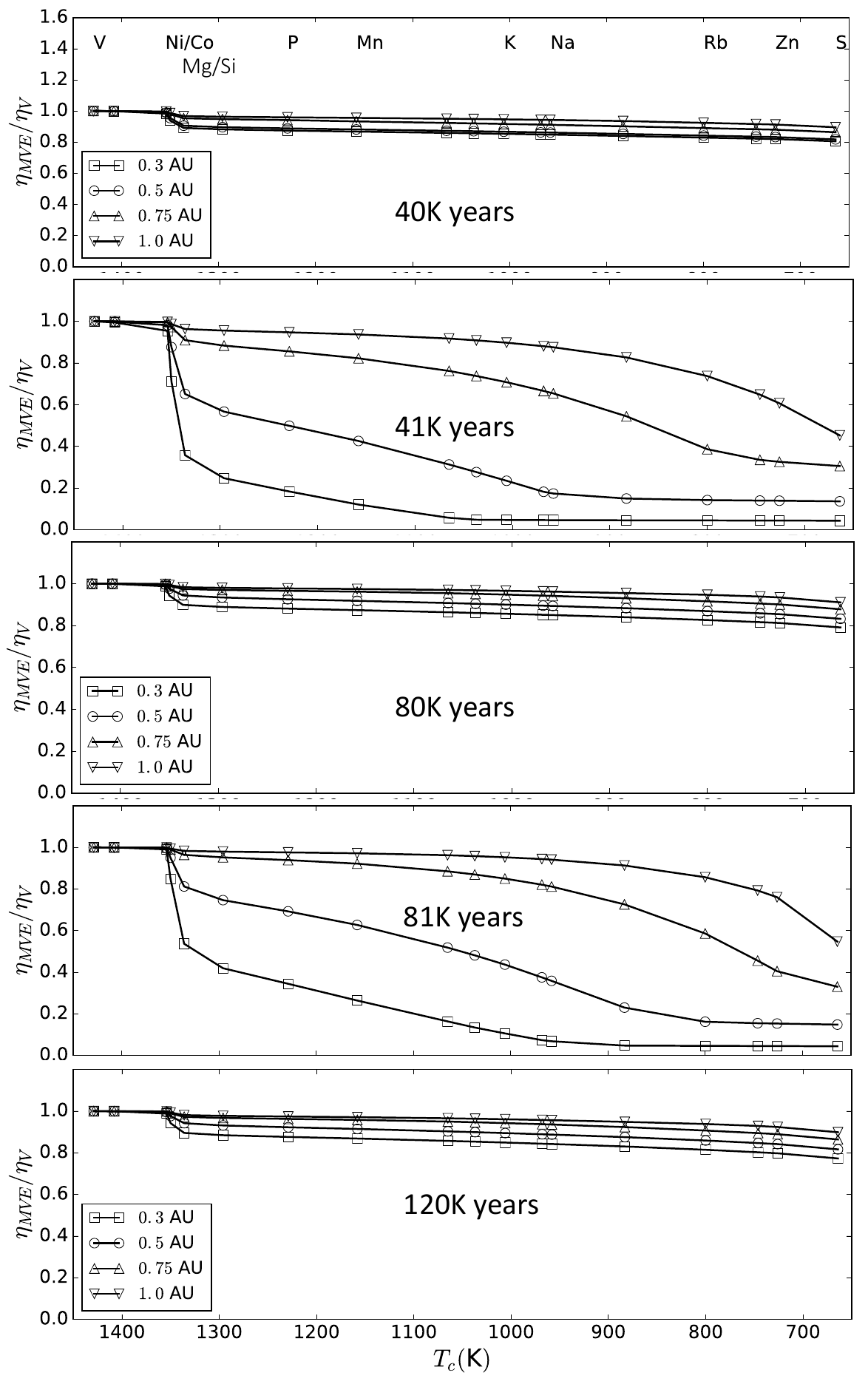}
\vspace{0 cm}
\caption{The MVE depletion signature for profile-4 mass loss with sequential  quiescent and burst phases (figure \ref{fig:windprofile}) consistent with a disk build-up process in the early class-0/class-I phase \citep{Zhuetal2010}. The depletion signatures are shown (from top to bottom) at $40$k (end of first quiescent phase), $41$k (end of first burst), $80$k (end of second quiescent phase), $81$k (end of second burst) and $120$k (end of third quiescent phase) years. The depletion signatures here show a repeating pattern, with rapid depletion at the inner boundary (here, 0.3 AU) going all the way to less than 0.1 after each burst phase. After the end of an intense burst phase, the profile is diluted by inward radial drift of nominally undepleted CI material to nearly its starting undepleted value. However, in reality, material at larger radii would not remain undepleted if it becomes admixed with depleted material during repeated, extensive outward mixing events that may terminate the rapid burst phases. Such a situation would lead to a  secular increase in the depletion signature (see section \ref{sec:fastmixing}). } 
\label{fig:profile4}
\end{figure}

It is now evident from the previous results (section \ref{sec:fiducialmodel} to \ref{subsec:varyalpha}) that a single mass loss profile (top panel figure \ref{fig:windprofile}) either fails to produce a proper depletion signature (profile-1), or reaches the desired depletion level in a fairly short time ($\sim 30$k years for profile-2 and $< 3$k years for profile-3). However, these results essentially represent the end stage of a {\it single} high-mass loss event, and from observations and theory, it appears that a given system can undergo  many such events in the early stages, which constitutes of repeated outbursts, followed by rapid outward expansion of the disk separated by quiescent periods of longer duration  \citep{Zhuetal2010, kadametal2020}.
As a good example, see figure 1 of  \citet{Zhuetal2010}. 

To better understand the situation, we combined profile-1 and profile-3 to mimic this multiple-outburst stage (as shown in the bottom panel of figure \ref{fig:windprofile}).  In figure \ref{fig:profile4}, we present the results of the simulation A3W4 with snapshots taken right before and after two simulated burst events, and before a third. In this simulation there is {\it no rapid expansion phase} associated with each outburst; the mass loss rate just cycles from low to high and back again. The figures represent (from top to bottom) the snapshots for depletion profiles at $40$k years (immediately before the first burst), $41$k years (after first burst), $80$k years (before second burst), $81$k years (after second burst) and $120$k years (after the third quiescent phase).

Similar to the results of the simulation A3W1, at the end of $40$k years, the depletion profile for the MVEs has almost flattened out due to the combined effects of the low mass loss rate and inward radial drift (see section \ref{sec:fiducialmodel} for details). However, at the end of the first burst the signature at $0.3$~AU goes down to $0.02$  in just $1$k years for all the MVEs from Na to S (see section \ref{sec:a3w2-a3w3} for details about the flat profile). Once the burst phase is turned off and the second quiescent phase starts, within $\sim 40$k years, a wash-out is again observed due to the effect of inward drift. This cycle repeats itself at $81$k years at which time a similar depletion, followed by a wash-out, is observed. The interesting feature of the simulation A3W4 is the periodicity in the depletion profile with the repeated burst and quiescent phases. We do not continue the simulation after $120$k years. 

%\textcolor{red}{Not sure what the following means: However, the relatively slow and quiet nature of our fiducial model (simulation A3W1, figure \ref{fig:alpha3-profile1}, section \ref{sec:fiducialmodel}) for an extended period of time.. do we mean that the observations and theory of others gives us this confidence?} gives confidence that this periodic behavior will continue for several cycles. 

According to the \citet{Nanneetal2019} scenario, and some theoretical work \citep{kadametal2020}, the bursts are associated with rapid outward expansion, mixing the MVE depleted material with  material of CI composition in the outer disk. With several such bursts as observed in A3W4 \citep[or][figure 1]{Zhuetal2010}, several such mixing events would take place, each one homogenizing the mixed region, diluting the depletion signature from the inner nebula,  and {\it increasing} the MVE depletion signature for the disk material out to some distance (say, $2-5$~AU). The results of figure \ref{fig:profile4} {\it do not} include this mixing process, which is beyond the scope of this paper, so each cycle starts from scratch. In section \ref{sec:fastmixing} we present a better, but still simplified, discussion and sanity check on the effects of multiple outbursts.

\subsection{Solid vs Porous particles: Effects on depletion signature.}

\begin{figure}
%\centering
%\vspace{-.7cm}
%\includegraphics[width=\textwidth]{Braukmuller_fig.pdf}
\includegraphics[width=0.48\textwidth]{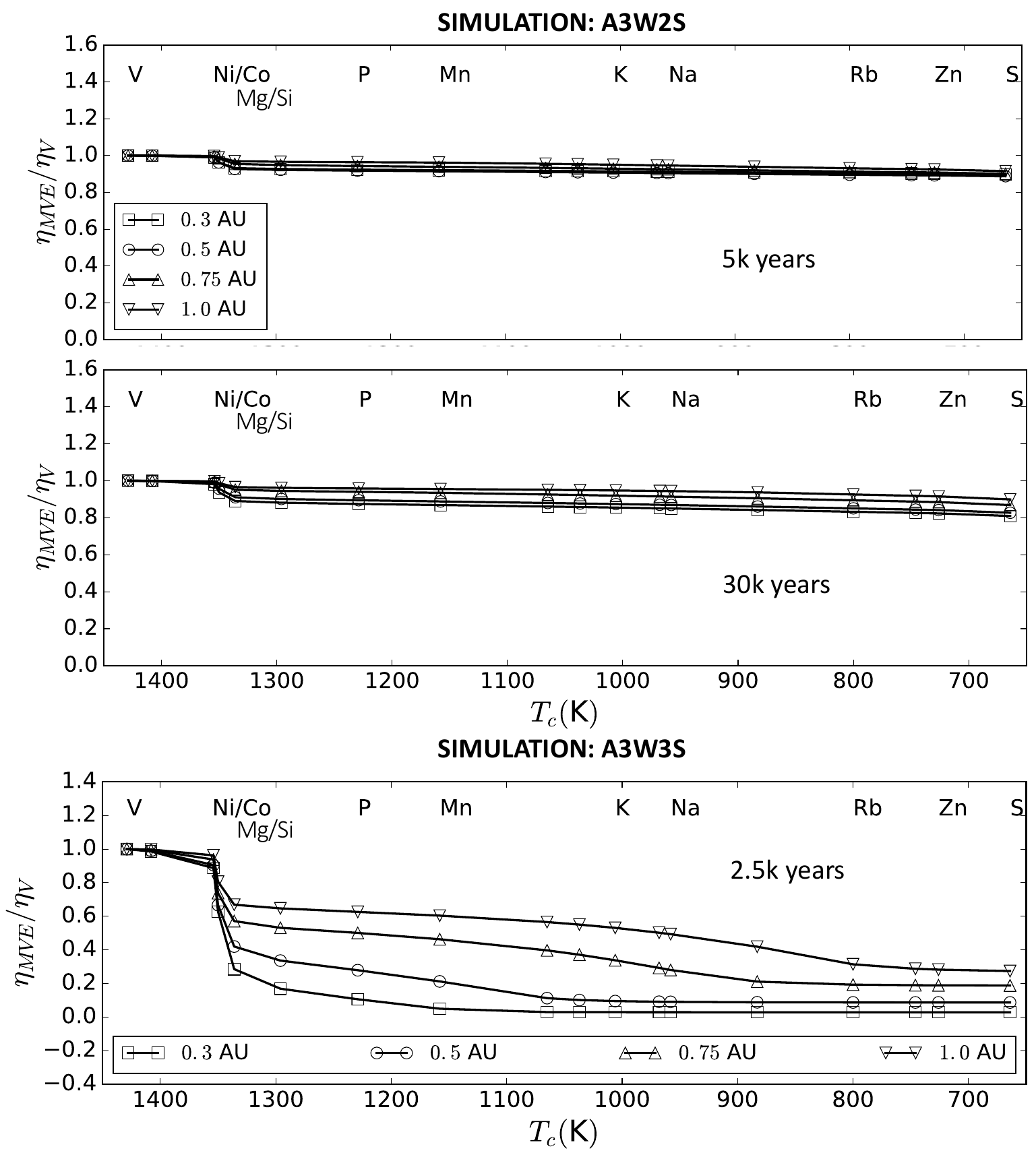}
\vspace{0 cm}
\caption{MVE depletion profile for simulations A3W2S and A3W3S with solid particles and $\apr=10^{-3}$. With profile-2 mass loss rate (top two figures), two snapshots at $5$k and $30$k years are presented. Their depletion profiles overall are fairly flat, stagnating long before reaching the observed level of depletion. The reason seems to be that the solid particles, with higher St, diffuse less easily and drift more rapidly than the porous ones, washing out the depletion signature more readily. The panel at the bottom is from simulation A3W3S where a depletion profile similar to A3W3 is observed, and after about the same time ($2.5$k years).} 
\label{fig:solid}
\end{figure}

Most of the simulations in this paper employ the new porous aggregate particle growth treatment by \citet{Estradaetal2021}. This treatment uses the physics developed by \citet{Okuzumietal2012}, and leads to particles with very high porosity ($\sim$0.99 or larger), with Stokes numbers smaller than for particles that are allowed to grow as solid. It is unclear just how realistic these models are, as the particle porosities are hard to measure observationally given other uncertain parameters (temperature, optical depth, and actual particle size). Multi-wavelength ALMA observations are becoming increasingly powerful along these lines however \citep{Carrascogonzalezetal2019, Maciasetal2021, Sierraetal2021} and eventually the particle porosity might be better constrained. For the purpose of this paper, we  bracket the influence of porosity by running some comparison cases at the opposite end of the spectrum, assuming the particles grow as solids (zero porosity) as in E16. The two runs described in this section use the full self-consistent particle growth code of \citet{Estradaetal2016, Estradaetal2021}, only assuming the particles grow as solids rather than as porous, fractal aggregates.

In figure \ref{fig:solid}, we show the results of our simulation assuming zero-porosity solid particles with $\apr=10^{-3}$. The top two panels are depletion profiles at $5$k and $30$k years from simulation A3W2S (with profile-2 mass loss rate). These results can be directly compared to the ones from A3W2 with porous particles (figure \ref{fig:alpha3-profile2}). At  $5$k years and even at 30k years,  the MVE depletion stays above $0.8$. For the equivalent porous particle case (A3W2) the 5k year values hover between $0.6$ and $0.8$, and at $30$k years, the porous grains approach the desired depletion signature. % in contrast to the solid particles, where nothing significant takes place except for a little more depletion of the MVE species, similar for all radii shown. 
The solid particles, compared to their porous counterparts, affect the signature in two ways. First, the solid particles provide a smaller overall Rosseland mean opacity, accelerating the cooling of the disk.  Second, solid particles are more susceptible to radial drift compared to porous aggregates with same mass, because of their higher St \citep{Estradaetal2021}. This allows the solid particles to decouple from the gas and drift inwards faster, diluting the MVE depletion signature. These two effects, working in conjunction, keep the solid particle depletion profile from reaching its desired level under mass loss profile-2. 

In simulation A3W3S however, which assumes the larger profile-3 mass loss rates (lower panel of figure \ref{fig:solid}), we see a similar depletion profile as in simulation A3W3, in which  significant depletion is achieved in a very short time for all radii. When compared with figure \ref{fig:alpha3-profile3}, it becomes evident that for the profile-3 mass loss rate, particle porosity hardly makes any difference, and  suggests  that a combination of intermediate porosity particles, with mass loss rates between profiles 2 and 3, could also produce acceptable results (see section \ref{sec:discussionparams} for more discussion).

\subsection{Is refractory ``enrichment" in CCs  merely an artifact of MVE depletion?}\label{sec:normalization}

In figure \ref{fig:braukmuller} we showed a typical plot of elemental abundances as a function of volatility (condensation temperature $T_{ci}$), with the one slightly novel feature of being normalized to CI at Vanadium (V), a refractory element. Throughout this paper we follow the same convention; for the same purpose, it could also be normalized to the average of the elements from, say, Hf to V, or the elements more refractory than Ca.  Usually plots like this are normalized at more common elements like Fe, Mg, or Si. When the more typical normalization is used (or if {\it no} normalization to any particular element is used but merely ratios to CI abundances), the relative abundance of refractories appears to be enhanced by 20-40\% over CI \citep[see, eg.][]{Braukmulleretal2018}. This enhancement  is generally envisioned as an actual physical addition of refractory elements (allegedly from CAIs) to CCs, relative to CI. This aspect of CC composition is deeply embedded in the community's thinking about the MVEs, and plays an important role in some models \citep{Deschetal2018}. 

To our knowledge there is no fundamental reason why one would not {\it prefer} to normalize to one of the refractory elements, or to their average.  While less abundant, they are the least likely to be removed in what our results suggest could be a large-scale  volatility-dependent loss process. The common elements (Fe, Si, Mg) are usually used for normalization because uncertainties in their measurement are thought to be less likely to lead to uncertainties in the normalized ratios. However, because the common elements also represent by far the largest mass fraction of the chondrites (80\%), if they were {\it actually removed from} all chondrites except CI, while the more refractory elements were left behind, the refractory elements will represent a larger mass fraction of the remaining host rock, and if any of the {\it depleted} elements is  used as a normalization, an overabundance of refractories will appear as an artifact. Indeed the same result is obtained even if ratios of sample abundances to CI abundances are left un-normalized by any specific elements - refractories appear to be enhanced and volatiles depleted, plausibly for the same reason. 

 A comment along this line was  made by \citet{Connollyetal2000}, and several sample calculations in support were shown by \citet{Connollyetal2001}. Also,  \citet{Hussetal2003} appear to caution against exactly this problem regarding CR chondrites (their section 4.2.1), which is reiterated in \citet[their section 5.2]{Huss2004} and extended to LL's as well (their section 5.1).  
 
 \begin{figure*}[t]
\centering
%\vspace{-.7cm}
%\includegraphics[width=\textwidth]{Braukmuller_fig.pdf}
\includegraphics[width=\textwidth]{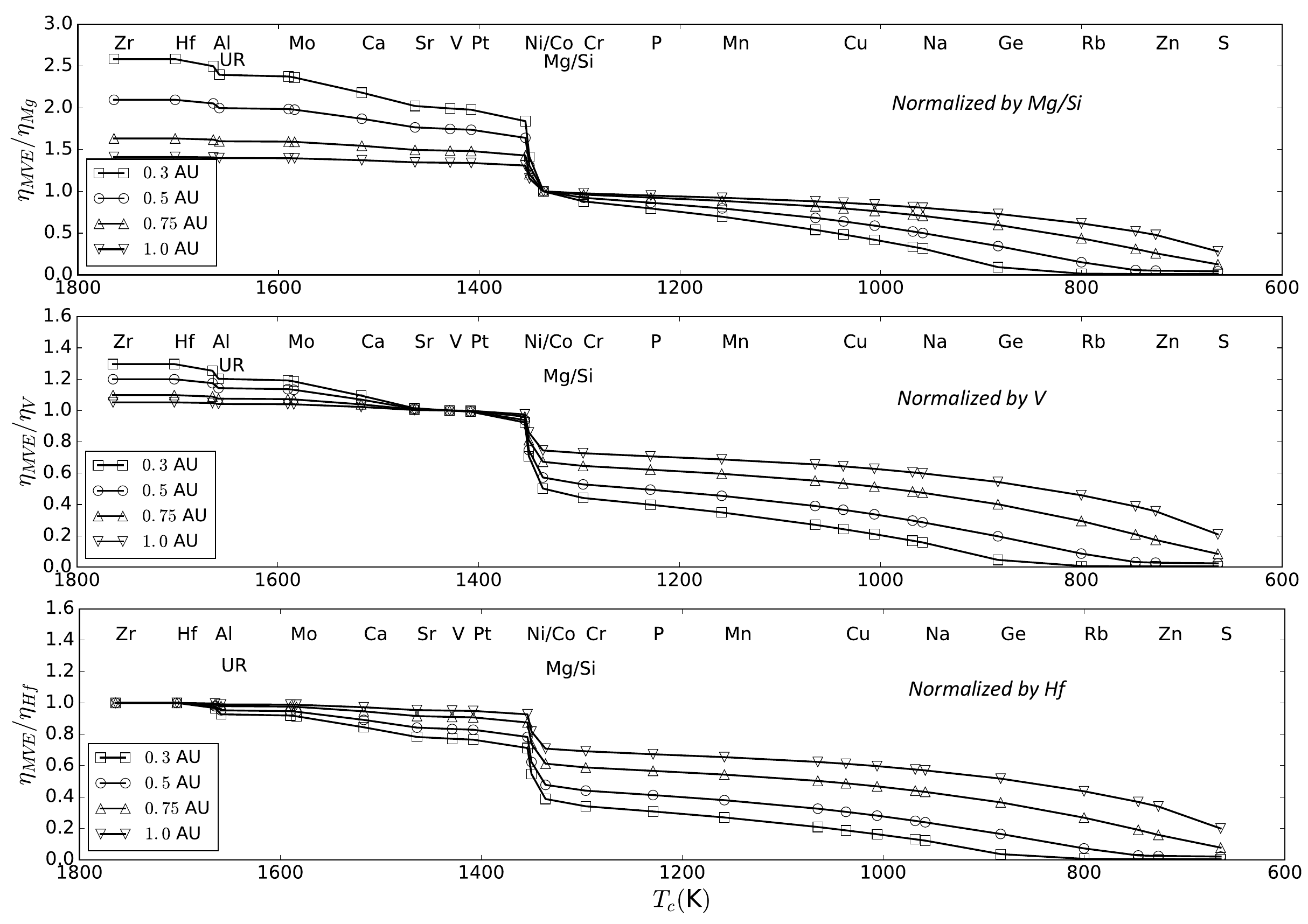}
\vspace{0 cm}
\caption{This figure shows the MVE depletion patterns as normalized by CI and different elements,from model A3W3 at 3.5k years. {\textbf {top:}} normalized by (Mg,Si); {\textbf {Middle:}} normalized by V; {\textbf {bottom:}} normalized by Hf. The  refractories with $T_{ci} > 1450$K show strong enhancement signatures if normalized to the common elements (top panel); however, if we normalize by the most refractory elements, the trend is a simpler one of montonically increasing depletion to lower $T_{ci}$, with an opacity-related  step in the (Fe,Mg) range (see sections \ref{sec:normalization} and  \ref{sec:discussionopacity}). In this latter perspective, the hot inner nebula open system model may be capable of explaining two puzzles at once. } 
\label{fig:refrac_enhancement}
\end{figure*}

We illustrate this point in figure \ref{fig:refrac_enhancement}, where the depletion signatures from simulation A3W3 of all species tabulated in table \ref{tbl:mves} are presented  at $3.5$~k years using three different normalizations. In the top panel of figure \ref{fig:refrac_enhancement} the elemental abundances are presented normalized in the usual way by (Mg,Si)  with $T_{ci} = 1336$~K. In the middle panel, the normalization is done with respect to V with $T_{ci}=1429$~K. The bottom panel shows the signatures normalized by Hf with $T_{ci}=1703$~K. Depending on the normalization, one can conclude either that elements more refractory than the common rock-forming elements are {\it enhanced} to varying degrees, or that everything less refractory than Hf is {\it depleted} to varying degrees. 

These two different ways of looking at the normalized data are only semantic, until one advocates physical processes to cause them. The open system loss mechanism described here, operating in a hot early inner nebula, seems to be a plausible physical mechanism naturally leading to the second conclusion. 

This conclusion is also consistent with petrology.  \citet{Hussetal2003} mention a ``step" in elemental abundances possibly identifiable with CAI minerals per se; they take this as grounds for treating the refractory enhancements as real - in the physical form of CAIs - rather than an artifact.  \citet{Huss2004} further go on to  attribute the ``excess" of refractories in COs to their 2\% of CAIs. \citet{Hezeletal2008} find a very good correlation between total refractory ``enrichment" and the observed modal abundance of CAIs. We believe this direct identification {\it of} enrichment in refractory elements {\it with} actual CAIs and the like is also consistent with natural predictions of our models, as described below.

The strongest differential refractory element enrichments are originally seen, in our models, in the hottest regions, at small radii. Compare figures \ref{fig:alpha3-profile2} and \ref{fig:alpha3-profile3} with figure \ref{fig:refrac_enhancement}, that extends the plots to more refractory elements.  Figure \ref{fig:refrac_enhancement} shows that in the very innermost regions, even V is depleted and the depletion is most extreme at 0.1AU (see also figure 16). Note in the central columns of the plots of figures \ref{fig:alpha3-profile2} and \ref{fig:alpha3-profile3}, especially in figure \ref{fig:alpha3-profile3}, that the temperature in the inner nebula exceeds the evaporation temperature of the silicate-rock-forming elements over extensive ranges of radius and some duration of time. 

Actually, it is buffered at these temperatures due to the thermostat effect of large decreases in opacity as the mass-dominant, silicate solids evaporate (section \ref{sec:discussionopacity}).  In these hot regions, there are no solid silicates and the only minerals that {\it can} be solid, are CAIs. This is why we see CAIs at all. They are the mineral hosts that form and survive in those regions, but escaped those regions before they could re-equilibrate at lower temperatures when the nebula cooled (some material did evolve more gradually, maybe through the AOA stage). So, when material from these inner regions mixes rapidly outwards, along with MVEs and even rock-forming elements that have been variously {\it depleted}, we get a mixture in which refractory elements {\it appear} differentially enriched (because a significant amount of the mass in common rock-forming elements has been {\it removed}) but {\it these refractory elements are being mixed and redistributed {\it in the form of CAIs}}. 

So the physical manifestation of the differential refractory element enrichments our models predict {\it is} the presence of CAIs, which are their primary hosts. The scenario we have in mind – a rapid expansion during outburst events – would provide a rapid quenching of CAI mineralogy and preserve them in their original form, rather than allow them to slowly equilibrate with rock-forming elements as they subsequently condense in a cooling nebula. For these reasons we believe our scenario naturally explains an observation that previously was a puzzle (apparent enrichment of refractories relative to CI), as only one more aspect of the hot inner nebula MVE depletion process itself.

In the data from \citet{Braukmulleretal2018}, there may be a  ``step" between V and (Ni,Co,Mg,Fe), and then a ``slope" range down to Bi and Ag showing volatility-dependent depletion, with another flat ``plateau" range for higher volatility elements. That is, the ``refractory  step'' lies between (Ni,Co) and (Pt,V) in figure \ref{fig:braukmuller}, rather than  between (Si,Fe,Mg) and (Ni,Co) as in our various models. We expect the specific location and steepness of the ``step" in our models depends on our assumptions of $T_{ci}$ (perhaps on the mineral hosts) for elements between (Ni,Co,Rh) and the common elements. The subject of the ``step" is discussed more below in section \ref{sec:discussionopacity}. Moreover, the depletion profiles at small radii in figure \ref{fig:refrac_enhancement} show differential depletion even across the range from Zr to V, which is not seen in the data; this could be a matter of the scatter, the details of the model radial temperature structure, or of the subsequent radial mixing. In general though, although there is scatter, the common elements Fe, Si, Mg can be looked at as lying on the MVE slope in figure  \ref{fig:braukmuller}, causing us to suspect that even some of the common rock-forming elements providing most of the silicate mass have been irreversibly depleted (relative to CI) by the same hot inner nebula process as appears capable of depleting the MVEs. 
 
%\textcolor{blue}{Normalization by the abundance of elements other than V, however, can illustrate the point we intend to present here. In figure \ref{fig:refrac_enhancement}, the depletion signature from simulation A3W3 of all species tabulated in table \ref{tbl:mves} are presented using two other normalization at $3.5$~k years. In the top panel of figure \ref{fig:refrac_enhancement} the elemental abundances are presented normalized by Mg/Si with $T_{ci} = 1336$~K. In the middle panel, the normalization is done with respect to V, the usual normalizing species adopted in this paper. Finally, the bottom panel shows the signatures normalized by Hf with $T_{ci}=1703$~K. The figure demonstrates the relative dependence of refractory abundances on the choice of normalization, and can be as high as $2.5$ when normalized by Mg/Si.  }

\section{DISCUSSION}\label{sec:disc}

In this paper, we addressed the longstanding puzzle of the depletion of MVEs in the solid bodies in the inner solar system by using a global nebula evolution model with a parameterized disk wind mass loss. In our model, MVEs are systematically and  irreversibly lost from the inner disk through `open-system' processes leaving more refractory solids behind. As shown in our results in section \ref{sec:results}, in some cases we have found promising depletion profiles, whereas in others, the profiles become static and saturate well before reaching levels that match the meteoritic observations. In this section, we discuss why different situations provide different outcomes based on nebula physical processes, along with their implications for the evolution of the early solar system.

\subsection{Two distinct pathways to MVE depletion}\label{depletionpathways}

In view of the results presented in the previous sections, we can identify two distinct pathways to the depletion of MVEs that conform to our `open-system' hypothesis. It is now evident that under no circumstances does  profile-1 mass loss  leads to the desired depletion signature. However,  profile-1 is more consistent with a disk wind in the class-II classical T-Tauri phase. Both profile-2 and profile-3 mass loss rates, on the other hand, have shown promising results which can lead to the MVE depletion within the framework of open-system processes, and at least profile-2 is quite plausible for the Class 0/I stages we model here. 

In the context of our hot early inner nebula model, profile-2 and profile-3 mass loss rates imply quite different durations of MVE depletion. In the profile-2 scenario, the MVEs are depleted during the $\sim 10^4$ year long quiescent phases between    short outbursts. With the results of profile-4 as elaborated in section \ref{sec:profile4} and in figure \ref{fig:profile4}, we see that a significant depletion can be achieved during each quiescent phase, while the signature will be more pronounced with the higher rate of profile-2.

With profile-3, however, the overall scenario of depletion, transport and mixing is different.  Profile-3 has a radially integrated mass loss rate of $10^{-5} M_{\odot}$~yr$^{-1}$.  At these high loss rates, a depleted MVE profile is achieved in a much shorter timescale than in the profile-1 or profile-2 simulations in figure \ref{fig:alpha3-profile3} and \ref{fig:alpha4-prof3}. These short timescales are consistent with outbursts, which can have mass loss rates even ten times higher \citep{Zhuetal2010,kadametal2020}. Depending on the physics of the outbursts, these events can last for several hundred to few thousand years \citep{kadametal2020}. In most models, these outbursts may be associated with rapid outward transport of the depleted material and mixing with the more primitive (CI-like, not ISM-like) outer disk material. Physically this suggests that the depletion of MVEs {\it and} outward radial transport and mixing would take place {\it together} during a single outburst.  

The actual processes taking place during these outburst events are far from being settled and still being debated. \citet{calvetetal93} argued that these outburst events are associated with high rates ($10^{-6} - 10^{-5}$~$\dot{M_{\odot}}$~yr$^{-1}$) of mass loss in the form of disk winds, as an extension of the increases we have argued are expected going from the Class II T-Tauri stage to the Class 0/I stage. These mass loss rates encompass both profile-2 and profile-3 rates in our model.  On the other hand, profile-3 mass loss rates would be generally regarded as inconsistent with a layered accretion scenario \citep{Turneretal2014}. Meanwhile, \citet{Zhuetal2010} and \citet{ kadametal2020} did not have a wind solution in their model, and their high mass loss rate outburst mechanism is either gravitational or magnetic in nature, affecting the entire disk thickness at once.

Recall that we envision `open-system' processes as mechanisms where the MVEs are able to escape the disk in vapor form, leaving all the refractories behind as solids, resulting in an MVE-depleted inventory. In this context, as we have mentioned earlier, both disk wind and layered accretion constitute open-system and are capable of losing MVEs irreversibly and retaining the refractories at the same time. On the other hand, if the whole vertical extent  of the inner disk is accreted, as is the case in the outburst phases of \citet{Zhuetal2010}, \citet{kadametal2020}, or \citet{VorobyovBasu2005,VorobyovBasu2006}, both the MVEs and the refractories are drained into the central star at the same rate. Such processes and stages are not `open-system' as we see it, and probably not capable of fractionating MVEs.

Hence, although it is true that the profile-3 scenario cannot be ruled out as yet, the scenario involving profile-2 seems more realistic regarding our current understanding of the outburst process. We can only confirm the actual picture regarding the outburst mechanism when a proper 3-D MHD-simulation is performed with wind solutions included as a natural consequence, which is unfortunately not yet present in the existing literature.

 \subsection{Effects of different parameters and mass loss processes} \label{sec:discussionparams}
 
 In our hypothesis, MVEs are lost in their vapor phase by some `open-system' mass loss process (disk wind and/or layered accretion). The loss for a particular MVE species $i$ depends on its condensation temperature $T_{ci}$ and its radial position in the disk. A local disk temperature exceeding $T_{ci}$ will vaporize MVE$_i$, allowing it to escape the disk, decreasing its abundance. Inward transport of MVEs from cooler regions in the solid form, at something like CI abundances, continually works to compensate the loss.  It is true that the desired depletion is achieved in all cases of profile-3 and only for the porous particles for profile-2. However, it is evident from the results presented in section \ref{sec:results}, that the overall depletion profile of the MVEs is not a matter of mass loss profile alone - other disk radial transport processes also play important roles in determining the overall signature.
 
 In a protoplanetary disk as modeled in this work, mass-loss, disk cooling,   inward drift, and radial diffusion all work in tandem to guide the disk evolution. The MVE depletion signature, according to our hypothesis, depends on the fact that each species $i$ is lost systematically according to the radial placement of their respective $T_{ci}$ so that a less volatile MVE will have its EF further from the star allowing it to escape through a larger portion of the inner disk compared to some other species with a higher $T_{ci}$. So, disk cooling and its rate would inevitably affect the process. We have seen that as long as the temperature stays above the condensation temperature of the least volatile, most refractory MVE, all species are lost equally, resulting into a flat profile (e.g., figure \ref{fig:alpha3-profile1} or figure \ref{fig:alpha3-profile2} at $5$k years). The depletion becomes monotonically increasing with decreasing $T_{ci}$ only when the disk starts to cool. However, once the inner disk boundary temperature goes below the $T_{ci}$ of the most volatile MVE (S in our case), depletion ceases and the depletion profile starts to get washed out until eventually it becomes flat again. 
 
 In addition to the cooling effect, the inward drift of solids also  washes out the depletion signature, by allowing solid, undepleted MVE material to move inward, replenishing the inventory in the inner disk as it  evolves towards the star by drift and advection. As in  simulation A4W2 (figure \ref{fig:alpha4-prof2}), we initially see a nice depletion profile, but it then goes away due to the effect of inward drift and replacement of solids. 
 
 It is thus evident that in the presence of robust processes like disk cooling and inward drift, the open system loss process (wind/layered accretion) leading to the loss of MVEs in the vapor form has a certain window of time within which the depletion signature needs to be achieved. This is the main reason why none of our simulations with the profile-1 mass loss succeed to achieve the observed depletion profile, even after $150$k years of evolution. In the case of simulation A4W1, we even see local $T_{ci}$-dependent enhancement due to a combination of factors %due to the increased radial drift 
 (see figure \ref{fig:alpha4-profile1} in section \ref{subsec:varyalpha}). In all these cases, the mass loss rate is too low to carry off enough vapor before cooling and radial drift take effect. However, once the loss rates are increased with profile-2, we get the signature at least with the porous particles. In the solid particle case, even  profile-2 does not remove vaporized MVEs fast enough to keep radial drift from weakening or erasing the depletion signature. However, in profile-3, irrespective of the values of $\apr$ and the particle porosity, the desired depletion is always achieved. The loss rates in profile-3 are too high to allow cooling and drift any opportunity to wash the porous aggregate particle signature away. This also explains why even with the solid particles a nice signature is seen. It is  worth noting that in some recent simulations of high mass loss rate events, the loss rate is even higher than used in our profile-3 \citep{Zhuetal2010, Lietal2021}. Such higher mass loss rates would achieve the desired signatures  in an even shorter timescale. On the other hand, the range of current uncertainty also seems to allow mass loss rates of profile-2 during ``quiescent" periods in the early stages of evolution. 
 
 %\textcolor{red}{What does the following mean?: When the depletion profile goes down rapidly within the time-window} adjusted by cooling and radial drift, the porosity of the solids becomes largely irrelevant. 
 
%\textcolor{red}{Here the interplay of disk cooling, MVE loss and inward drift will be discussed.}

\subsection{Rationale for equilibrium mineralogy}\label{sec:equilib} In this work, we have assumed that each element is hosted in one mineral, with a well-defined condensation temperature $T_c$, taking our $T_c$ values from \citet{Lodders2003} who calculated them using thermochemical equilibrium approaches, as is routine in the field \citep{GrossmanLarimer1974, Wasson1985}.  As we noted in section 1.1 some of these MVEs can actually be hosted within more than one mineral, with different $T_c$ values, which may cause some of the fluctuations in figure 1 but would not distort the general trend we seek to explain. We also noted that for most elements, condensation temperatures are slightly pressure sensitive, leaving an uncertainty in $T_c$ of a few hundred degrees K over a very large pressure range. A tangible argument supporting our simple $T_c$ model can be found in the results of \citet{Lietal2020}, who conduct a full-up pressure dependent, multi-host, condensation (while still based on thermochemical equilibrium) with partial removal of elements into previous higher $T_c$ hosts, but with a far simpler nebula model. Figure 8 in that paper implies that all the complex, history-dependent chemistry produces no significant difference in the results relative to a model using a single $T_c$ for each element, as we do. 

Equilibrium is the state in which the evaporation and condensation rates are in balance. When cooling from a hot nebula, the condensation rate exceeds the evaporation rate but it need only do so by a tiny bit to assure (slow) grain growth. However, one can wonder what the implications might be of neglecting kinetics and nonequilibrium factors. Once major silicates have condensed, lower temperature reactions become difficult to model with rigor, because they can only proceed if elements from the vapor can diffuse within already-formed solids to create equilibrium minerals. Kinetic limitation of the uptake of MVEs on mineral surfaces could enhance the observed depletions beyond that indicated by thermochemical equilibrium calculations \citep{Larimer1967, Humayun2012}. Diffusivities of trace elements are not generally measured at the temperatures of interest (600-1100K) and the uncertainties on the activation energies measured at higher temperatures are too large to reliably extrapolate to relevant temperatures \citep{Righteretal2005}. However, the slow cooling timescales of nebular processes (hundreds of years to tens of thousands of years) may be long enough for diffusion to proceed adequately on relevant mineral grain lengthscales \citep{Kennedyetal1993}. As one example, experimental studies of the reaction of sulfur between H$_2$ in nebular gas and Fe metal at the relevant temperatures indicate that mineral transformations (Fe-Ni metal to FeS) will run to completion in 20 $\mu$m diameter grains on timescales of 200 years \citep{Laurettaetal1996}. 

For practical studies in protoplanetary nebulae, the prior MVE models dating back to \citet{Cassen1996} are best seen as equilibrium fractionation-by-condensation \citep{HumayunCassen2000}. Some of the grains and particles of interest either originally resided in the hot regions, were evaporated, and underwent a plausibly equilibrium condensation process. In addition, we handle Evaporation Fronts (EFs) of each element, where that element is on the cusp of condensation-evaporation equilibrium. At these locations small particles and vapor are being diffused from one side to another on timescales longer than the 200 years noted above. Moreover, inwardly drifting, more primitive particles, possibly originally condensed in a non-equilibrium fashion at the lower densities and temperatures of the ISM, will get gradually warmed up as they drift inwards slowly over thousands or tens of thousands of years, during which time thermal diffusion within the grains could relax whatever odd mineral distribution {\it might} have been present into something close to an equilibrium mineral distribution by the time they get close to the EFs.  Given this evolution over the $100-10^4$ year time scales involved, we believe the underlying equilibrium assumptions are plausible. 

So from several lines of argument, we think our treatment using a single equilibrium $T_c$ for each element is appropriate for our purposes of comparing with the general trend of abundance with volatility

\subsection{Effect of opacity on depletion pattern}\label{sec:discussionopacity}

\begin{figure}
\includegraphics[width=0.5\textwidth]{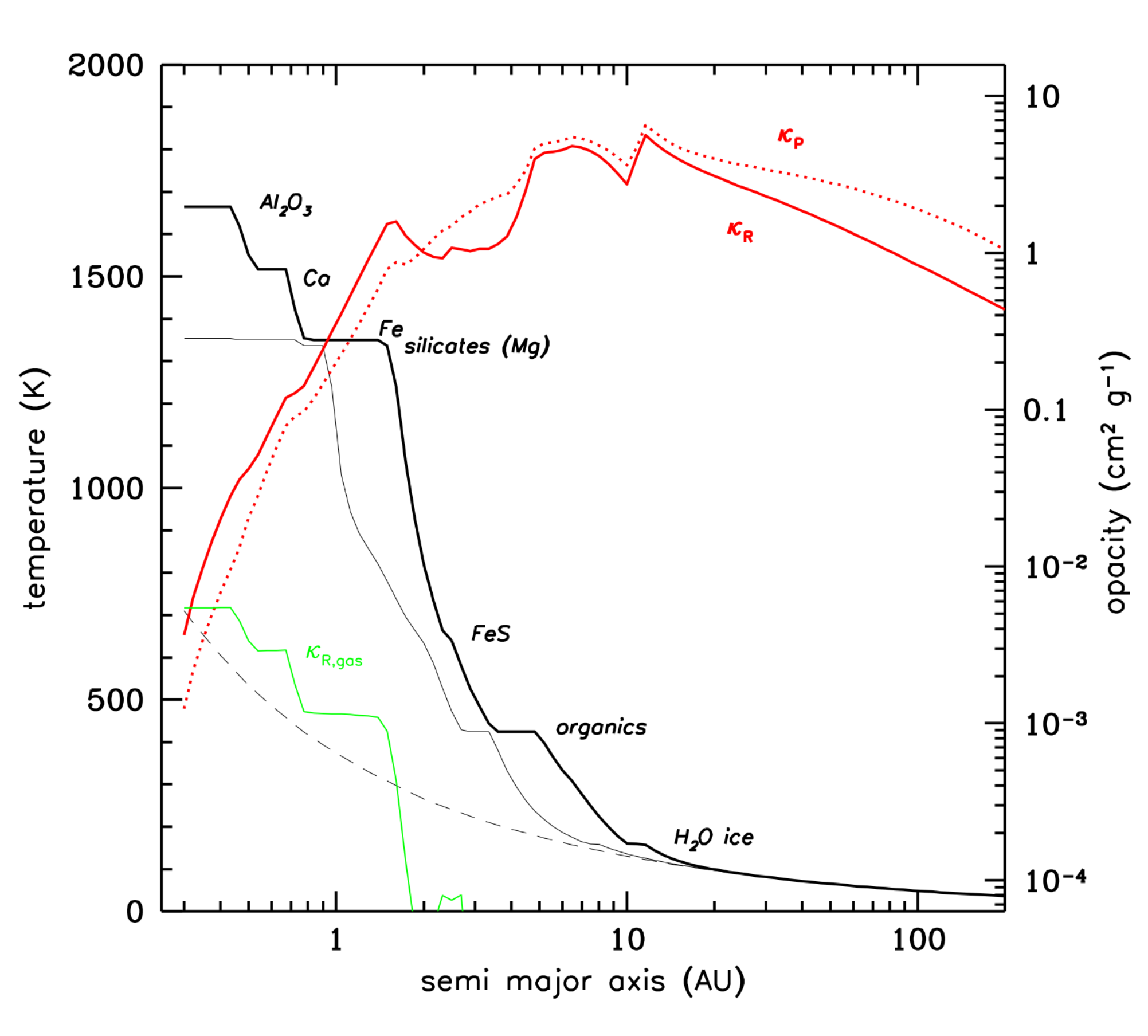}
\vspace{0 cm}
\caption{Temperature and opacity snapshots of a young nebula ($\alpha=10^{-3}, M_{\rm disk}=0.2 M_{\odot}$ shortly after its initial condition, merely to show the radial changes in opacity due to evaporation of major species. The photosphere temperature is the grey dashed line. The midplane temperature at the earliest time (100 yrs,  black line) shows inner nebula evaporation fronts for silicates, iron metal, and even CAI minerals hosting Ca and Al. Note that, inside the EF for Fe and silicates at about $1.5$~AU, the opacity (red curves) drops by an order of magnitude in less than an AU, buffering the midplane temperature. For a very short time, the refractories are also vaporized (over smaller ranges of radii) so can also be depleted by open system processes. Only $200$ years later (grey solid curve) the midplane temperature has decreased sufficiently for refractory solids (Ca- and Al- bearing) to exist, but the entire innermost nebula inside 1 AU remains hot enough for all silicates to remain evaporated, and to continue to be lost by open system processes. While the depletion of the common rock-forming elements is more widespread and long-lasting than that of the refractories, it is less widespread and shorter than for the MVEs; nevertheless the most abrupt step in depletion vs. $T_{ci}$ happens at the $T_{ci}$ for those elements whose condensed forms provide most of the opacity. }
\label{fig:opac_jnc_300}
\end{figure}

There is an abrupt change in the modeled relative abundances, at $T_{ci}$ close to that of Fe and the silicates (figures \ref{fig:alpha3-profile1}-\ref{fig:alpha4-prof3} and \ref{fig:refrac_enhancement}). There appears to be a noticeable change in the observed values as well (figure \ref{fig:braukmuller}), which we take as general support for the effect, but it is more subdued and differs in detail. For instance, the observed values of Ni and Co agree more with the depletion levels of the common rock-forming elements but in our models, they agree more with the depletion levels of the refractories. This may be simply a matter of having $T_{ci}$ not quite correct for Ni and Co in our models. Overall though, one asks, why is the abundance change so dramatic, over such a small range of $T_{ci}$? We think the explanation is related to the buffering of midplane temperature by evaporation of Fe and silicates themselves, as discussed previously by \citet{Morfill1988} and others since. These species contain most of the solids mass in the inner nebula, and when they evaporate the opacity declines dramatically (see figure \ref{fig:opac_jnc_300}) This smaller opacity buffers the midplane temperature inwards of the EFs for Fe and silicates, for a wide range of radii. This region of the nebula is where things looking like CAIs or AOAs (Ameboid Olivine Aggregates) are the only stable solids, and their mass is only a few percent of all rock-forming solids. Indeed there are times very early on, when the temperature can even exceed $T_{ci}$ for some of the refractories like Ca and Al (figure \ref{fig:opac_jnc_300}); during these times, and inwards of these radii, even these refractory elements can be depleted in our model relative to the most extreme refractories Zr and Hf (figure \ref{fig:refrac_enhancement}) which are essentially always solid. It is not clear that the observations show evidence for Ca and Al depletion however (figure \ref{fig:braukmuller}), suggesting that the times and ranges suitable for depletion of these elements were limited and putting constraints on nebula conditions. %Future improvements of this model, such as repeated cycles of depletion and radial mixing as in Profile 4, may provide better constraints on nebula conditions. 

%>>>>>>>>>>>>>>>>>>>>>>>>>>>>>>>>>>>>>>>>>>>>>>>>>>>>
%>>>>>>>>>>>>>>>>>>>>>>>>>>>>>>>>>>>>>>>>>>>>>>>>>>>>
\subsection{Limitations of 1-D model: Comparing material loss and replenishment timescales.}\label{sec:1Dlimitations}

For the purpose of this exploratory study of open system loss, we adopt a  {\it 1-D} (radial) numerical model which rests on several assumptions,  which we discuss in this section along with their possible implications. The open system mass loss concept assumes that when a particular MVE species in its solid form crosses its evaporation front  \citep[generally due to inward radial drift of macroscopic particles,][]{Estradaetal2016, Estradaetal2021}, it turns into vapor and escapes the disk through a disk wind (section \ref{subsec:diskwind}) or layered accretion (section \ref{sec:other_proc}),  leaving more refractory solid particles at lower altitudes behind. Our numerical model assumes the mass fraction associated with each MVE, now in vapor form, is always fully distributed vertically in the disk.  In reality of course, it takes a finite time for the vapor to mix through the full vertical column. For our assumption to be reasonable, this vertical mixing time must be comparable to or less than the mass loss timescale, which is bounded by the cooling time of the nebula to a stage where the specific MVE no longer evaporates anywhere in the nebula (its evaporation front recedes into the star or at least inside our inner grid boundary). For the MVEs in our evolutionary model, this cooling time is on the order of  $10^4$ to $10^5$ years as discussed earlier. In some cases, the appropriate bounding timescale can be shorter than the total cooling time - either for the more refractory elements, or for depletion signatures to reach an asymptotic state (see section \ref{sec:results}).

The vertical diffusion timescale to altitude $z$ can be estimated as $z^2 \sim D t_{\rm diff}$, where $D  = \apr c H_g$ is the diffusion coefficient. Assuming the turbulence is homogeneous and isotropic, $\apr$ characterizes diffusion in the vertical direction as well as the radial direction. A conservative assumption is that the vapor has to diffuse all the way from the disk midplane to the altitude of wind launch, at $N$ gas scale heights $H_g$. In this case the diffusion time can be roughly estimated as $t_{\rm diff}\sim (NH_g)^2/(\apr c H_g) %\sim N^2 H_g/(\apr c) 
\sim N^2/(\apr \Omega)$. Assuming that the wind is launched from a height $z=z_{wl}\sim 3 H_g$, the timescale for a complete mixing of vapor in the vertical column at $1$ AU can be estimated as $\sim 10^3 (10^{-3}/\apr)$ years. This estimate is conservatively large, because instead of all lying at the midplane,  pre-existing  particles of realistic sizes \citep{Estradaetal2016, Estradaetal2021} are sufficiently small to be vertically mixed to about $H_g$ even as they drift inside their evaporation front. In the layered accretion case, the active layer is at about the same altitude and has a significantly higher $\apr$ \citep{Zhuetal2010, Senguptaetal2019}, so upwardly mixed material reaching it is quickly mixed throughout it. Indeed, these estimated timescales would change as we move outward along the disk. However, the inner nebula is where the MVE depletion is taking place in this model, so $1$~AU serves as a reasonable reference radius. 
%\textcolor{red}{Let's just be sure that 1AU is the most appropriate radius to take this vertical mixing timescale at.}

As shown in figure \ref{fig:windprofile} and discussed in more detail in section \ref{sec:results}, the duration of low-to-moderate mass loss processes, perhaps more typical of disk winds and midplane turbulence with $\apr \sim 10^{-3}$, is several times $10^4$ years, comfortably longer than the above vertical mixing time estimates.  The higher mass loss rate scenarios  are characterized by durations that are roughly an order of magnitude shorter, so even if most of the disk below 1-2$H_g$ retains the same $\apr$ during these outburst phases, the vertical mixing time is still comparable to the mass loss duration, not significantly longer. Only in the case combining the largest mass loss rate with a low main disk $\apr$ (A4W3) may there be a noticeable shortfall of supply of evaporated MVEs to the loss region (see also below).

A different way to assess this assumption is to compare the vertical mixing time with the timescale of mass removal $\Sigma_{a}/\dot{\Sigma}$ from the upper layers through disk wind (or layered accretion). With $z_{wl}$ lying between $3 - 4 H_g$, the surface mass density contained in the active layers is approximately a few $\times 10^{-3}$ of that of the total column. In our weakest wind model\footnote{at $1$ AU, with $\apr=10^{-3}$, $\Sigma \sim 10^4$ g-cm$^{-2}$ and wind mass loss following  profile 1 of figure \ref{fig:windprofile} corresponding to a mass loss rate of $\sim 10^{-9}$ g-cm$^{-2}$-s$^{-1}$} with profile-1, it would take a few thousand years to remove all the mass above the launch height. This timescale is a few times the diffusion timescale estimated above, allowing for substantial replenishment of the material in the upper layers. For profile-2, the timescales are comparable.   For the highest mass loss rate profile we have explored  (figure \ref{fig:windprofile} and section \ref{sec:a3w2-a3w3}), the mass loss rate into the star somewhat exceeds the nominal replenishment rate by vertical mixing, by a factor of several, so we advise treating these results with caution. 

In general though, while more sophisticated models are certainly appropriate for future work, we believe based on these two assessments that assuming the MVE species in vapor form to be well mixed throughout the vertical column as mass is lost from high altitudes by open system processes is acceptable for the purpose of this exploratory paper, at least for the most plausible and successful mass loss rate (profile-2).  %In section ?? we address another wrinkle in the process, namely the fate of tiny MVE grains that may recondense at very high altitudes above their nominal evaporation fronts \citep[discussion of vertical thermal structure in ][]{Estradaetal2016}. 

\subsection{Open system loss of MVEs in Solid form}\label{sec:solidmveloss}

In our present model, it is assumed that MVEs that are evaporated at or near the midplane and diffused upwards remain vapor at the wind base and are lost in vapor form, leaving more refractory solid particles behind. However, the validity of this underlying assumption relies on two fundamental questions. First, do the vapor phase MVEs condense in the upper layers into tiny grains if a realistic vertical temperature stratification is considered? Second, if so, do they remain well mixed in those upper layers or settle quickly downwards, out of reach of the removal process? 

\begin{figure}
\includegraphics[width=0.45\textwidth]{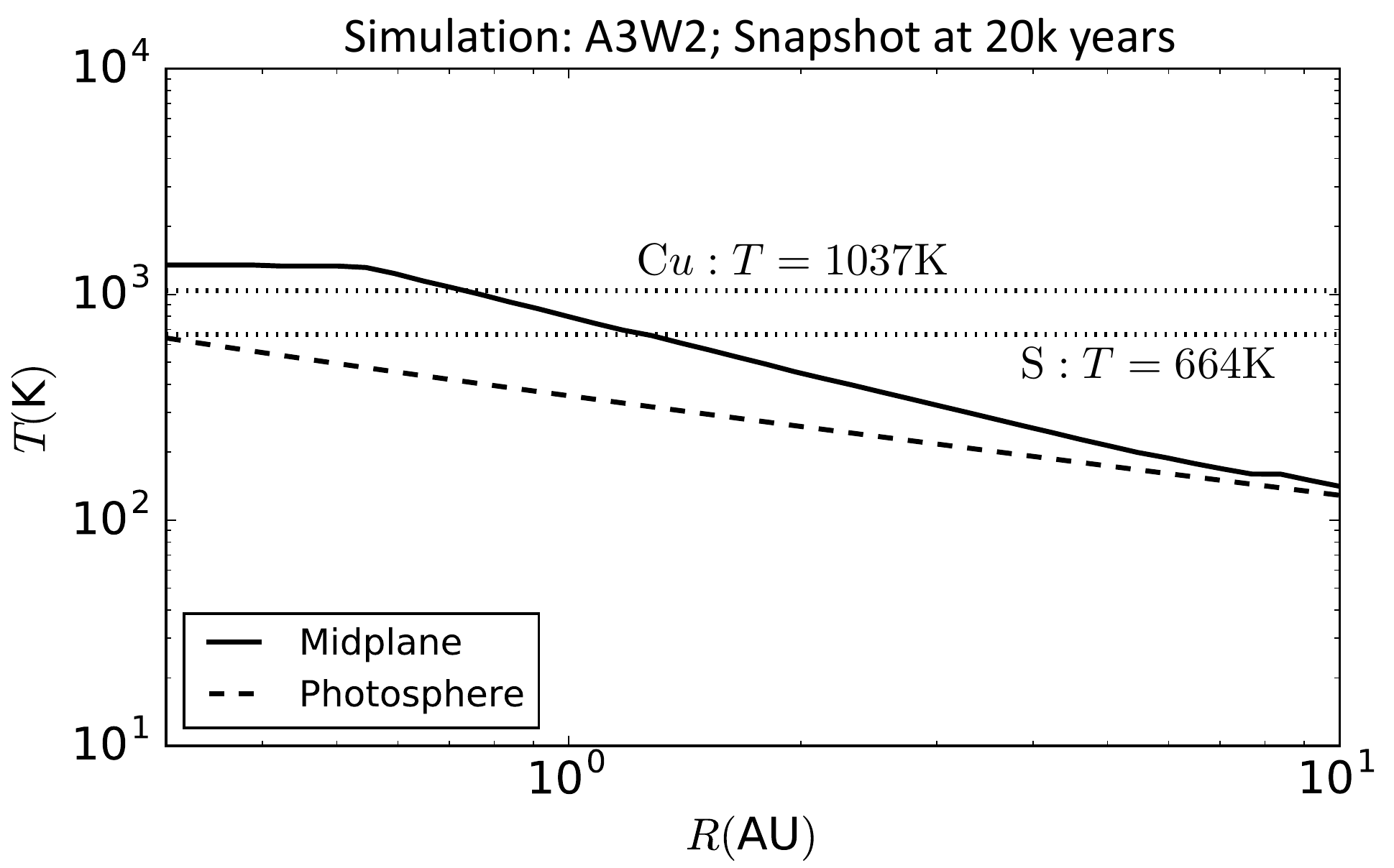}
\vspace{0 cm}
\caption{The radial variation of midplane (solid black) and photosphere (dashed black) temperatures inside $10$~AU. The dotted horizontal lines denote the condensation temperatures of two representative MVE species: Cu and S.}
\label{fig:temp_mp_phot}
\end{figure}

Addressing the question of MVE  condensation into tiny solid grains in the upper disk warrants a detailed vertical temperature stratification. We first note that the ``midplane cold finger" scenario of \citet{Krijtetal2015} where the midplane regions are colder than the photosphere, and volatiles can be cold-trapped on large particles near the midplane, is likely to be more valid for the outer nebula. A more plausible situation for our region of interest (inside 1.5 AU or so), where viscous heating is significant, is shown in figure \ref{fig:temp_mp_phot}. Here, the temperatures at the midplane (solid black) and photosphere (dashed black) are shown for simulation A3W2 with $\apr=10^{-3}$ and profile-2 for wind mass loss (See E16 for more details on vertical thermal structure). The dotted horizontal lines correspond to the condensation temperatures for two representative MVE species: Cu (top) and S (bottom). It is evident that in our numerical model, where we set the phases of MVEs (solid or vapor) according to the midplane temperature, S would be in vapor form out to roughly $1.5$~AU. However, the temperature of the photosphere everywhere (at this time) is below the condensation temperature of S. So, it would seem that the S-vapor will condense into or onto small grains in the upper layers (somewhere below the photosphere) as it diffuses upward. A similar case is shown for Cu with $T_{Cu}=1037$~K, for which the midplane EF is closer to the star. Indeed there may be tiny particles of refractory material permanently sustained at high altitudes and low densities for MVEs to condense upon, but (a) they must be tiny to be sustainable there (see below), and so (b) they would  represent an insignificant fraction of the mass in refractories, which primarily reside in larger particles near the midplane (E16).

Addressing the second point above requires estimates of how fast the new (sub)-micron  condensate particles, or pre-existing tiny refractory grains,  grow in the upper disk layers to sizes which can settle back down, out of the active layers and closer to the midplane. The combined effect of growth and settling has been studied by \citet{Senguptaetal2019} through Monte Carlo simulations, and indeed they found that for $\alpha \sim 10^{-3}$ or higher, micron and sub-micron size solid particles are always present in the upper disk atmosphere (and much larger fluffy aggregate particles would act the same way). Below we sketch those general conclusions. 

Assuming that the solid grains are well mixed with the gas, the growth timescale can be written as 
\begin{equation}
    t_g \sim \left|\frac{\,d \ln m}{\,dt}\right|=\frac{m}{\dot{m}},
\end{equation}
with
\begin{equation}
    \dot{m}=\frac{\,dm}{\,dt}\sim \frac{\Delta m}{\Delta t}\sim \frac{m}{1/n\sigma v}.
\end{equation}
Here, $n$ is the particle number density, and $v$ is the relative velocity of collision. Assuming that at the smaller size limit particles grow through sticking by colliding with particles of similar sizes \citep{Estradaetal2016, Estradaetal2021}, the growth timescale $t_g$ can be written as 
\begin{equation}
    t_g \sim \frac{m}{\rho_d \sigma v}=\frac{4}{3\chi}\frac{a\rho_m}{\sqrt{\alpha}\rho_g c}.
\end{equation}
Here, $a$ is the radius of the particles, $\chi$ is the solid-to-gas mass ratio, and $\rho_m$ and $\rho_d$ are the material (internal) density, and the volume mass density of the particles. As an estimate, the particle growth timescale  from $1\upmu$m to $100\upmu$m, above $3 H_g$ at $1$~AU in our disk model, is in excess of $10^4$~years. Hence, at the upper layers of the disk, due to the low number density and small collision cross-section, the growth timescale for particles is comparable to or longer than the timescales of interest in our simulations.

In addition, small particles are  well coupled to the gas, hindering their overall vertical settling. This coupling is further enhanced when porous particles are considered, as considered here and by \citet{Estradaetal2021}. We can estimate the time small particles will take to settle to the midplane by 
\begin{equation}\label{eqn:settlingtime1}
   t_{\rm sett}=\left|\frac{\,d\ln z}{\,d t}\right|^{-1}\sim \frac{z}{v_z},
\end{equation}
where $z$ is the vertical height and $v_z$ is the vertical terminal velocity of the particles. In terms of the Epstein regime particle gas drag stopping time $t_s$, $v_z=t_s \Omega^2z = (a \rho_m/c\rho_g)\Omega^2z$, and equation \ref{eqn:settlingtime1} becomes
\begin{equation}\label{eqn:settlingtimescale}
    t_{\rm sett}\sim\frac{c \rho_g}{a \rho_m}\Omega^{-2} \sim \frac{1}{\Omega {\rm St}},
\end{equation}
where $\rm St = t_s \Omega$ is the particle Stokes number. Comparison with the results of \citet{Estradaetal2016, Estradaetal2021} shows that for the mass-dominant particles inside the H$_2$O snowline, $\rm St \sim 10^{-3}$ (or even smaller, for porous aggregates). For solid particles at $1$~AU, equation \ref{eqn:settlingtimescale} gives a settling timescale of $\sim 10^4(100\mu{\rm m}/a)$~years from an altitude of $H_g$; however, at  altitudes of several $H_g$, lower gas densities and stronger vertical solar gravity allow only grains up to a few microns in radius to remain suspended. This effect becomes more pronounced with higher intensity of the turbulence in the disk gas. So, sufficiently tiny grains can be supported at high altitudes for long times and do not grow or settle out easily. 

The final element of this situation is the degree to which disk winds and/or layered accretion can carry off sufficiently tiny solid grains or highly porous, fractal aggregate particles, and of course this will eventually require a more detailed study. In recent work, \citet{Giacaloneetal2019} presented an estimate of the maximum grain size $a_{max}$ that can be lifted to the base of the wind, from where the grains can be carried away by advection in the wind. The wind strengths they studied were typical of older, T-Tauri systems ($3.5\times10^{-8}$ M$_{\odot}$/yr), far weaker than those we contemplate for the  early stages of interest for the MVE depletion. In these systems,  they found that even grains that {\it could} be lofted by disk winds fell back into the outer reaches of the nebula disk at large radii because of the decreasing gas density and increasing vertical solar gravity affecting the particles at increasingly high altitudes. In such a case, some outer nebula  primordial materials such as cometary IDPs could be enriched in MVEs - but observations remain too noisy (probably because of the tiny sample sizes) to discern any volatility trend at all, not to mention determining whether  IDPs are enriched or depleted in MVEs at levels complementary to MVE depletions in chondrites  \citep{Jessbergeretal2001}.  \citet{Giacaloneetal2019} also commented that for disk winds of   mass loss rate $10^{-6}-10^{-5}M_{\odot}$/yr, only somewhat larger than our profile-2 rates,  0.1-1.0$\mu$m solid grains (or obviously, larger porous grains) could be lofted without returning.  Indeed, inspection of equation (6) of \citet{Giacaloneetal2019} indicates that our profile-2 disk wind could even loft grains as large as tens of microns. Such large grains return, of course, but such large grains are also somewhat irrelevant as they are unable to reside in the rarefied upper layers where the base of the wind lies. Thus, it is not an obvious problem for tiny solid grains, or porous aggregates, of MVEs to be removed irreversibly by a disk wind such as we model here.  %\textcolor{red}{Debanjan, I changed the wording a little, to remove the connection with FUOri events. $10^{-6}$ is basically our profile 2 rate, so grain loss is not easily ruled out.}

Meanwhile, the open system loss associated with layered accretion is directly into the star, so tiny grains persisting at high altitudes will merely be advected inwards along with the  evolving accretion layer material, possibly evaporating as the material moves to regions of higher photospheric temperature. This may be an advantage for layered accretion as the open system loss process. Finally, also regarding the photospheric temperature, our radiative transfer models are not capable of capturing the excess photospheric heating implied by the high luminosity of the early sun ($\sim 10L_{\odot}$; see section \ref{sec:diskmodel}). More sophisticated radiative transfer models may find a smaller contrast between midplane and photospheric temperature than we assume in our models.

%It appears that the limited growth and slow  settling will keep solid particles smaller than a few $\mu m$ radius, or porous aggregates,  suspended in the upper layers of the disk for at least the timescale of interest in our work. The same result has been demonstrated through Monte Carlo simulations by \cite{Senguptaetal2019}. At this point, it is important to remark that hydrodynamic simulations with dust particles routinely show efficient settling of particles near the midplane and subsequent planetesimal formation through streaming instability. These simulations use substantially large particles with $St \sim 0.1$ and higher with $\apr < 10^{-5}$. However, \cite{Estradaetal2016} showed that in global simulations with realistic  disk parameters, it is hard to achieve such high $St$ for particles which can efficiently settle near the midplane. Moreover, our choice of $\apr$ is motivated by the recent simulations of several robust hydrodynamic instabilities in the disk, routinely producing $\apr \sim 10^{-4}$ and higher, where settling effect of small particles is significantly mitigated \citep{Senguptaetal2019}.  

\subsection{Does the sun get enriched in MVEs?}

 It has long been suspected that planetary accretion is inefficient. The concept of the MMSN has obscured this, since it is, by definition, 100\% efficient at gathering all the nebula solids into planets. But more recent models that avoid the several known growth barriers (which are also consistent with observations) suggest the nebula was more massive than the MMSN. \cite{Cuzzietal2010} (sections 2.1 and 3.4.2) outline some examples leading to estimates of planetesimal formation efficiency in the few percent range, in both the inner and outer solar system,  based on primordial planetesimal masses estimated by others. That is, 96-98\% of nebula solids are lost to the sun. From the standpoint of our model, this means that both MVE-depleted solids remaining in solar system bodies, and the complementary MVE-enhanced vapors that were either removed from upper nebula layers into the sun (Layered accretion option) the solar system entirely (Disk wind option), are a negligible fraction of all the elements of both types that {\it did} go into the sun. So the MVE depletion signature is expected to be indistinguishable in the sun. A simple quantitative way to estimate this is as follows:

Taking the worst case that the ``MVE-enriched vapors" are irreversibly lost from the solar system in a disk wind at some loss rate $\dot{\Sigma}$ (g/cm$^2$/sec) and not just transferred directly to the sun as in layered accretion, we can estimate their total loss, taking very conservatively the entire metallicity of the nebula $Z$ as their mass fraction in the wind, as $ \pi R^2 \dot{\Sigma} Z \Delta T$ where the zone of loss covers out to about 1 AU (see figure 4, for profile 2). In section 5.4 (footnote 4) we estimate the surface mass density loss rate in the flow for profile 1 as $10^{-9}$ g/cm$^2$/sec, and we multiply by a factor of 5 to get the profile-2 rate. The duration of the loss is roughly $10^5$ years. We can ratio this loss to the total mass of the MVEs in the sun, or $ZM_{\odot}$, so the mass fraction lost is $ \pi R^2 \dot{\Sigma}\Delta T / M_{\odot} \approx 0.02$. If this is regrded as an estimate of the ``complementary" MVE-depleted solids left behind, it can be seen as consistent with the estimates above regarding the inefficiency of accretion overall. Either estimate is undetectable in current estimates of solar abundances.

\section{Global distribution of processed material}\label{sec:globaldist}

\subsection{Radial Diffusion and mixing of processed and unprocessed material over short periods of time} \label{sec:fastmixing}

\begin{figure}
\centering
%\vspace{-.7cm}
%\includegraphics[width=\textwidth]{Braukmuller_fig.pdf}
\includegraphics[width=0.45\textwidth]{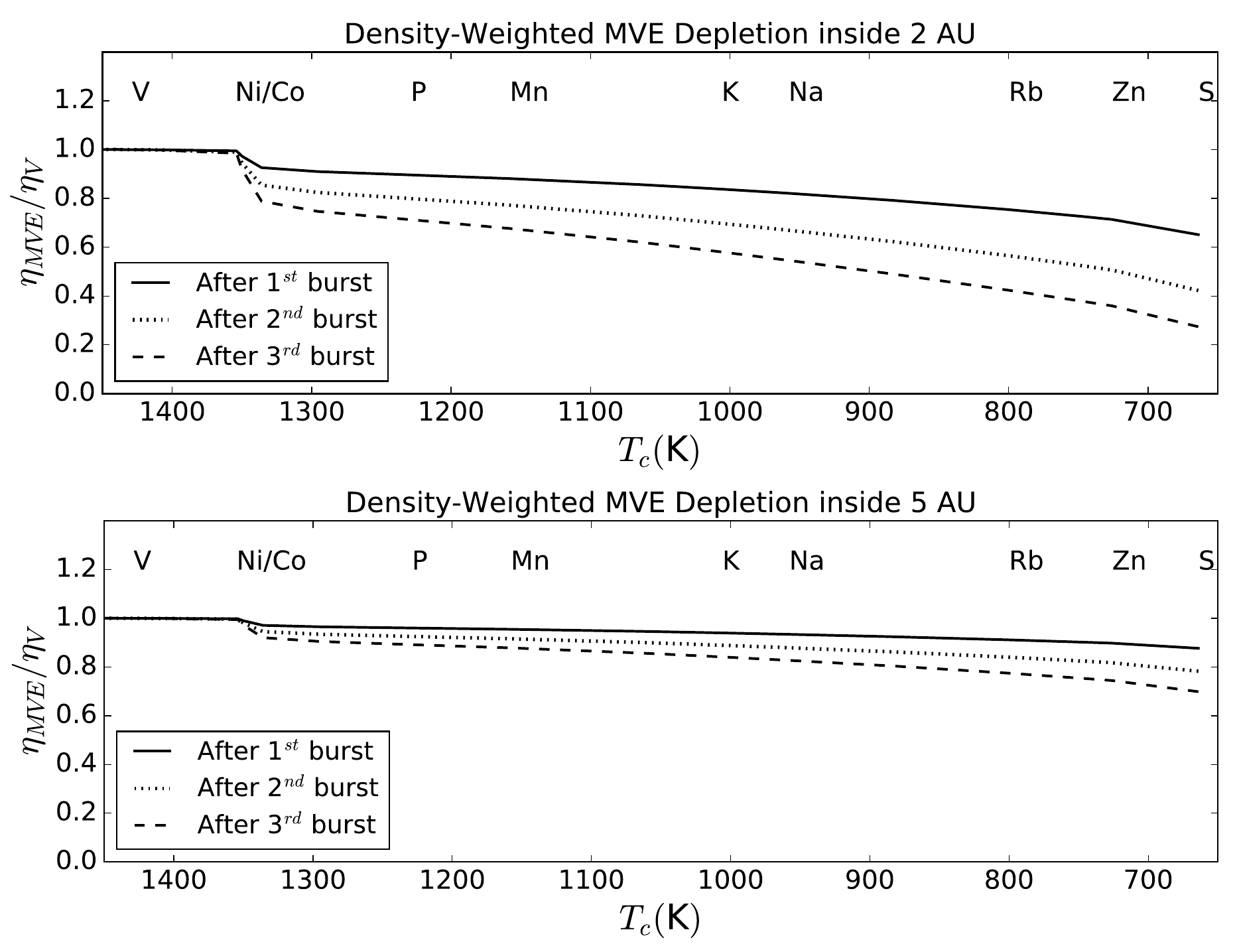}
\vspace{0 cm}
\caption{The depletion of MVEs presented in a density weighted form inside $2$~AU (top) and $5$~AU (bottom). After each burst with the profile-4 mass loss the depletion profile approaches the reference level of figure \ref{fig:braukmuller}. The number of bursts required to have the correct amount of depletion can be estimated from the figure, depending on the radial location (see section \ref{sec:fastmixing}).}
\label{fig:fastmixing}
\end{figure}

The simulation A3W4 with a profile-4 mass loss consisting of repeated bursts and quiescent phases portrays the initial disk build-up phase when infall is still going on. As we have noticed in figure \ref{fig:profile4}, after each burst the MVEs are depleted to the level depicted in figure \ref{fig:braukmuller}, but mostly effective in disk regions inside $1$~AU. However, according to \citet{Zhuetal2010} these are the stages where the disk becomes gravitationally unstable due to the accumulation of infalling material and exhibit rapid outward expansion building the disk, and mixing the MVE depleted material with the CI material in the outer region beyond $1$~AU. 

Although we do not have the fast mixing through outward expansion modeled in our numerical code as yet, we constructed a sanity check based on the A3W4 simulations as a proof-of-concept in this regard. Figure \ref{fig:fastmixing} shows the outcome of the depletion profile when the MVE depleted material from inside $1$~AU is mixed in a (solid: $\Sigma_d$) density-weighted fashion with the unprocessed material out to $2$~AU (top figure) or to $5$~AU (bottom figure). This test is valid even if the depletion signature results from a longer (10-15kyr) profile-2 mass loss instead of the more rapid, profile-3 mass loss, which may be more realistic. That is, the burst phase in $\dot{M}$ may not produce the MVE depletion, but instead produce the rapid expansion and inflow to the star, driven perhaps by gravitational instability. In the $2$~AU case the depletion profile deepens after each burst and the depletion for S reaches $\sim 0.2$ after the third burst event. However, in the $5$~AU case, for our assumed $\Sigma_d$ profile, the depletion  marches more slowly towards the desired level after each successive burst event because it must be blended with more undepleted mass. Note that, although we have simulated only three burst events in A3W4, the actual number of such events needed can be estimated as 5-10, depending on mixing distance, from the monotonic deepening seen in figure \ref{fig:fastmixing}. The number of burst events during the disk formation phase when the infall is going on, as seen by \citet{Zhuetal2010} and \citet{kadametal2020}  exceeds the number required to achieve a depletion level similar to figure \ref{fig:braukmuller} within $5$~AU. 

It is also important to note that the picture we get in figure \ref{fig:fastmixing} is on the conservative side of the actual progress towards the desired depletion levels. After each burst event which produces rapid mixing, the material {\it outside} $1$~AU will be depleted in MVEs relative to the composition before mixing, and the effect will be imprinted in the subsequent composition of  inward drifting solid material. However, because this effect is not included in our numerical model, the composition of the solids undergoing radial drift and washing out the depletion signature is CI-like. As a result, it can be safely stated that the average depletion suggested in figure \ref{fig:fastmixing} will probably deepen more quickly  when this actual physical process is included in the model. 

Moreover, the ability of a strongly depleted inner nebula to affect the {\it average} depletion of a mixed region extending out to some larger radius depends strongly on the radial variation of the solids surface mass density distribution. The models here can be approximated by a fairly flat distribution. For example, $\Sigma_d$ varies roughly as $R^{-1/2}$ out to 5 AU in figure 5, so that annuli at larger radii contain most of the mass. However, steeper powerlaws are not implausible. For instance the ``decretion disk" model of \citet{Desch2007} can be approximated by $\Sigma_d \propto R^{-2}$, and most of the mass is in {\it inner} annuli. In disks like this, the above  mixing process will rapidly lead to depleted signatures at larger radii. Intermediate cases are also possible, as the surface density variation of solids, at these radii and early times, especially in the presence of bursts and reorganization, are not yet well understood.  

%As a result, it can be safely stated that the effect of mixing as depicted in figure \ref{fig:fastmixing} will \textcolor{red}{probably} be more effective when the actual physical process is included in the model.

%\textcolor{red}{Here we put the new material on mixing during the very early, and probably repeated, outburst stages. that is, using the new simple model.}

\subsection{Radial Diffusion and mixing of processed and unprocessed material over long periods of time}\label{sec:slowmixing}

We venture a glance ahead, to place the solutions of this paper in the context of recent meteoritical observations and processes subsequent to the stage studied here, as  hypothesized by \citet{Nanneetal2019}. In their scenario, it was hypothesized that early inner nebula material (the MVE-depleted material we are modeling here) rapidly expanded out past the snowline where a growing Jupiter core then  prevented it from mixing with inner nebula material to satisfy the discrete clumping of isotopic properties between CC and NC chondrites \citep{Warren2011}.  However, one {\it would} expect this newly emplaced material to mix subsequently with pre-existing, unprocessed,  {\it outer} nebula (CI-like) material  by turbulent diffusion even if the particles remained too small for significant particle drift. Indeed, some chondrite properties strongly suggest that matrix material in CM chondrites in particular  {\it is} CI material \citep[see section \ref{sec:matrix}]{Zandaetal2011_Paris, Zandaetal2018}. Enforcing negligible radial diffusion led  \citet{Deschetal2018} to adopt a very low $\alpha = 10^{-5}$ in the outer disk, but here we  explore a wider range of parameter space.

As a quick way of assessing the possible outcomes of this process, we have used a Green's function solution developed by  \citet{ClarkePringle1988}. It can handle (in principle) arbitrary radial distributions of the nebula gas density and the turbulent viscosity, as well as arbitrary Schmidt number \citep[ratio of  viscosity to diffusivity, which we will assume is unity; see Appendix B of ][]{Estradaetal2016}. Because it neglects several physical processes, we will use this approach only as a rough guide to how much mixing the nebula outside the snowline might experience. Processes it neglects would include gas-drag drift for larger particles with ${\rm St}/\alpha > 1$ \citep{Jacquetetal2012, Estradaetal2016}, as well as outward torques and even nebula gas truncation by a growing Jupiter. 

\citet[henceforth CP88]{ClarkePringle1988} assume a nebula with surface gas mass density $\Sigma = \Sigma_o R^{-a}$. Moreover, they  assume a ``steady disk" in which the mass accretion rate $\dot{M} = 3 \pi \nu \Sigma = 2 \pi R v_g \Sigma(R)$ is constant. This constrains the turbulent viscosity to the form $\nu = \nu_o R^a$, and also allows them to substitute for the radial advection velocity term $v_g= -3 \nu/2R = -3\nu_o R^{a-1}/2$ in the advection-diffusion equation. Note that $\Sigma_o$ and $\nu_o$ are not dimensionally physical, but this is easily resolved by rewriting, for instance, $\nu = \nu_o'(R/R_o)^{-a}$ where $\nu_o' = \alpha c H$ in the usual way. The complete (dimensional) solution of CP88 is their equation 3.2.4, and a nondimensionalized version is given in their equation 3.2.6. We will not repeat all the definitions here. The solutions are expressed in terms of a modified Bessel function $I_{\beta}$ of arbitrary order $\beta$, where $\beta=3/(4-2a)$. Fortunately it is not hard to select a few reasonable values of $a$ that leave $\beta$ an integer, for which solutions for $I_{\beta}$ are not hard to find ($a=0.5,\beta=1; \, a=5/4, \beta=2; \, a=3/2, \beta=3$).\footnote{We used a fortran code from http://jean-pierre.moreau.pagesperso-orange.fr/Fortran/tbessi\_f90.txt. An ostensibly similar Green's function approach of \citet{Cuzzietal2003} was tailored to an operator-splitting approach in which the advection by nebula gas, drift of the particles due to gas drag, and diffusion by turbulence were handled separately. The solutions are functionally {\it similar}, not surprisingly, but not identical.} . 
For completeness, we give below our derivation of the dimensional form of the CP88 Green's function $G(R,t)$, which is the radial distribution after time $t$ of the concentration (fractional abundance) of some ``contaminant" that was initialized as a delta-function at $R_o,t=0$. CP88 defined a term $\alpha_o \equiv 2/(2-a)$ which we denote $A_o$ to avoid confusion with nebula turbulent intensity parameter $\alpha$, and we have rewritten the viscosity in the dimensional form given above.  

\begin{equation}\label{eq:CP88}
\begin{aligned}
C(R,t)= & \frac{C_o A_o R_o}{2 \nu_o' t}\left( \frac{R_o}{R}\right)^{3/4} 
       I_{\beta}\left(\frac{A_o^2 R^{(2-a)/2} R_o^{(2+a)/2}}{2\nu_o' t} \right)\times   \\
& {\rm{exp}}\left( -\frac{A_o^2}{4 \nu_o' t} \left(R_o^a\left(R_o^{2-a} +R^{2-a} \right) \right)\right).
\end{aligned}
\end{equation}

\begin{figure}
%\centering
%\vspace{-.7cm}
%\includegraphics[width=\textwidth]{Braukmuller_fig.pdf}
\includegraphics[width=0.45\textwidth, height=0.3 \textheight]{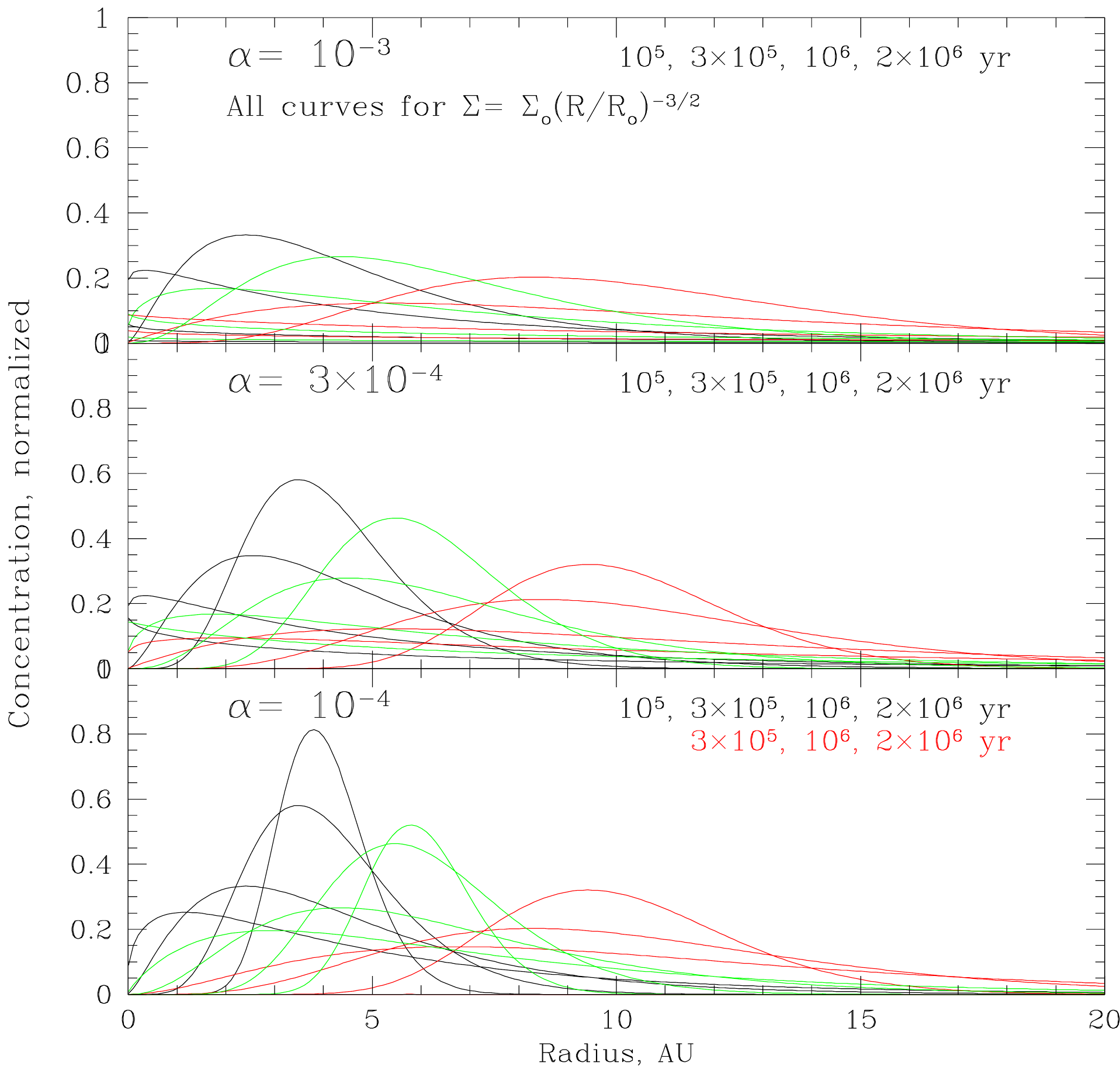}
\vspace{0 cm}
\caption{Plots of the concentration function $C(R,t)$ for a variety of times from $10^5$ to $2\times 10^6$ years after the final expansion of the inner nebula, for three different values of nebula turbulent intensity $\alpha$. The different color sets started at $R_o$= 4 (black), 6 (green), and 10 (red) AU. The profiles become flatter with increasing time. }
\label{fig:diffusion3alpha}
\end{figure}

\vspace{0.05 in}
The constant $C_o$ is defined by the initial mass of the tracers: $C(R,0) = C_o \delta(R-R_o)$. We checked our dimensionalized version by also calculating the published nondimensionalized equation 3.2.6 of CP88, with perfect agreement. 
The results are shown in figure \ref{fig:diffusion3alpha}, for three different values of nebula $\alpha$. Initial delta function tracers started at 4, 6, and 10 AU, and $\nu_o'$ was calculated for those locations. Numerical issues arise for the  smallest $\alpha$ and shortest $t$.

These solutions may serve as rough indications of mixing  in the emplacement region hypothesized by \citet{Nanneetal2019} of the CC-to-be material outside a subsequent Jupiter truncation radius, here merely assumed to be at 4 AU), after an early expansion phase that follows the MVE depletion phase. The material the MVE-depleted material mixes {\it with} would be less processed, plausibly CI-like in nature \citep{Zandaetal2011_Paris, Zandaetal2018}.  

It can be seen that after 1-2Myr (the time between expansion/truncation/achondrite formation and the formation of the chondrites) there has been significant mixing of  ``unprocessed"  material from 10AU in to 4AU, but the unprocessed material never constitutes more than about half of the total - consistent with, for instance, CI-like matrix in CM chondrites. If disk truncation stalled the inward advective motion of the gas and particles, the CI fraction might be even  smaller, as for instance in the NC chondrites.  The different $\alpha$ values do, however, produce different {\it radial  variations} in the mixing ratios, with $\alpha=10^{-3}$ leading to essentially no radial variation. As we know, there is  some variation in the amount of total matrix, and the plausible CI content {\it of} the matrix, between CC groups (see section \ref{sec:matrix}).  

Overall, the results of figure \ref{fig:diffusion3alpha} at least suggest that planetesimals accreting outside the snowline  would be expected to retain some significant fraction of their initial (MVE-depleted, inner nebula) material and also contain some minimally processed, nominally CI material - but will not necessarily be {\it totally dominated} by CI material, even after 2Myr. 

An important unknown is how far outward previous outbursts might have emplaced {\it processed} material, and displaced {\it unprocessed} material. Models of this sort are important to pursue but well beyond the scope of this paper. We provided a simplified analysis of the possible consequences in section \ref{sec:fastmixing} above. 
%The underlying solids distribution becomes well mixed over a Myr or so, during which time chondrule production might be going on, with the final mixing in of still-unprocessed material, the plateau MVEs, etc, happening over some shorter time, between chondrule formation and planetesimal accretion.  
Missing pieces in the puzzle include how far outward the hypothetical early expansion phase can emplace originally inner nebula material,   how many outbursts occurred before the final one which emplaced the material we see today, and the presence or absence of other barriers to diffusion and drift. 

As an alternate to a very low outer-nebula $\alpha$ \citep{Deschetal2018}, the fact that any CI-like contribution to  CV or CO chondrules remains small (figure \ref{fig:braukmuller})  might be telling us that a rather large radial extent of the disk even outside the snowline had been {\it already} processed, radially expanded and mixed, and processed again, multiple times, and only material residing beyond 5-10AU at this stage remained even approximately ``unprocessed". Note that even CI material is not completely ``unprocessed" - it is significantly depleted in carbon  \citep{Jessbergeretal1988, MummaCharnley2011, Woodwardetal2020}, and perhaps even water  \citep[see below]{Tenneretal2015,Marrocchietal2018, Hertwigetal2018, Hertwigetal2019,McCubbinBarnes2019,  Alexander2019CC,  Woodwardetal2020, Fukudaetal2021}. Clearly, the situation might be complicated, and actual models of the inner nebula processing, expansion, mixing, reprocessing, remixing, and so on, as perhaps manifested in  multiple FU-Orionis type bursts, will need to be modeled more carefully for a complete test of this scenario. The problem   surely deserves considerable future study.

%\subsection{Water and carbon} \label{subsec:watercarbon}  
%It has not escaped our attention that the scenario described here has broader implications worthy of followup. Specifically, carbon (even carried in the moderately refractory organics assumed by \citet{Pollacketal1994} and \citet{Estradaetal2016} and water are volatiles that would be depleted along with the MVEs by this process, and indeed over a wider range of radii. It may be that what most people consider as ``unprocessed" CI material already bears the signature of this depletion, relative to truly ``primitive" material \citep{Marrocchietal2018,  Alexander2019CC, McCubbinBarnes2019, Woodwardetal2020}. A single early process that could provide   significant depletions of all such major materials at once is worth further investigation. 

 \section{Summary of Results:}
 
 %Overall, we believe that the ubiquity of the MVE depletion, its associated high temperatures, and the bulk of the meteorite evidence to date, leave early hot inner nebula processes, acting over a finite radial extent, attractive. 
 \subsection{Depletion of MVEs}\label{sec:sec7.1} We suggest that the problems encountered by previous `closed system' hot early nebula  models of MVE depletion \citep{Cassen1996, Cassen2001, Ciesla2008} may be avoided by  `open system' behavior (irreversible loss of MVEs), so that whenever or wherever the depleted gas finally {\it does} cool, there is selectively less of the more volatile material in the region  {\it to} condense. Thus, chondrite parent body formation can be delayed into the chondrule formation era. Achondrite parent bodies indeed formed much earlier \citep{Kruijeretal2017}, but because of their melting, may have incurred other forms of MVE losses which would be hard to untangle. 
 
  %\textcolor{blue}{A lower rate of $10^{-6} M_{\odot}$ is also capable of getting a desired level of depletion, however, whether that loss rate will be associated with fast radial mixing is questionable. A loss rate higher than that depicted in profile-3 is however will always achieve required depletion, perhaps in a shorter timescale.}
 
 %\sout{The model parameters most likely to satisfy the observations suggest short (100-1000yr), probably repeated, bursts of high mass loss with radially integrated rates on the order of
% $10^{-5} M_{\odot}$ per year, separated by longer periods of more quiescent viscous evolution with $\alpha \sim 10^{-4}-10^{-3}$, not unlike expectations for early ExOr or FUori stages.} 
 
 The scenario we currently think most reasonably likely to satisfy the observations is {\it a series of repeated depletions} by moderate strength (profile-2) disk winds (or layered accretion) acting during quiescent viscous evolution with $\alpha \sim 10^{-4}-10^{-3}$, and lasting for a period of some tens of thousands of years, {\it terminated by short (100-1000yr) bursts} of high mass loss and dramatic radial mixing outwards beyond the MVE active loss region, with radially integrated rates on the order of
 $10^{-5} M_{\odot}$ per year
 (not unlike expectations for early ExOr or FUori stages). {\it The actual MVE depletions arise mostly during the quiescent periods (section \ref{sec:a3w2-a3w3}), and the depleted material is radially expanded and mixed during the outbursts.}
 It has previously been suggested by others that outburst events are associated with rapid radial mixing events, perhaps associated with gravitational instabilities,  extending to several AU beyond the loss region \citep{Boss2008,Boss2012, VorobyovBasu2005, VorobyovBasu2006}. Such a rapid expansion and mixing stage (or stages) would represent an important aspect of the larger scenario in which the work of this paper is embedded \citep[eg.,][]{Nanneetal2019}, but modeling it properly is beyond the scope of this paper. Our preliminary estimates  lead us to suggest 5-10 such depletion-then-mixing events in all (section \ref{sec:fastmixing}), each leaving the radial range {\it outside the hot inner nebula region where direct loss occurs} also depleted. 
 
 An  early nebula process like this could  be overwritten or modified, perhaps only in part, by subsequent processes such as ongoing, radially variable  infall from the parent cloud,  radial mixing with less- or un-processed material, and/or ultimately chondrule formation.

\subsection{Enrichment of refractories} We also suggest that these same hot inner nebula, open system processes may naturally produce what is commonly regarded as an ``enrichment'' of refractory elements in CCs in a natural way, by quantitative {\it depletion} of the major rock-forming elements (to which the refractory abundances are usually normalized) in the same process by which the trace MVE elements are depleted. In explaining multiple puzzling observations at once, the hot inner nebula model may be more appealing than alternate models such as MVE depletions associated with chondrule formation. The processes so far hypothesized  for chondrule formation would seem not to have an intensity or duration sufficient to irreversibly remove significant quantities of the common rock-forming materials entirely and irreversibly  from the nebula region in which they occur \citep{Deschetal2012, Chaumardetal2018}. 
 
 \subsection{Water and carbon} \label{subsec:watercarbon}  
It has not escaped our attention that the scenario described here has broader implications worthy of followup. Specifically, carbon (perhaps primarily carried in the moderately refractory or ``CHON" organics assumed by \citet{Pollacketal1994} and \citet{Estradaetal2016}) and water are even more volatile materials that would be depleted along with the MVEs by this process, and indeed over a wider range of radii. It may be that what most people consider as ``unprocessed" CI material already bears the signature of this depletion, relative to truly ``primitive" material \citep{Tenneretal2015,Marrocchietal2018, Hertwigetal2018, Hertwigetal2019,McCubbinBarnes2019,  Alexander2019CC,  Woodwardetal2020, Fukudaetal2021}. A single early process that could provide significant depletions of all such major materials at once is worth further investigation. 
 
\subsection{Future studies of value}

It will be important to better understand how later stage processes such as ongoing infall, radial mixing, and/or chondrule formation by various processes  augment or modify the observed MVE depletions.  
 %Could chondrule formation be an alternate, or additive MVE depletion mechanism? %{\bf  How ubiquitous was it? Is chondrule formation required for planetesimal formation? } 
 %Are there aspects of chondrite MVE depletion that {\it  require} chondrule formation?  

Amongst observations that might  discriminate between alternate processes, we might suggest some attempt to unravel whether the as-accreted MVE signature of the early-formed achondrites (the irons, the Angrites, etc) is either CI-like or MVE-depleted, thus better age-dating the MVE-depletion process.  Additionally, can more MVE measurements be made of the matrix in the most primitive CV, CO and CR chondrites \citep{AbreuBrearley2010, Zandaetal2012} or in IDPs \citep{Jessbergeretal2001}? And finally, it would be exciting if  astronomical observations of winds or jets could detect, and constrain, the composition of the ejected material. Can the missing MVEs be detected in outflows?

%\textcolor{red}{Indeed, a hot early nebula open system loss process may have even more unexpected and intriguing implications, regarding what is commonly interpreted as ``enhancement" of the elements more refractory than the common rock-forming elements (Ni, Fe, Mg, Si). We discuss this somewhat radical idea in the Appendix (?? or in results??). }

\section*{ACKNOWLEDGMENTS} We are grateful to Steve Desch, Uma Gorti, Gary Huss, Sasha Krot, Justin Simon, and Diane Wooden for highly illuminating conversations, and to Scott Sandford and Jeff Scargle for thoughtful internal reviews.  We are also grateful to the two anonymous reviewers for the insightful comments which improved the quality of this manuscript. DS was supported by the NASA Postdoctoral Program (NPP) fellowship, and partly by the NASA Astrobiology Institute. PE, JC, and MH were supported partly by grants from the NASA Emerging Worlds program (16-EW16-2-0094 and 80NSSC18K0595). JC and PE were partly supported by an ISFM grant from NASA's Planetary Science Division, and partly by a grant from the NASA Astrobiology Institute.   All the simulations presented in this paper are performed on the NASA Advanced Supercomputing  (NAS) facility with generous computational resources provided through NPP and ISFM allocations.  

\section*{Appendix: Consistency of Numerical Model.}

\begin{figure*}[t]
\centering
%\vspace{-.7cm}
%\includegraphics[width=\textwidth]{Braukmuller_fig.pdf}
\includegraphics[width=\textwidth]{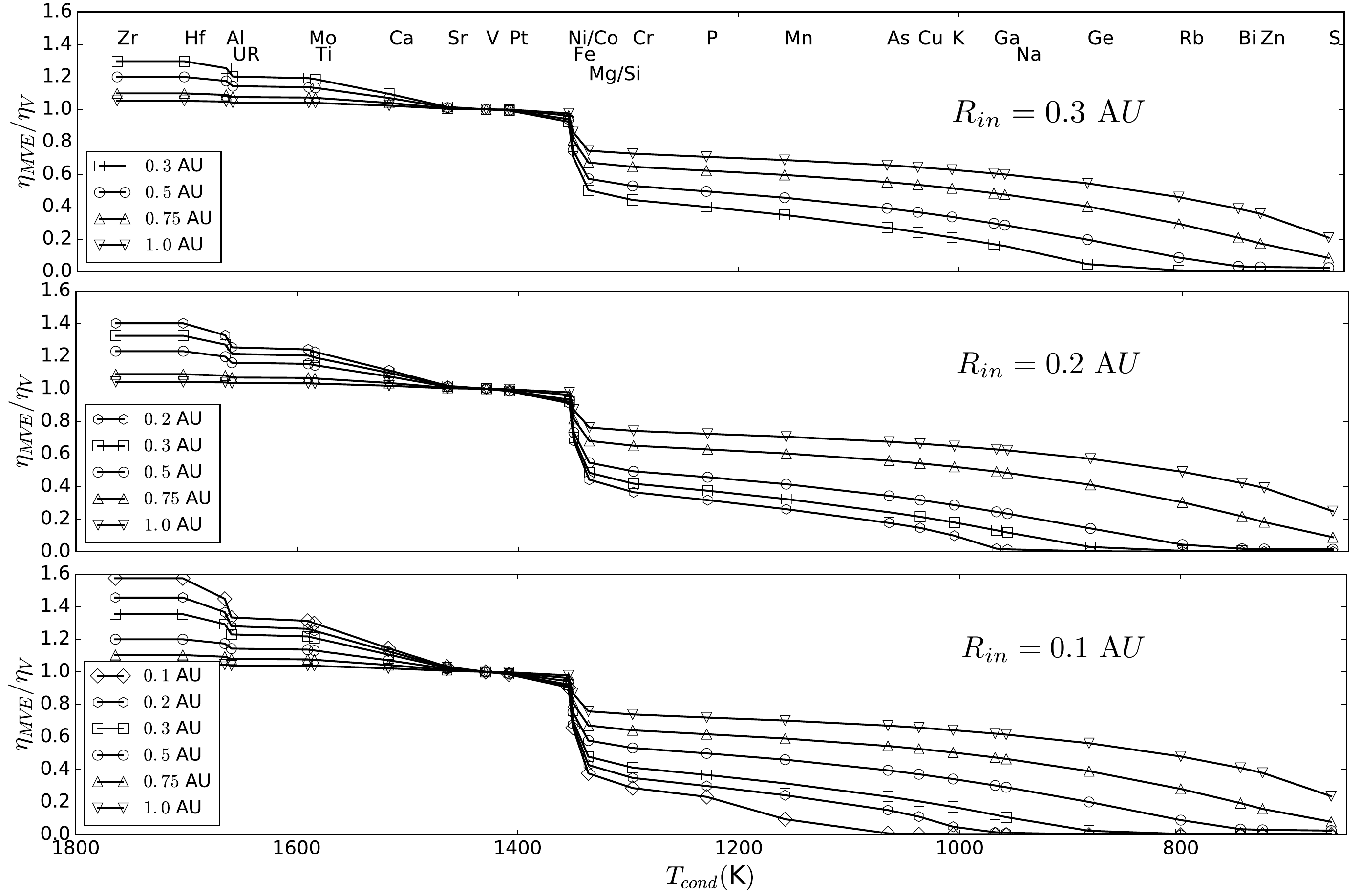}
\vspace{0 cm}
\caption{Figure showing the consistency of our numerical model and the implementation of boundary condition. Three simulation snapshots at $3.5$~k years, with three different inner boundaries are presented here with $\apr=10^{-3}$ and profile-3 mass loss: {\textbf {top:}} $R_{in}=0.3$~AU (A3W3), {\textbf {middle:}} $R_{in}=0.2$~AU (A3W3R02) and {\textbf {bottom:}} $R_{in}=0.1$~AU (A3W3R01). The MVE depletion profile at $0.3$~AU is identical in all three cases. Similarly, the signature is same for $R_{in}=0.2$~AU for the middle and bottom plot. Similar trends can be observed at $0.5$, $0.75$ and $1$~AU in all three snapshots.} 
\label{fig:code_check}
\end{figure*}

In the simulations presented throughout this paper, all the interesting depletion in MVE abundances takes place in the very inner part of the nebula, close to the innermost radius of our computational setup. Hence, it is crucial that we test the consistency of our numerical model, especially in the context of the inner boundary condition. In figure \ref{fig:code_check}, we present the results from essentially the same simulation A3W3, but with different $R_{in}$. The top panel presents the usual setup with $R_{in}=0.3$~AU, the middle panel is from simulation A3W3R02 with $R_{in}=0.2$~AU and the bottom panel shows the results from simulation A3W3R01 with $R_{in}=0.1$~AU. In all three cases the snapshots are taken at $3.5$k years. In the top panel, four different radii at $0.3$, $0.5$, $0.75$ and $1$~AU are shown. The innermost radius of $0.2$~AU in the middle panel is added for simulation A3W3R02. An additional radius at $0.1$~AU is added in the bottom figure for simulation A3W3R01. The depletion profiles at $0.3$~AU from all three simulations are same within the computational error. Similarly, the profiles for $0.2$~AU for simulations A3W2R02 and A3W3R01 exhibit exactly the same pattern. This gives us confidence that the depletion signatures we find in the inner part of the disk are not affected by the choice of inner boundary condition.

\bibliographystyle{aasjournal}

\bibliography{reference}

\begin{thebibliography}{}
\expandafter\ifx\csname natexlab\endcsname\relax\def\natexlab#1{#1}\fi

\bibitem[{{Abreu} \& {Brearley}(2010)}]{AbreuBrearley2010}
{Abreu}, N.~M., \& {Brearley}, A.~J. 2010, \gca, 74, 1146

\bibitem[{{Alexander}(2005)}]{Alexander2005}
{Alexander}, C.~M.~O. 2005, Meteoritics and Planetary Science, 40, 943

\bibitem[{{Alexander}(2019{\natexlab{a}})}]{Alexander2019CC}
{Alexander}, C. M.~O. 2019{\natexlab{a}}, \gca, 254, 277

\bibitem[{{Alexander}(2019{\natexlab{b}})}]{Alexander2019NC}
---. 2019{\natexlab{b}}, \gca, 254, 246

\bibitem[{{Amelin} {et~al.}(2002){Amelin}, {Krot}, {Hutcheon}, \&
  {Ulyanov}}]{Amelinetal2002}
{Amelin}, Y., {Krot}, A.~N., {Hutcheon}, I.~D., \& {Ulyanov}, A.~A. 2002,
  Science, 297, 1678

\bibitem[{{Anders}(1964)}]{Anders1964}
{Anders}, E. 1964, \ssr, 3, 583

\bibitem[{{Anders}(1977)}]{Anders1977}
---. 1977, Earth and Planetary Science Letters, 36, 14

\bibitem[{{Audard} {et~al.}(2014){Audard}, {{\'A}brah{\'a}m}, {Dunham},
  {Green}, {Grosso}, {Hamaguchi}, {Kastner}, {K{\'o}sp{\'a}l}, {Lodato}, \&
  {Romanova}}]{Audardetal2014}
{Audard}, M., {{\'A}brah{\'a}m}, P., {Dunham}, M.~M., {et~al.} 2014, in
  Protostars and Planets VI, ed. H.~{Beuther}, R.~S. {Klessen}, C.~P.
  {Dullemond}, \& T.~{Henning}, 387

\bibitem[{{Bai}(2013)}]{Bai2013}
{Bai}, X.-N. 2013, \apj, 772, 96

\bibitem[{{Bai}(2016)}]{Bai2016}
---. 2016, \apj, 821, 80

\bibitem[{{Balbus} \& {Hawley}(1991)}]{balbus1991}
{Balbus}, S.~A., \& {Hawley}, J.~F. 1991, \apj, 376, 214

\bibitem[{{Bland} {et~al.}(2005){Bland}, {Alard}, {Benedix}, {Kearsley},
  {Menzies}, {Watt}, \& {Rogers}}]{Blandetal2005}
{Bland}, P.~A., {Alard}, O., {Benedix}, G.~K., {et~al.} 2005, Proceedings of
  the National Academy of Science, 102, 13755

\bibitem[{{Bond} {et~al.}(2010){Bond}, {Lauretta}, \& {O'Brien}}]{Bondetal2010}
{Bond}, J.~C., {Lauretta}, D.~S., \& {O'Brien}, D.~P. 2010, \icarus, 205, 321

\bibitem[{{Boss}(2008)}]{Boss2008}
{Boss}, A.~P. 2008, Earth and Planetary Science Letters, 268, 102

\bibitem[{{Boss}(2012)}]{Boss2012}
---. 2012, Annual Review of Earth and Planetary Sciences, 40, 23

\bibitem[{{Braukm{\"u}ller} {et~al.}(2019){Braukm{\"u}ller}, {Wombacher},
  {Funk}, \& {M{\"u}nker}}]{Braukmulleretal2019}
{Braukm{\"u}ller}, N., {Wombacher}, F., {Funk}, C., \& {M{\"u}nker}, C. 2019,
  Nature Geoscience, 12, 564

\bibitem[{{Braukm{\"u}ller} {et~al.}(2018){Braukm{\"u}ller}, {Wombacher},
  {Hezel}, {Escoube}, \& {M{\"u}nker}}]{Braukmulleretal2018}
{Braukm{\"u}ller}, N., {Wombacher}, F., {Hezel}, D.~C., {Escoube}, R., \&
  {M{\"u}nker}, C. 2018, \gca, 239, 17

\bibitem[{{Budde} {et~al.}(2016){Budde}, {Kleine}, {Kruijer}, {Burkhardt}, \&
  {Metzler}}]{Buddeetal2016}
{Budde}, G., {Kleine}, T., {Kruijer}, T.~S., {Burkhardt}, C., \& {Metzler}, K.
  2016, Proceedings of the National Academy of Science, 113, 2886

\bibitem[{Burshtein {et~al.}(1993)Burshtein, Shimony, \& Levy}]{Burshtein:93}
Burshtein, Z., Shimony, Y., \& Levy, I. 1993, J. Opt. Soc. Am. A, 10, 2246

\bibitem[{{Calvet} {et~al.}(1993){Calvet}, {Hartmann}, \&
  {Kenyon}}]{calvetetal93}
{Calvet}, N., {Hartmann}, L., \& {Kenyon}, S.~J. 1993, \apj, 402, 623

\bibitem[{{Carrasco-Gonz{\'a}lez} {et~al.}(2019){Carrasco-Gonz{\'a}lez},
  {Sierra}, {Flock}, {Zhu}, {Henning}, {Chandler}, {Galv{\'a}n-Madrid},
  {Mac{\'\i}as}, {Anglada}, {Linz}, {Osorio}, {Rodr{\'\i}guez}, {Testi},
  {Torrelles}, {P{\'e}rez}, \& {Liu}}]{Carrascogonzalezetal2019}
{Carrasco-Gonz{\'a}lez}, C., {Sierra}, A., {Flock}, M., {et~al.} 2019, \apj,
  883, 71

\bibitem[{{Carter} \& {Stewart}(2020)}]{CarterStewart2020}
{Carter}, P.~J., \& {Stewart}, S.~T. 2020, \psj, 1, 45

\bibitem[{{Cassen}(1996)}]{Cassen1996}
{Cassen}, P. 1996, Meteoritics and Planetary Science, 31, 793

\bibitem[{{Cassen}(2001)}]{Cassen2001}
---. 2001, Meteoritics and Planetary Science, 36, 671

\bibitem[{{Cassen} \& {Moosman}(1981)}]{CassenMoosman1981}
{Cassen}, P., \& {Moosman}, A. 1981, \icarus, 48, 353

\bibitem[{{Chaumard} {et~al.}(2018){Chaumard}, {Humayun}, {Zanda}, \&
  {Hewins}}]{Chaumardetal2018}
{Chaumard}, N., {Humayun}, M., {Zanda}, B., \& {Hewins}, R.~H. 2018, \maps, 53,
  984

\bibitem[{{Ciesla}(2008)}]{Ciesla2008}
{Ciesla}, F.~J. 2008, Meteoritics and Planetary Science, 43, 639

\bibitem[{{Ciesla} \& {Cuzzi}(2006)}]{CieslaCuzzi2006}
{Ciesla}, F.~J., \& {Cuzzi}, J.~N. 2006, \icarus, 181, 178

\bibitem[{{Clarke} \& {Pringle}(1988)}]{ClarkePringle1988}
{Clarke}, C.~J., \& {Pringle}, J.~E. 1988, \mnras, 235, 365

\bibitem[{{Connolly} {et~al.}(2001){Connolly}, {Huss}, \&
  {Wasserburg}}]{Connollyetal2001}
{Connolly}, H.~C., {Huss}, G.~R., \& {Wasserburg}, G.~J. 2001, \gca, 65, 4567

\bibitem[{{Connolly} {et~al.}(2000){Connolly}, {Huss}, \&
  {Wasserburg}}]{Connollyetal2000}
{Connolly}, H.~C., J., {Huss}, G.~R., \& {Wasserburg}, G.~J. 2000, in Lunar and
  Planetary Science Conference, Lunar and Planetary Science Conference, 1437

\bibitem[{{Cuzzi} {et~al.}(2003){Cuzzi}, {Davis}, \&
  {Dobrovolskis}}]{Cuzzietal2003}
{Cuzzi}, J.~N., {Davis}, S.~S., \& {Dobrovolskis}, A.~R. 2003, \icarus, 166,
  385

\bibitem[{{Cuzzi} {et~al.}(2014){Cuzzi}, {Estrada}, \& {Davis}}]{Cuzzietal2014}
{Cuzzi}, J.~N., {Estrada}, P.~R., \& {Davis}, S.~S. 2014, \apjs, 210, 21

\bibitem[{{Cuzzi} {et~al.}(2010){Cuzzi}, {Hogan}, \& {Bottke}}]{Cuzzietal2010}
{Cuzzi}, J.~N., {Hogan}, R.~C., \& {Bottke}, W.~F. 2010, \icarus, 208, 518

\bibitem[{{D'Antona} \& {Mazzitelli}(1994)}]{DantonaMazzitelli1994}
{D'Antona}, F., \& {Mazzitelli}, I. 1994, \apjs, 90, 467

\bibitem[{{Desch}(2007)}]{Desch2007}
{Desch}, S.~J. 2007, \apj, 671, 878

\bibitem[{{Desch} {et~al.}(2017){Desch}, {Estrada}, {Kalyaan}, \&
  {Cuzzi}}]{Deschetal2017}
{Desch}, S.~J., {Estrada}, P.~R., {Kalyaan}, A., \& {Cuzzi}, J.~N. 2017, \apj,
  840, 86

\bibitem[{{Desch} {et~al.}(2018){Desch}, {Kalyaan}, \& {O'D.
  Alexander}}]{Deschetal2018}
{Desch}, S.~J., {Kalyaan}, A., \& {O'D. Alexander}, C.~M. 2018, \apjs, 238, 11

\bibitem[{{Desch} {et~al.}(2012){Desch}, {Morris}, {Connolly}, \&
  {Boss}}]{Deschetal2012}
{Desch}, S.~J., {Morris}, M.~A., {Connolly}, H.~C., \& {Boss}, A.~P. 2012,
  \maps, 47, 1139

\bibitem[{{Dullemond} {et~al.}(2018){Dullemond}, {Birnstiel}, {Huang},
  {Kurtovic}, {Andrews}, {Guzm{\'a}n}, {P{\'e}rez}, {Isella}, {Zhu}, {Benisty},
  {Wilner}, {Bai}, {Carpenter}, {Zhang}, \& {Ricci}}]{Dullemondetal2018}
{Dullemond}, C.~P., {Birnstiel}, T., {Huang}, J., {et~al.} 2018, \apjl, 869,
  L46

\bibitem[{{Elser} {et~al.}(2012){Elser}, {Meyer}, \& {Moore}}]{Elseretal2012}
{Elser}, S., {Meyer}, M.~R., \& {Moore}, B. 2012, \icarus, 221, 859

\bibitem[{{Estrada} \& {Cuzzi}(2008)}]{EstradaCuzzi2008}
{Estrada}, P.~R., \& {Cuzzi}, J.~N. 2008, \apj, 682, 515

\bibitem[{{Estrada} \& {Cuzzi}(2022)}]{Estradaetal2022}
---. 2022, \apj, in preparation

\bibitem[{{Estrada} {et~al.}(2016){Estrada}, {Cuzzi}, \&
  {Morgan}}]{Estradaetal2016}
{Estrada}, P.~R., {Cuzzi}, J.~N., \& {Morgan}, D.~A. 2016, \apj, 818, 200

\bibitem[{{Estrada} {et~al.}(2022){Estrada}, {Cuzzi}, \&
  {Umurhan}}]{Estradaetal2021}
{Estrada}, P.~R., {Cuzzi}, J.~N., \& {Umurhan}, O.~M. 2022, \apj, in
  preparation

\bibitem[{{Flaherty} {et~al.}(2015){Flaherty}, {Hughes}, {Rosenfeld},
  {Andrews}, {Chiang}, {Simon}, {Kerzner}, \& {Wilner}}]{flaherty2015}
{Flaherty}, K.~M., {Hughes}, A.~M., {Rosenfeld}, K.~A., {et~al.} 2015, \apj,
  813, 99

\bibitem[{{Flaherty} {et~al.}(2017){Flaherty}, {Hughes}, {Rose}, {Simon}, {Qi},
  {Andrews}, {K{\'o}sp{\'a}l}, {Wilner}, {Chiang}, {Armitage}, \&
  {Bai}}]{flaherty2017}
{Flaherty}, K.~M., {Hughes}, A.~M., {Rose}, S.~C., {et~al.} 2017, \apj, 843,
  150

\bibitem[{{Freedman} {et~al.}(2014){Freedman}, {Lustig-Yaeger}, {Fortney},
  {Lupu}, {Marley}, \& {Lodders}}]{Fre14}
{Freedman}, R.~S., {Lustig-Yaeger}, J., {Fortney}, J.~J., {et~al.} 2014, \apjs,
  214, 25

\bibitem[{{Fukuda} {et~al.}(2021){Fukuda}, {Tenner}, {Kimura}, {Tomioka},
  {Siron}, {Ushikubo}, {Chaumard}, {Hertwig}, \& {Kita}}]{Fukudaetal2021}
{Fukuda}, K., {Tenner}, T.~J., {Kimura}, M., {et~al.} 2021, in LPI
  Contributions, Vol.~84, 84th Annual Meeting of the Meteoritical Society, 6030

\bibitem[{{Gammie}(1996)}]{Gammie1996}
{Gammie}, C.~F. 1996, \apj, 457, 355

\bibitem[{{Giacalone} {et~al.}(2019){Giacalone}, {Teitler}, {K{\"o}nigl},
  {Krijt}, \& {Ciesla}}]{Giacaloneetal2019}
{Giacalone}, S., {Teitler}, S., {K{\"o}nigl}, A., {Krijt}, S., \& {Ciesla},
  F.~J. 2019, \apj, 882, 33

\bibitem[{{Grossman} \& {Larimer}(1974)}]{GrossmanLarimer1974}
{Grossman}, L., \& {Larimer}, J.~W. 1974, Reviews of Geophysics and Space
  Physics, 12, 71

\bibitem[{{Hartmann} {et~al.}(1998){Hartmann}, {Calvet}, {Gullbring}, \&
  {D'Alessio}}]{Hartmannetal1998}
{Hartmann}, L., {Calvet}, N., {Gullbring}, E., \& {D'Alessio}, P. 1998, \apj,
  495, 385

\bibitem[{{Hertwig} {et~al.}(2018){Hertwig}, {Defouilloy}, \&
  {Kita}}]{Hertwigetal2018}
{Hertwig}, A.~T., {Defouilloy}, C., \& {Kita}, N.~T. 2018, \gca, 224, 116

\bibitem[{{Hertwig} {et~al.}(2019){Hertwig}, {Kimura}, {Defouilloy}, \&
  {Kita}}]{Hertwigetal2019}
{Hertwig}, A.~T., {Kimura}, M., {Defouilloy}, C., \& {Kita}, N.~T. 2019, \maps,
  54, 2666

\bibitem[{{Hewins} {et~al.}(2014){Hewins}, {Bourot-Denise}, {Zand a}, {Leroux},
  {Barrat}, {Humayun}, {G{\"o}pel}, {Greenwood}, {Franchi}, {Pont}, {Lorand},
  {Courn{\`e}de}, {Gattacceca}, {Rochette}, {Kuga}, {Marrocchi}, \&
  {Marty}}]{Hewinsetal2014}
{Hewins}, R.~H., {Bourot-Denise}, M., {Zand a}, B., {et~al.} 2014, \gca, 124,
  190

\bibitem[{{Hezel} {et~al.}(2008){Hezel}, {Russell}, {Ross}, \&
  {Kearsley}}]{Hezeletal2008}
{Hezel}, D.~C., {Russell}, S.~S., {Ross}, A.~J., \& {Kearsley}, A.~T. 2008,
  \maps, 43, 1879

\bibitem[{{Hueso} \& {Guillot}(2005)}]{HuesoGuillot2005}
{Hueso}, R., \& {Guillot}, T. 2005, \aap, 442, 703

\bibitem[{{Humayun}(2012)}]{Humayun2012}
{Humayun}, M. 2012, Meteoritics and Planetary Science, 47, 1191

\bibitem[{{Humayun} \& {Cassen}(2000)}]{HumayunCassen2000}
{Humayun}, M., \& {Cassen}, P. 2000, in Origin of the Earth and Moon, ed. R.~M.
  {Canup}, K.~{Righter}, \& {et al.}, 3--23

\bibitem[{{Humayun} \& {Clayton}(1995)}]{HumayunClayton1995}
{Humayun}, M., \& {Clayton}, R.~N. 1995, \gca, 59, 2131

\bibitem[{{Huss}(2004)}]{Huss2004}
{Huss}, G.~R. 2004, Antarctic Meteorite Research, 17, 132

\bibitem[{{Huss} {et~al.}(2003){Huss}, {Meshik}, {Smith}, \&
  {Hohenberg}}]{Hussetal2003}
{Huss}, G.~R., {Meshik}, A.~P., {Smith}, J.~B., \& {Hohenberg}, C.~M. 2003,
  \gca, 67, 4823

\bibitem[{{Jacquet}(2014)}]{Jacquet2014}
{Jacquet}, E. 2014, Comptes Rendus Geoscience, 346, 3

\bibitem[{{Jacquet} {et~al.}(2012){Jacquet}, {Gounelle}, \&
  {Fromang}}]{Jacquetetal2012}
{Jacquet}, E., {Gounelle}, M., \& {Fromang}, S. 2012, \icarus, 220, 162

\bibitem[{{Jacquet} {et~al.}(2013){Jacquet}, {Paulhiac-Pison}, {Alard},
  {Kearsley}, \& {Gounelle}}]{Jacquetetal2013}
{Jacquet}, E., {Paulhiac-Pison}, M., {Alard}, O., {Kearsley}, A.~T., \&
  {Gounelle}, M. 2013, \maps, 48, 1981

\bibitem[{{Jacquet} {et~al.}(2019){Jacquet}, {Pignatale}, {Chaussidon}, \&
  {Charnoz}}]{Jacquetetal2019}
{Jacquet}, E., {Pignatale}, F.~C., {Chaussidon}, M., \& {Charnoz}, S. 2019,
  \apj, 884, 32

\bibitem[{{Jessberger} {et~al.}(1988){Jessberger}, {Christoforidis}, \&
  {Kissel}}]{Jessbergeretal1988}
{Jessberger}, E.~K., {Christoforidis}, A., \& {Kissel}, J. 1988, \nat, 332, 691

\bibitem[{{Jessberger} {et~al.}(2001){Jessberger}, {Stephan}, {Rost}, {Arndt},
  {Maetz}, {Stadermann}, {Brownlee}, {Bradley}, \&
  {Kurat}}]{Jessbergeretal2001}
{Jessberger}, E.~K., {Stephan}, T., {Rost}, D., {et~al.} 2001, {Properties of
  Interplanetary Dust: Information from Collected Samples}, 253

\bibitem[{{Johansen} {et~al.}(2015){Johansen}, {Jacquet}, {Cuzzi},
  {Morbidelli}, \& {Gounelle}}]{Johansenetal2015}
{Johansen}, A., {Jacquet}, E., {Cuzzi}, J.~N., {Morbidelli}, A., \& {Gounelle},
  M. 2015, {New Paradigms for Asteroid Formation; ArXiv 1703.05871}, 471--492

\bibitem[{{Johnson} {et~al.}(2014){Johnson}, {Bowling}, \&
  {Melosh}}]{Johnsonetal2014}
{Johnson}, B.~C., {Bowling}, T.~J., \& {Melosh}, H.~J. 2014, \icarus, 238, 13

\bibitem[{{Kadam} {et~al.}(2020){Kadam}, {Vorobyov}, {Reg{\'a}ly},
  {K{\'o}sp{\'a}l}, \& {{\'A}brah{\'a}m}}]{kadametal2020}
{Kadam}, K., {Vorobyov}, E., {Reg{\'a}ly}, Z., {K{\'o}sp{\'a}l}, {\'A}., \&
  {{\'A}brah{\'a}m}, P. 2020, \apj, 895, 41

\bibitem[{{Kataoka} {et~al.}(2013){Kataoka}, {Tanaka}, {Okuzumi}, \&
  {Wada}}]{kataokaetal2013}
{Kataoka}, A., {Tanaka}, H., {Okuzumi}, S., \& {Wada}, K. 2013, \aap, 554, A4

\bibitem[{{Kennedy} {et~al.}(1993){Kennedy}, {Lofgren}, \&
  {Wasserburg}}]{Kennedyetal1993}
{Kennedy}, A.~K., {Lofgren}, G.~E., \& {Wasserburg}, G.~J. 1993, Earth and
  Planetary Science Letters, 115, 177

\bibitem[{{Kita} {et~al.}(2013){Kita}, {Yin}, {MacPherson}, {Ushikubo},
  {Jacobsen}, {Nagashima}, {Kurahashi}, {Krot}, \& {Jacobsen}}]{Kitaetal2013}
{Kita}, N.~T., {Yin}, Q.-Z., {MacPherson}, G.~J., {et~al.} 2013, Meteoritics
  and Planetary Science, 48, 1383

\bibitem[{{Koike} {et~al.}(1995){Koike}, {Kaito}, {Yamamoto}, {Shibai},
  {Kimura}, \& {Suto}}]{Koikeetal1995}
{Koike}, C., {Kaito}, C., {Yamamoto}, T., {et~al.} 1995, \icarus, 114, 203

\bibitem[{{Krijt} {et~al.}(2015){Krijt}, {Ormel}, {Dominik}, \&
  {Tielens}}]{Krijtetal2015}
{Krijt}, S., {Ormel}, C.~W., {Dominik}, C., \& {Tielens}, A.~G.~G.~M. 2015,
  \aap, 574, A83

\bibitem[{{Krot} {et~al.}(2020){Krot}, {Nagashima}, {Lyons}, {Lee}, \&
  {Bizzarro}}]{Krotetal2020}
{Krot}, A.~N., {Nagashima}, K., {Lyons}, J.~R., {Lee}, J.-E., \& {Bizzarro}, M.
  2020, Science Advances, 6, eaay2724

\bibitem[{{Krot} {et~al.}(2009){Krot}, {Amelin}, {Bland}, {Ciesla}, {Connelly},
  {Davis}, {Huss}, {Hutcheon}, {Makide}, {Nagashima}, {Nyquist}, {Russell},
  {Scott}, {Thrane}, {Yurimoto}, \& {Yin}}]{Krotetal2009}
{Krot}, A.~N., {Amelin}, Y., {Bland}, P., {et~al.} 2009, \gca, 73, 4963

\bibitem[{{Kruijer} {et~al.}(2017){Kruijer}, {Burkhardt}, {Budde}, \&
  {Kleine}}]{Kruijeretal2017}
{Kruijer}, T.~S., {Burkhardt}, C., {Budde}, G., \& {Kleine}, T. 2017,
  Proceedings of the National Academy of Science, 114, 6712

\bibitem[{{Kunitomo} {et~al.}(2020){Kunitomo}, {Suzuki}, \&
  {Inutsuka}}]{Kunitomoetal2020}
{Kunitomo}, M., {Suzuki}, T.~K., \& {Inutsuka}, S.-i. 2020, \mnras, 492, 3849

\bibitem[{{Larimer}(1967)}]{Larimer1967}
{Larimer}, J.~W. 1967, \gca, 31, 1215

\bibitem[{{Larimer} \& {Anders}(1967)}]{LarimerAnders1967}
{Larimer}, J.~W., \& {Anders}, E. 1967, \gca, 31, 1239

\bibitem[{{Lauretta} {et~al.}(1996){Lauretta}, {Kremser}, \&
  {Fegley}}]{Laurettaetal1996}
{Lauretta}, D.~S., {Kremser}, D.~T., \& {Fegley}, Bruce, J. 1996, \icarus, 122,
  288

\bibitem[{{Li} {et~al.}(2020){Li}, {Huang}, {Petaev}, {Zhu}, \&
  {Steffen}}]{Lietal2020}
{Li}, M., {Huang}, S., {Petaev}, M.~I., {Zhu}, Z., \& {Steffen}, J.~H. 2020,
  \mnras, 495, 2543

\bibitem[{{Li} {et~al.}(2021){Li}, {Huang}, {Zhu}, {Petaev}, \&
  {Steffen}}]{Lietal2021}
{Li}, M., {Huang}, S., {Zhu}, Z., {Petaev}, M.~I., \& {Steffen}, J.~H. 2021,
  \mnras, 503, 5254

\bibitem[{{Lodders}(2003)}]{Lodders2003}
{Lodders}, K. 2003, \apj, 591, 1220

\bibitem[{{Lodders}(2021)}]{Lodders2021}
---. 2021, \ssr, 217, 44

\bibitem[{{Lodders} {et~al.}(2009){Lodders}, {Palme}, \& {Gail}}]{Lodders2009}
{Lodders}, K., {Palme}, H., \& {Gail}, H.~P. 2009, Landolt B\&ouml;rnstein, 4B,
  712

\bibitem[{{Lovelace} {et~al.}(1999){Lovelace}, {Li}, {Colgate}, \&
  {Nelson}}]{lovelace1999}
{Lovelace}, R.~V.~E., {Li}, H., {Colgate}, S.~A., \& {Nelson}, A.~F. 1999,
  \apj, 513, 805

\bibitem[{{Lynden-Bell} \& {Pringle}(1974)}]{LyndenBellPringle1974}
{Lynden-Bell}, D., \& {Pringle}, J.~E. 1974, \mnras, 168, 603

\bibitem[{{Lyra}(2014)}]{Lyra2014}
{Lyra}, W. 2014, \apj, 789, 77

\bibitem[{{Mac{\'\i}as} {et~al.}(2021){Mac{\'\i}as}, {Guerra-Alvarado},
  {Carrasco-Gonz{\'a}lez}, {Ribas}, {Espaillat}, {Huang}, \&
  {Andrews}}]{Maciasetal2021}
{Mac{\'\i}as}, E., {Guerra-Alvarado}, O., {Carrasco-Gonz{\'a}lez}, C., {et~al.}
  2021, \aap, 648, A33

\bibitem[{{MacPherson}(2017)}]{Macpherson2017}
{MacPherson}, G.~J. 2017, in Lunar and Planetary Science Conference, Lunar and
  Planetary Science Conference, 2719

\bibitem[{{Marcus} {et~al.}(2015){Marcus}, {Pei}, {Jiang}, {Barranco},
  {Hassanzadeh}, \& {Lecoanet}}]{Marcusetal2015}
{Marcus}, P.~S., {Pei}, S., {Jiang}, C.-H., {et~al.} 2015, \apj, 808, 87

\bibitem[{{Marrocchi} {et~al.}(2018){Marrocchi}, {Bekaert}, \&
  {Piani}}]{Marrocchietal2018}
{Marrocchi}, Y., {Bekaert}, D.~V., \& {Piani}, L. 2018, Earth and Planetary
  Science Letters, 482, 23

\bibitem[{{Marrocchi} {et~al.}(2019){Marrocchi}, {Villeneuve}, {Jacquet},
  {Piralla}, \& {Chaussidon}}]{Marrochietal2019}
{Marrocchi}, Y., {Villeneuve}, J., {Jacquet}, E., {Piralla}, M., \&
  {Chaussidon}, M. 2019, Proceedings of the National Academy of Science, 116,
  23461

\bibitem[{{McCubbin} \& {Barnes}(2019)}]{McCubbinBarnes2019}
{McCubbin}, F.~M., \& {Barnes}, J.~J. 2019, Earth and Planetary Science
  Letters, 526, 115771

\bibitem[{{McDonough} \& {Sun}(1995)}]{McDonoughSun1995}
{McDonough}, W.~F., \& {Sun}, S.~s. 1995, Chemical Geology, 120, 223

\bibitem[{{Morfill}(1988)}]{Morfill1988}
{Morfill}, G.~E. 1988, \icarus, 75, 371

\bibitem[{{Moriarty} {et~al.}(2014){Moriarty}, {Madhusudhan}, \&
  {Fischer}}]{Moriartyetal2014}
{Moriarty}, J., {Madhusudhan}, N., \& {Fischer}, D. 2014, \apj, 787, 81

\bibitem[{{Mumma} \& {Charnley}(2011)}]{MummaCharnley2011}
{Mumma}, M.~J., \& {Charnley}, S.~B. 2011, \araa, 49, 471

\bibitem[{{Nakamoto} \& {Nakagawa}(1994)}]{NakamotoNakagawa1994}
{Nakamoto}, T., \& {Nakagawa}, Y. 1994, \apj, 421, 640

\bibitem[{{Nanne} {et~al.}(2019){Nanne}, {Nimmo}, {Cuzzi}, \&
  {Kleine}}]{Nanneetal2019}
{Nanne}, J. A.~M., {Nimmo}, F., {Cuzzi}, J.~N., \& {Kleine}, T. 2019, Earth and
  Planetary Science Letters, 511, 44

\bibitem[{{Nelson} {et~al.}(2013){Nelson}, {Gressel}, \&
  {Umurhan}}]{Nelsonetal2013}
{Nelson}, R.~P., {Gressel}, O., \& {Umurhan}, O.~M. 2013, \mnras, 435, 2610

\bibitem[{{Norris} \& {Wood}(2017)}]{NorrisWood2017}
{Norris}, C.~A., \& {Wood}, B.~J. 2017, \nat, 549, 507

\bibitem[{{Okuzumi} {et~al.}(2012){Okuzumi}, {Tanaka}, {Kobayashi}, \&
  {Wada}}]{Okuzumietal2012}
{Okuzumi}, S., {Tanaka}, H., {Kobayashi}, H., \& {Wada}, K. 2012, \apj, 752,
  106

\bibitem[{{Ormel} \& {Cuzzi}(2007)}]{OrmelCuzzi2007}
{Ormel}, C.~W., \& {Cuzzi}, J.~N. 2007, \aap, 466, 413

\bibitem[{{Ormel} {et~al.}(2007){Ormel}, {Spaans}, \&
  {Tielens}}]{Ormeletal2007}
{Ormel}, C.~W., {Spaans}, M., \& {Tielens}, A.~G.~G.~M. 2007, \aap, 461, 215

\bibitem[{{Palme}(2001)}]{Palme2001}
{Palme}, H. 2001, Philosophical Transactions of the Royal Society of London
  Series A, 359, 2061

\bibitem[{{Palme} {et~al.}(1988){Palme}, {Larimer}, \&
  {Lipschutz}}]{Palmeetal88}
{Palme}, H., {Larimer}, J.~W., \& {Lipschutz}, M.~E. 1988, {Moderately volatile
  elements.}, ed. J.~F. {Kerridge} \& M.~S. {Matthews}, 436--461

\bibitem[{{Pignatale} {et~al.}(2019){Pignatale}, {Jacquet}, {Chaussidon}, \&
  {Charnoz}}]{Pignataleetal2019}
{Pignatale}, F.~C., {Jacquet}, E., {Chaussidon}, M., \& {Charnoz}, S. 2019,
  \apj, 884, 31

\bibitem[{{Pinte} {et~al.}(2016){Pinte}, {Dent}, {M{\'e}nard}, {Hales}, {Hill},
  {Cortes}, \& {de Gregorio-Monsalvo}}]{pinte2016}
{Pinte}, C., {Dent}, W.~R.~F., {M{\'e}nard}, F., {et~al.} 2016, \apj, 816, 25

\bibitem[{{Pollack} {et~al.}(1994){Pollack}, {Hollenbach}, {Beckwith},
  {Simonelli}, {Roush}, \& {Fong}}]{Pollacketal1994}
{Pollack}, J.~B., {Hollenbach}, D., {Beckwith}, S., {et~al.} 1994, \apj, 421,
  615

\bibitem[{{Pringle} \& {Moynier}(2017)}]{PringleMoynier2017}
{Pringle}, E.~A., \& {Moynier}, F. 2017, Earth and Planetary Science Letters,
  473, 62

\bibitem[{{Pringle}(1981)}]{Pringle1981}
{Pringle}, J.~E. 1981, \araa, 19, 137

\bibitem[{{Righter} {et~al.}(2005){Righter}, {Campbell}, \&
  {Humayun}}]{Righteretal2005}
{Righter}, K., {Campbell}, A.~J., \& {Humayun}, M. 2005, \gca, 69, 3145

\bibitem[{{Sanders} \& {Scott}(2012{\natexlab{a}})}]{SandersScott2012a}
{Sanders}, I.~S., \& {Scott}, E.~R.~D. 2012{\natexlab{a}}, Meteoritics and
  Planetary Science Supplement, 75, 5177

\bibitem[{{Sanders} \& {Scott}(2012{\natexlab{b}})}]{SandersScott2012b}
{Sanders}, I.~S., \& {Scott}, E. R.~D. 2012{\natexlab{b}}, Meteoritics and
  Planetary Science, 47, 2170

\bibitem[{{Scott}(2006)}]{Scott2006}
{Scott}, E. R.~D. 2006, \icarus, 185, 72

\bibitem[{{Sengupta} {et~al.}(2019){Sengupta}, {Dodson-Robinson}, {Hasegawa},
  \& {Turner}}]{Senguptaetal2019}
{Sengupta}, D., {Dodson-Robinson}, S.~E., {Hasegawa}, Y., \& {Turner}, N.~J.
  2019, \apj, 874, 26

\bibitem[{{Sheikh} \& {Humayun}(2021)}]{SheikhHumayun2021}
{Sheikh}, D., \& {Humayun}, M. 2021, in 52nd Lunar and Planetary Science
  Conference, Lunar and Planetary Science Conference, 1279

\bibitem[{{Sierra} {et~al.}(2021){Sierra}, {P{\'e}rez}, {Zhang}, {Law},
  {Guzm{\'a}n}, {Qi}, {Bosman}, {{\"O}berg}, {Andrews}, {Long}, {Teague},
  {Booth}, {Walsh}, {Wilner}, {M{\'e}nard}, {Cataldi}, {Czekala}, {Bae},
  {Huang}, {Bergner}, {Ilee}, {Benisty}, {Le Gal}, {Loomis}, {Tsukagoshi},
  {Liu}, {Yamato}, \& {Aikawa}}]{Sierraetal2021}
{Sierra}, A., {P{\'e}rez}, L.~M., {Zhang}, K., {et~al.} 2021, \apjs, 257, 14

\bibitem[{{Siess} {et~al.}(2002){Siess}, {Livio}, \& {Lattanzio}}]{Siessetal02}
{Siess}, L., {Livio}, M., \& {Lattanzio}, J. 2002, \apj, 570, 329

\bibitem[{{Simon} {et~al.}(2011){Simon}, {Hutcheon}, {Simon}, {Matzel},
  {Ramon}, {Weber}, {Grossman}, \& {DePaolo}}]{Simonetal2011}
{Simon}, J.~I., {Hutcheon}, I.~D., {Simon}, S.~B., {et~al.} 2011, Science, 331,
  1175

\bibitem[{{Simon} {et~al.}(2019){Simon}, {Ross}, {Nguyen}, {Simon}, \&
  {Messenger}}]{Simonetal2019}
{Simon}, J.~I., {Ross}, D.~K., {Nguyen}, A.~N., {Simon}, S.~B., \& {Messenger},
  S. 2019, \apjl, 884, L29

\bibitem[{{Simon} {et~al.}(2018){Simon}, {Cuzzi}, {McCain}, {Cato},
  {Christoffersen}, {Fisher}, {Srinivasan}, {Tait}, {Olson}, \&
  {Scargle}}]{Simonetal2018}
{Simon}, J.~I., {Cuzzi}, J.~N., {McCain}, K.~A., {et~al.} 2018, Earth and
  Planetary Science Letters, 494, 69

\bibitem[{{Speck} {et~al.}(2011){Speck}, {Whittington}, \&
  {Hofmeister}}]{Speck_etal_2011}
{Speck}, A.~K., {Whittington}, A.~G., \& {Hofmeister}, A.~M. 2011, \apj, 740,
  93

\bibitem[{{Stoll} \& {Kley}(2014)}]{Stoll_Kley_2014}
{Stoll}, M. H.~R., \& {Kley}, W. 2014, \aap, 572, A77

\bibitem[{{Suyama} {et~al.}(2012){Suyama}, {Wada}, {Tanaka}, \&
  {Okuzumi}}]{Suyamaetal2012}
{Suyama}, T., {Wada}, K., {Tanaka}, H., \& {Okuzumi}, S. 2012, \apj, 753, 115

\bibitem[{{Suzuki} {et~al.}(2016){Suzuki}, {Ogihara}, {Morbidelli}, {Crida}, \&
  {Guillot}}]{Suzukietal2016}
{Suzuki}, T.~K., {Ogihara}, M., {Morbidelli}, A.~r., {Crida}, A., \& {Guillot},
  T. 2016, \aap, 596, A74

\bibitem[{{Teitler}(2011)}]{teitler2011}
{Teitler}, S. 2011, \apj, 733, 57

\bibitem[{{Tenner} {et~al.}(2015){Tenner}, {Nakashima}, {Ushikubo}, {Kita}, \&
  {Weisberg}}]{Tenneretal2015}
{Tenner}, T.~J., {Nakashima}, D., {Ushikubo}, T., {Kita}, N.~T., \& {Weisberg},
  M.~K. 2015, \gca, 148, 228

\bibitem[{{Turner} {et~al.}(2014){Turner}, {Fromang}, {Gammie}, {Klahr},
  {Lesur}, {Wardle}, \& {Bai}}]{Turneretal2014}
{Turner}, N.~J., {Fromang}, S., {Gammie}, C., {et~al.} 2014, in Protostars and
  Planets VI, ed. H.~{Beuther}, R.~S. {Klessen}, C.~P. {Dullemond}, \&
  T.~{Henning}, 411

\bibitem[{{van Kooten} {et~al.}(2019){van Kooten}, {Moynier}, \&
  {Agranier}}]{vanKooten2019}
{van Kooten}, E. M.~M.~E., {Moynier}, F., \& {Agranier}, A. 2019, Proceedings
  of the National Academy of Science, 116, 18860

\bibitem[{{Vernazza} {et~al.}(2014){Vernazza}, {Zanda}, {Binzel}, {Hiroi},
  {DeMeo}, {Birlan}, {Hewins}, {Ricci}, {Barge}, \&
  {Lockhart}}]{Vernazzaetal2014}
{Vernazza}, P., {Zanda}, B., {Binzel}, R.~P., {et~al.} 2014, \apj, 791, 120

\bibitem[{{Visser} {et~al.}(2009){Visser}, {van Dishoeck}, {Doty}, \&
  {Dullemond}}]{Visseretal2009}
{Visser}, R., {van Dishoeck}, E.~F., {Doty}, S.~D., \& {Dullemond}, C.~P. 2009,
  \aap, 495, 881

\bibitem[{{Vorobyov} \& {Basu}(2005)}]{VorobyovBasu2005}
{Vorobyov}, E.~I., \& {Basu}, S. 2005, \apjl, 633, L137

\bibitem[{{Vorobyov} \& {Basu}(2006)}]{VorobyovBasu2006}
---. 2006, \apj, 650, 956

\bibitem[{{Wai} \& {Wasson}(1977)}]{WaiWasson1977}
{Wai}, C.~M., \& {Wasson}, J.~T. 1977, Earth and Planetary Science Letters, 36,
  1

\bibitem[{{Warren}(2011)}]{Warren2011}
{Warren}, P.~H. 2011, Earth and Planetary Science Letters, 311, 93

\bibitem[{{Wasson}(1977)}]{Wasson1977}
{Wasson}, J.~T. 1977, Earth and Planetary Science Letters, 36, 21

\bibitem[{{Wasson}(1985)}]{Wasson1985}
---. 1985, {Meteorites : their record of early solar-system history}

\bibitem[{{Wasson} \& {Chou}(1974)}]{WassonChou1974}
{Wasson}, J.~T., \& {Chou}, C.-L. 1974, Meteoritics, 9, 69

\bibitem[{{Williams} {et~al.}(2020){Williams}, {Sanborn}, {Defouilloy}, {Yin},
  {Kita}, {Ebel}, {Yamakawa}, \& {Yamashita}}]{Williamsetal2020}
{Williams}, C.~D., {Sanborn}, M.~E., {Defouilloy}, C., {et~al.} 2020,
  Proceedings of the National Academy of Science, 117, 23426

\bibitem[{{Wood} \& {Halliday}(2005)}]{WoodHalliday2005}
{Wood}, B.~J., \& {Halliday}, A.~N. 2005, \nat, 437, 1345

\bibitem[{{Woodward} {et~al.}(2020){Woodward}, {Wooden}, {Harker}, {Kelley},
  {Russell}, \& {Kim}}]{Woodwardetal2020}
{Woodward}, C.~E., {Wooden}, D.~H., {Harker}, D.~E., {et~al.} 2020, arXiv
  e-prints, arXiv:2011.06943

\bibitem[{{Woolum} \& {Cassen}(1999)}]{WoolumCassen1999}
{Woolum}, D.~S., \& {Cassen}, P. 1999, Meteoritics and Planetary Science, 34,
  897

\bibitem[{{Yang} \& {Ciesla}(2012)}]{YangCiesla2012}
{Yang}, L., \& {Ciesla}, F.~J. 2012, Meteoritics and Planetary Science, 47, 99

\bibitem[{{Yin}(2005)}]{Yin2005}
{Yin}, Q. 2005, in Astronomical Society of the Pacific Conference Series, Vol.
  341, Chondrites and the Protoplanetary Disk, ed. A.~N. {Krot}, E.~R.~D.
  {Scott}, \& B.~{Reipurth}, 632

\bibitem[{{Yoshizaki} \& {McDonough}(2020)}]{YoshizakiMcDonough2020}
{Yoshizaki}, T., \& {McDonough}, W.~F. 2020, \gca, 273, 137

\bibitem[{{Zanda} {et~al.}(2009){Zanda}, {Bland}, {Le Guillou}, \&
  {Hewins}}]{Zandaetal2009}
{Zanda}, B., {Bland}, P.~A., {Le Guillou}, C., \& {Hewins}, R.~H. 2009, in
  Lunar and Planetary Science Conference, Lunar and Planetary Science
  Conference, 1810

\bibitem[{{Zanda} {et~al.}(2006){Zanda}, {Hewins}, {Bourot-Denise}, {Bland}, \&
  {Albar{\`e}de}}]{Zandaetal2006}
{Zanda}, B., {Hewins}, R.~H., {Bourot-Denise}, M., {Bland}, P.~A., \&
  {Albar{\`e}de}, F. 2006, Earth and Planetary Science Letters, 248, 650

\bibitem[{{Zanda} {et~al.}(2011{\natexlab{a}}){Zanda}, {Humayun}, {Barrat},
  {Bourot-Denise}, \& {Hewins}}]{Zandaetal2011_Paris}
{Zanda}, B., {Humayun}, M., {Barrat}, J.~A., {Bourot-Denise}, M., \& {Hewins},
  R. 2011{\natexlab{a}}, in Lunar and Planetary Science Conference, Lunar and
  Planetary Science Conference, 2040

\bibitem[{{Zanda} {et~al.}(2011{\natexlab{b}}){Zanda}, {Humayun}, {Barrat},
  {Bourot-Denise}, \& {Hewins}}]{Zandaetal2011_complementarity}
{Zanda}, B., {Humayun}, M., {Barrat}, J.~A., {Bourot-Denise}, M., \& {Hewins},
  R.~H. 2011{\natexlab{b}}, Meteoritics and Planetary Science Supplement, 74,
  5358

\bibitem[{{Zanda} {et~al.}(2012){Zanda}, {Humayun}, \&
  {Hewins}}]{Zandaetal2012}
{Zanda}, B., {Humayun}, M., \& {Hewins}, R.~H. 2012, in Lunar and Planetary
  Science Conference, Lunar and Planetary Science Conference, 2413

\bibitem[{{Zanda} {et~al.}(2018){Zanda}, {Lewin}, \& {Humayun}}]{Zandaetal2018}
{Zanda}, B., {Lewin}, {\'E}., \& {Humayun}, M. 2018, {The Chondritic
  Assemblage. Complementarity Is Not A Required Hypothesis}, ed. S.~S.
  {Russell}, J.~{Connolly}, Harold~C., \& A.~N. {Krot}, 122--150

\bibitem[{{Zhu} {et~al.}(2010){Zhu}, {Hartmann}, \& {Gammie}}]{Zhuetal2010}
{Zhu}, Z., {Hartmann}, L., \& {Gammie}, C. 2010, \apj, 713, 1143

\end{thebibliography}
%\bibliography{reference}{}
%\bibliographystyle{aasjournal}

\end{document}